%% file: ms_pow.tex
\documentclass[11pt,a4paper]{article}

\usepackage{jcappub}

\allowdisplaybreaks

\usepackage{ifthen}
\usepackage{graphicx}% Include figure files
\usepackage{dcolumn}% Align table columns on decimal point
\usepackage{bm}% bold math
\usepackage{color}
\usepackage{amsmath}
\usepackage{amssymb}
\usepackage{bbold}
\usepackage{wrapfig}
\usepackage[titletoc]{appendix}
\usepackage{multirow}
\usepackage{adjustbox}
\usepackage{epstopdf}
\input{variables}

\newcommand{\no}[1]{}%coments section of text
\newcommand{\Lya}{Ly$\alpha$\ } 
\newcommand{\pow}{power spectrum\ }
\begin{document}

\graphicspath{ {./plots/} }

\title{The Non-Linear Power Spectrum of the Lyman Alpha Forest}

\input{authors}

\input{abstract}

\maketitle

\input{introduction}
\input{biasfactors}
\input{method}
\input{tests}
\input{physical}

\input{discussion}

\input{conclusions}

\input{acknowledgments}

\bibliographystyle{apj}
\newpage
\bibliography{ms_pow}
\newpage
\input{appendix}

%\nocite{*}

%\appendix
%\input{Correction}

\end{document}

%% file: variables.tex
\newcommand{\mpc}{\, {\rm Mpc}}

\newcommand{\hmpc}{\, h^{-1} \mpc}

\newcommand{\kms}{\, {\rm km\, s}^{-1}}

\newcommand{\lya}{Ly$\alpha$\ }

%% file: authors.tex
\author[a]{Andreu Arinyo-i-Prats}
\author[b,a]{, Jordi Miralda-Escud\'e}
\author[c,d]{, Matteo Viel}
\author[e]{ and Renyue Cen}

\affiliation[a]{Institut de Ci\`encies del Cosmos, Universitat de Barcelona, IEEC-UB, Barcelona 08028, Catalonia.}
\affiliation[b]{Instituci\'o Catalana de Recerca i Estudis Avan\c{c}ats, Barcelona, Catalonia.}
\affiliation[c]{INAF, Astronomical Observatory of Trieste, 34131 Trieste, Italy}
\affiliation[d]{INFN, Sezione di Trieste, 34100 Trieste, Italy}
\affiliation[e]{Princeton University Observatory, Princeton, NJ 08544, USA}

\emailAdd{andreuaprats@gmail.com }
\emailAdd{miralda@icc.ub.edu}
\emailAdd{viel@oats.inaf.it}
\emailAdd{cen@astro.princeton.edu}

%% file: abstract.tex
\abstract{
The Lyman alpha forest power spectrum has been measured on large scales by
the BOSS survey in SDSS-III at $z\sim 2.3$, has been shown to agree well
with linear theory predictions, and has provided the first measurement of
Baryon Acoustic Oscillations at this redshift. However, the power at small
scales, affected by non-linearities, has not been well examined so far.
We present results from a variety of hydrodynamic simulations to predict
the redshift space non-linear power spectrum of the \lya transmission
for several models, testing the dependence on resolution and box
size. A new fitting formula is introduced to facilitate the comparison of
our simulation results with observations and other simulations. The
non-linear power spectrum has a generic shape determined by a transition
scale from linear to non-linear anisotropy, and a Jeans scale below which
the power drops rapidly. In addition, we predict the two linear bias
factors of the \lya forest and provide a better physical interpretation of
their values and redshift evolution. The dependence of these bias factors
and the non-linear power on the amplitude and slope of the primordial
fluctuations power spectrum, the temperature-density relation of the
intergalactic medium, and the mean \lya transmission, as well as the
redshift evolution, is investigated and discussed in detail. A preliminary
comparison to the observations shows that the predicted redshift distortion
parameter is in good agreement with the recent determination of Blomqvist
et al., but the density bias factor is lower than observed. We make all our
results publicly available in the form of tables of the non-linear power
spectrum that is directly obtained from all our simulations, and parameters
of our fitting formula.

}

\keywords{Cosmology, intergalactic medium, cosmological simulations, quasars: absorption spectra}

%% file: introduction.tex
\section{Introduction}
\label{sec:intro}

  The \lya forest is the main observational probe for
studying the structure and evolution of the intergalactic medium (IGM). 
The transmitted \lya flux observed in the spectrum of a distant source
(typically a quasar) provides us with a one-dimensional map of
absorption along the line of sight (LOS), with the observed wavelength
corresponding to the redshift of the intervening neutral hydrogen
causing the \lya scattering (e.g. \citep{rauch98,meiksin09}).

  The way in which the flux transmission fraction is related to
the underlying density field,
temperature and peculiar velocity gradient of the gas on small scales is
non-linear, and can be modelled in detail only from hydrodynamic
cosmological simulations \citep{Cen94, Zhang95, Hern96, MCOR96, Theuns98, McD00, Croft02, VHS04, Lukic15}.
However, in the limit of large scales, the transmission fraction
averaged over a large region depends linearly on the mean overdensity
and peculiar velocity gradient in the region, and the power spectrum of
the transmission is simply proportional to the power spectrum of mass
fluctuations, with the standard redshift distortions that were predicted
initially for galaxy surveys \citep{kaiser87,Hamilton92}.

This simple linear treatment of the \lya forest applicable on large
scales has led to its use as a cosmological tool to measure the power
spectrum, first from few tens of quasar spectra \citep[where only the
projected one-dimensional power as a function of the parallel Fourier
component is measured; see][]{Croft98,McD00,Croft02,VHS04,McD06}
and then in full redshift space, where the correlation in the
transmission among parallel lines of sight is used \citep{Slosar11}.
As proposed in \cite{McD07}, and implemented in the Baryon Oscillation
Spectroscopic Survey \citep[BOSS; see][]{Dawson13} of the Sloan Digital
Sky Survey-III \citep[SDSS-III, see][]{eisenstein11}, the \lya power
spectrum is proving to be a powerful tool to measure the general
large-scale matter power spectrum, and in particular to
measure the baryon acoustic oscillation scale, at a relatively high
redshift which has not so far been probed by other observables.
This measurement provides geometrical constraints on the expansion rate
and the angular diameter distance as a function of redshift
\citep{busca13,slosar13,font14,delubac14}.

In contrast to the large scales, the \lya power spectrum at small
scales is affected by a variety of non-linear physical processes governing
the evolution of the IGM. These physical processes are highly
complex, and they may include several phenomena related to the formation
of stars and quasars in galaxies that can perturb the IGM. Some examples
are reionization and the inhomogeneous heating caused by it, and the
hydrodynamic effects from galactic winds and quasar jets. There is,
however, a more simple assumption that can be made for the evolution of
the IGM: that the ionization of the IGM is caused only by a nearly
uniform radiation background, which produces a nearly uniform heating, and
that shock waves arise only from the gravitational collapse of
structure formation and not from the ejection of any gas from galaxies due to
supernovae-driven winds or quasars. Even though it is known that quasar
jets and galaxy winds are present in the universe and they have some
impact on the IGM, the volume they affect may in practice be very small \citep{theuns02,mcdonald05,viel13},
and it is useful to test first the most simple assumption for the
evolution of the IGM against the observations. This simple model
should be mostly described by only five parameters,
which determine the statistical properties of the \lya forest: 
\begin{itemize}
\item The mean transmission fraction, $\bar F(z)$, which depends on the
intensity of the cosmic ionizing background and is directly measured
in the observations (except for uncertainties related to continuum fitting).
\item The density-temperature relation, usually parameterized by
the two parameters $T_0$ and $\gamma$ in the power-law relation
$T=T_0 (1+\delta)^{\gamma}$, where
$\delta=\rho/\bar\rho -1$ is the gas overdensity. When the IGM
is heated in photoionization equilibrium and cools adiabatically due to
Hubble expansion, one expects this power-law relation to hold with
$\gamma\simeq 0.6$, but the relation may be altered by the heating due
to HeII reionization \citep{MR94,Hui97,mcquinn09}.% We will discuss this below.
\item The mass power spectrum of primordial perturbations near the
Jeans scale of the IGM, $\lambda_J=2\pi/k_J$, which we
can parameterize also with two parameters as a power-law with free
amplitude and index, $P(k)=A_\alpha (k/k_J)^{n_{\alpha}}$. The
characteristic Jeans scale at the mean density is related to the IGM
temperature, although in detail it depends
also on the entire thermal history \citep{Gnedin98}, and therefore may
be considered as a sixth parameter.
\end{itemize}

  Even though the large-scale properties of the \lya forest are
simply understood from linear theory, there is a strong interest in
understanding the small-scale, non-linear properties as well. There
are several motivations for this: first, we need to test if our
understanding of the IGM in terms of a simple uniform photoionization
as mentioned above is essentially correct, or if there are important
modifications due to a strong impact of galactic winds and jets
\cite[e.g.,][]{Kollmeier06}, or large inhomogeneities due to HeII
reionization
\citep{mcquinn09,Compostella13}. Second, the \lya forest linear power
spectrum depends on two bias factors, with values that can be measured
and can be predicted from an understanding of the small-scale physics.
Finally, the \lya power spectrum transmission power spectrum is being
measured to increasingly high accuracy, both in projection
\citep[from single lines of sight, see][]{McD06,Palanque13}, and in its
full shape in redshift space \citep{Blomqvist15}, and a
detailed comparison of the observations
%(determined from a combination of large-scale surveys like BOSS, which
%provide large numbers of sources, and spectra of quasar pairs at small
%angular separations observed with high signal-to-noise to better probe
%the small-scale transverse correlations), 
with predictions from numerical simulations of the fully non-linear
power spectrum may offer us new clues to essential questions in
cosmology, such as the impact of neutrino masses on the growth of
structure \citep{nathalie15}, or limits on models of warm dark matter
\citep{vielwdm}, or other possible variations on the nature of the dark
matter. There is therefore a need to obtain reliable
theoretical predictions for the non-linear power spectrum of the \lya
transmission fraction as a function of redshift from numerical
simulations of a large array of cosmological models, in terms of the
most important \lya forest parameters mentioned above.

  The goal for the theory of the non-linear \lya forest is comparable to
that of numerical simulations of the hot, X-ray emitting gas in
clusters of galaxies. Detailed determinations of the gas density and
temperature distributions from X-ray observations and the
Sunyaev-Zeldovich effect, together with the mass distribution from
gravitational lensing and the kinematic distribution of galaxies,
have spurred advances in the theoretical modelling of clusters, the
comparison of numerical codes for cosmological simulations, and tests of
the convergence of the results. At present, the abundance of clusters of
galaxies can be used to infer the normalization of the mass power
spectrum, but this determination depends on the uncertain relation
between the observable properties from X-rays, gravitational lensing and
the Sunyaev-Zeldovich temperature decrement to the cluster mass. This
relation needs to be predicted from numerical
simulations, and the theoretical modelling affects the comparison with
the power spectrum normalization derived from CMB observations
\citep[e.g.,][]{Hasselfield13}. Similarly, the \lya forest is sensitive
to the amplitude of the power spectrum and several other cosmological
parameters and physical properties of the IGM, but constraints on these
quantities can only be inferred once we have a reliable understanding
and modelling of non-linear effects on the observed properties of the
\lya forest.

  The aim of this work is to study several cosmological simulations of
the \lya forest for a variety of models, to analyze the non-linear power
spectrum of the \lya transmission that they predict, and to test the
conditions that the simulations must satisfy, in terms of resolution and
simulation volume, to reach convergence of the results. This problem was
first addressed in the pioneering paper of \cite{McD03} (hereafter M03),
and here we attempt to continue this study by examining a large number
of hydrodynamic simulations, characterizing the power spectrum with a
new, simpler fitting formula with several non-linear parameters, and
studying the dependence of the linear bias factors on the IGM
properties. We start in \S 2 by reviewing the definition of the linear
bias factors of the \lya forest and the derivation of the linear power
spectrum, and we introduce alternative bias factors for the \Lya
effective optical depth. The simulations and our technique for measuring
and fitting the power spectrum are explained in \S 3. The results for
the non-linear power spectrum in one specific model, for which we have
run our largest simulation, are presented in \S 4. Several tests of
convergence with the box size and resolution of the simulations are
performed in \S 5, and the results for the power spectrum fits for a
variety of different physical models are presented in \S 6. Finally, the
results are discussed in \S 7 and conclusions are given in \S 8.

%% file: biasfactors.tex
\section{The Bias Factors of the Lyman Alpha Forest}
\label{sec:biasf}

  Before proceeding to describe our analysis of hydrodynamic
simulations, we review here the standard definitions of the power
spectrum and bias factors of the \Lya forest. We also introduce a new
definition for the optical depth bias factors, already discussed in the
context of metal lines in \cite{FM12}, which is useful to
better interpret their values and compare results at
different redshifts and for different types of absorption systems.

  The transmission fraction $F$ in the \Lya forest is the ratio of the
observed flux $f$ that is transmitted from the source to the continuum
flux, $f_c$, when there is no intervening absorption. Usually a model of
the source continuum spectrum is used to calculate $F=f/f_c$ from the
observed flux. The fluctuation in the transmission is defined as
\begin{equation}
 \delta_F = {F\over \bar F(z)} - 1 ~,
\end{equation}
where $\bar F(z)$ is the mean value of the transmission, evaluated as a
function of redshift. The power spectrum $P_F(k,\mu)$ is that of
$\delta_F$ in redshift space, with $\mu$ being the cosine of the angle
of the Fourier wavevector of modulus $k$ from the LOS to the observed
source. On small scales, the value of $\delta_F$ at a certain point
depends on the complex non-linear evolution of the IGM. However, the
average of $\delta_F$ over a large enough scale is related to the mass
density fluctuations according to a linear expression of the form
$\delta_F = b_{F\delta} \delta + b_{F\eta} \eta$. Here, $\delta$ is
the mass density fluctuation, and
\begin{equation}
 \eta = - {1\over aH} {\partial v_p\over \partial x_p}
\label{eq:etadef}
\end{equation}
is the dimensionless gradient of the peculiar velocity $v_p$ along the
LOS, both averaged over the same large region that is used to average
$\delta_F$; $x_p$ is the comoving LOS coordinate, $H$ is the Hubble
constant, and $a$ the scale factor. This is the most general linear
expression for any tracer of large-scale structure. The reason is that
the only scalar quantities that can be constructed from the deformation
tensor, $\phi_{,ij}$ (equal to the second derivatives of the
gravitational potential), and the LOS unit vector, $n_i$, are the trace
of $\phi$, which is proportional to $\delta$, and $n_i n_j \phi_{,ij}$,
which is proportional to $\eta$. The bias factors are simply defined as
the partial derivatives of $\delta_F$ with respect to $\delta$ and
$\eta$, after the large-scale averaging is done:
\begin{equation}
\label{eq:biasf}
 b_{F\delta} = 
%{1\over \bar F}
 {\partial \delta_F \over \partial \delta} ~,
 \qquad\qquad
 b_{F\eta} = 
%{1\over \bar F}
 {\partial \delta_F \over \partial \eta} ~,
\end{equation}
where each partial derivative is understood to be done by holding the
other variable ($\delta$ or $\eta$) constant.

  The bias factors in equation (\ref{eq:biasf}) are not defined in the
same way as for point objects like galaxies. In general, denser regions
or regions where the Hubble expansion rate has slowed down have stronger
absorption, and therefore lower transmission $F$, so the values of
$b_{F\delta}$ and $b_{F\eta}$ are negative. Moreover, when $\bar F$ is close
to unity (which occurs at low redshift), the absolute values of the bias
factors are very small simply because they express the absorption
fluctuation compared to the total transmitted fraction, when the mean
absorption is very small. A physical bias factor should reflect the
fluctuation of a quantity that is zero when the mass density is zero,
whereas the transmitted fraction is obviously one when the gas density
is zero. The quantity that reflects the relative
amount by which the \Lya absorption fluctuates when the mass or the
peculiar velocity gradient fluctuate is obtained from the fluctuation
of an effective optical depth, $\tau_e = - \log F$ (where $F$ has been
averaged over a large, linear scale before taking the logarithm), which
has a relative fluctuation $\delta_\tau = \delta_F/\log\bar F = 
- \delta_F/\bar\tau_e$. The bias factors for this effective optical
depth fluctuation are:
\begin{equation}
\label{eq:btau}
 b_{\tau\delta} = {\partial \delta_\tau \over \partial \delta} =
 {b_{F\delta}\over \log\bar F } ~,
 \qquad\qquad
 b_{\tau\eta} = {\partial \delta_\tau \over \partial \eta} =
 {b_{F\eta}\over \log\bar F } ~.
\end{equation}

  The usefulness of these new definitions is made apparent by
considering various simple models for the \Lya forest. Let us first
imagine that the \Lya absorption systems are clouds of gas distributed
in space with a density bias factor $b_c$; in other words, wherever the
mass density fluctuates by $\delta$, the number density of gas clouds
fluctuates by $b_c \delta$, but their individual absorption line
profiles do not vary. If the internal dynamics of these clouds are not
aligned in a correlated way with the LOS depending on the value of
$\eta$, the effective optical
depth they produce in the spectrum fluctuates as $\delta_\tau= b_c\delta
+ \eta$, because a peculiar velocity gradient is simply squeezing the
absorption lines of the clouds in the spectrum by the factor $1-\eta$,
without altering the absorption line profiles. Hence, for these
population of clouds we have $b_{\tau\delta}=b_c$ and $b_{\tau\eta}=1$:
the density bias factor reflects the true physical bias of the
population of clouds, and the peculiar velocity gradient bias factor is
equal to one, just like for any population of objects that is selected
isotropically, i.e., independently of the LOS direction.

  Next, consider the case where the IGM absorption is optically thin
everywhere. The optical depth is proportional to the density of hydrogen
atoms, which on large scales must again behave like a biased population
of objects with some bias factor $b_a$, and $b_{\tau\eta}$ must again be
unity because the column density of hydrogen atoms in a given interval
of the \Lya spectrum is proportional to the integrated optical depth,
which cannot be modified by any shifts due to peculiar velocities.

  Therefore, if we consider that the true \Lya forest is a combination
of optically thin absorption by the IGM, plus a population of clouds
with optically thick absorption lines that follow a linear bias factor
but do not change their internal properties as a function of $\delta$
and do not have correlated orientations with the principal axes of the
large-scale deformation tensor, we conclude that $b_{\tau\eta}=1$, and
$b_{\tau\delta}$ reflects a true, physical bias factor that results from
a weighted average of the optically thin, intergalactic neutral hydrogen
bias $b_a$, and the gas clouds bias $b_c$.

  This model is of course not exactly correct, because there is an
intermediate density range of gas that is not optically thin and is not
in a population of clouds that have lost any alignment of their internal
dynamics with the surrounding large-scale structure. Nevertheless we can
expect it to provide a first approximation to the reality of the \Lya
forest, and in this way the value of the two bias factors can have a
physical interpretation. Note also that these optical depth bias factors
can be defined for any other set of Lyman series lines and metal lines,
simply by using the appropriate value of $\bar F$ in equation
(\ref{eq:meanF}), as proposed in \cite{FM12}, and then the bias factors
of any hydrogen or metal absorption lines associated with a certain
population of clouds or galaxies are equal to the true bias factors of
these objects.

  The linear power spectrum is derived as in \cite{kaiser87}, from the
simple fact that in the linear regime, $\eta = f(\Omega)\mu^2\delta$ in
Fourier space, where $f(\Omega)=d\log G/d\log a$, and $G(a)$ is the
growth factor. For the linear power spectrum of the transmission
fluctuation $\delta_F$, we have
\begin{equation}
 P_F(k,\mu)=b_{F\delta}^2 (1+\beta \mu^2)^2 P_L(k) + N_0 ~,
\label{eq:lpow}
\end{equation}
where the redshift distortion parameter is
\begin{equation}
\label{eq:rdp}
 \beta= {b_{F\eta}\, f(\Omega)\over b_{F\delta} } ~,
\end{equation}
and $P_L$ is the matter density fluctuation power spectrum. We have
included the intrinsic shot noise term $N_0$, which is a shot noise
that should remain even when $P_F$ is measured from an arbitrarily
dense set of absorption sightlines due to the random nature of the
formation of absorption line systems as the evolution of the IGM
becomes non-linear \citep{PMcD06}, although this is believed to be
very small and we shall ignore it in this paper. This linear power
spectrum is of course valid only in the limit of large scales, and
the models we shall use to fit our simulations include a non-linear
multiplicative term, as described below in equation (\ref{eq:P3Deq}.
The linear power spectrum of the effective optical depth $\delta_\tau$
is the same, except for the normalization, which changes by replacing
$b_{F\delta}$ by $b_{\tau\delta}$. Note that $\beta$ is the same for
the transmission or effective optical depth power spectra, because the
ratio of the two bias factors remains unaltered.

  Finally, we mention also the radiation bias factor $b_{F\Gamma}$,
defined as the variation of $\delta_F$ when the photoionization rate of
hydrogen $\Gamma$, determined by the intensity of the ionizing
background, varies. If $\delta_\Gamma$ is the relative fluctuation of
this photoionization rate, then the total transmission fluctuation
averaged on a large scale is $\delta_F = b_{F\delta}\delta +
b_{F\eta}\eta + b_{F\Gamma} \delta_\Gamma$. The modification of the
power spectrum $P_F$ due to fluctuations in the ionizing radiation
intensity caused by sources that are tracers of the mass density
fluctuations was discussed in \cite{pontzen14,GMB14}. We ignore here the
possible additional effect from HeII reionization of a large-scale
modification of the temperature-density relation. Under the
assumption that the intensity of the ionizing background does not
appreciably affect the temperature and hydrodynamic evolution of the
IGM, and changes the optical depth at every spectral
pixel in inverse proportion to the photoionization rate, then the
variation of the transmission fluctuation with $\delta_\Gamma$ can
be computed in terms of the probability distribution of the transmission
fraction $P(F)$, as
\begin{equation}
\delta_F = {1\over \bar F}\int_0^1 dF\, P(F) 
\exp\left( {\log F \over 1+\delta_\Gamma} \right) -1 =
 - {\int_0^1 dF\, P(F)\, F\log F \over \bar F }\, \delta_\Gamma =
 b_{F\Gamma} \delta_\Gamma ~.
\end{equation}
Just as before, we define the optical depth radiation bias factor
as
\begin{equation}
\label{eq:bfgam}
b_{\tau\Gamma} = {b_{F\Gamma}\over \log \bar F } = 
 - {\int_0^1 dF\, P(F)\, F\log F \over \bar F \log \bar F } ~.
\end{equation}

  An alternative simple model for computing the peculiar velocity
gradient bias factor is to assume that all the fluctuations determining
the \lya forest absorption spectrum can be treated in the linear regime.
Using this assumption, \cite{seljak12} showed that $b_{F\eta}$ should be
given by the same expression for $b_{F\Gamma}$ in equation
(\ref{eq:bfgam}). The reason is easy to understand: for linear
fluctuations, the optical depth at any spectral pixel is simply
multiplied by the factor $(1-\eta)^{-1}\simeq 1+\eta$ under the effect
of a peculiar velocity gradient, so the same derivation shows that
$b_{F\eta} = b_{F\Gamma}$. For the case of radiation fluctuations,
however, the derivation of equation (\ref{eq:bfgam}) does not need to
assume that the \lya forest fluctuations are linear even on small
scales, so the prediction for the radiation bias factor is much more
reliable. We will show in \S \ref{subs:fid} that in fact, $b_{F\eta}$ is
quite different from $b_{F\Gamma}$ because of the non-linearities that
affect the change of small scale \lya forest fluctuations under a
variation of the large-scale peculiar velocity gradient.

%% file: method.tex
\section{Method of Analysis of the Simulations}
\label{sec:Method}

  Our goal in this paper is to use cosmological hydrodynamic simulations
of the IGM to predict the three-dimensional power
spectrum of the Lyman alpha forest in redshift space, $P_F(k,\mu;z)$,
where $k$ and $\mu$ are the modulus and the cosine of the angle from the
LOS of the Fourier mode vector, and $z$ is the redshift. The simulations
used are described in \S \ref{subs:simulations}. The method of analysis
is inspired in that of M03 and is based on the following steps:
(1) starting from a grid of cells containing the hydrodynamic quantities
of gas density, ionized fraction, temperature and velocity at a certain
redshift output of a simulation, the corresponding spectra of \Lya
transmission are computed for the entire grid, using one of the
simulation axes as the assumed LOS direction, and the three-dimensional
Fast Fourier Transform of this transmission field is
obtained, as described in \S \ref{subs:pows};
(2) the mean value of $P_F(k,\mu)$ is computed in bins of
$(k,\mu)$, and errorbars are assigned which take into account
the variance due to the finite simulation volume (\S \ref{subs:bins});
(3) a parameterized fitting function for $P_F(k,\mu)$ is chosen to
obtain best-fit values of the parameters for several simulations
(\S \ref{subs:fiteq}).

\subsection{Simulation characteristics}
\label{subs:simulations}

  Two types of hydrodynamic simulations will be used in this chapter.
Most of the simulations rely on the Tree-PM (Particle Mesh) Smoothed
Particle Hydrodynamics (SPH) {\sc GADGET-II} code \cite{Springel05},
and the bulk of our analysis will be performed on the outputs of these
Lagrangian simulations. One simulation that is based on a fixed-grid
Eulerian code is also used, to allow for a first comparison of the
results for the two types of hydrodynamic numerical methods. This
simulation is described in \cite{Cen10}.

 Table \ref{tab:sims} shows a list of all the simulations that will be
used in this work, including variations in the spatial grid size and the
spectral pixel size for the analysis of the \Lya forest. The first two
columns give the comoving box size, $L$, and the number of dark matter
particles in the SPH simulations (the number of dark matter and gas
particles in the SPH simulations being the same). 
The third one gives the number of cells, $N_c$, in the uniform grid that
is constructed to compute the density, temperature and velocity in real
space. Note that the simulation labelled Euler does not use particles,
and the cells used to run the simulation are $N_c^3$ and are directly
used as this spatial grid. The \Lya spectra are
computed for each of the three axes of the simulation playing the role
of the LOS, with the number of pixels in each spectrum from
each row of $N_c$ cells of length $L$ given in the fourth column; 
generally there are as many pixels in the spectra as cells in the
spatial grid, except in the analysis labelled P1024 where the number of
pixels is doubled. The other columns give values of physical
parameters used in the simulations: the variable $\sigma_8$ parameterizing
the present amplitude of linear perturbations on a sphere of $8\hmpc$,
the mean temperature at
the mean density $T_0$, and the power-law index that fits the
density-temperature relation at low densities, which is described below
in more detail. In general, models have variations of different
parameters around the values of the fiducial model in the first row of
Table \ref{tab:sims}, and they are labelled with
names that refer to the parameter that is being varied.
\input{./tables/sims}

\subsubsection{SPH simulations}\label{subsubs:simmatt}

  All simulations in Table \ref{tab:sims} except for the one denoted as
Euler were run using the publicly available Tree-Particle Mesh Smoothed
Particle Hydrodynamics (SPH) {\sc{GADGET-II}} code \cite{Springel05}.

  The fiducial simulation uses a box of 60 comoving $h^{-1}\, {\rm Mpc}$
and $2\times 512^3$ particles (for the total of gas and dark matter).
Other simulations are run with larger boxes of 80 and 120 $h^{-1}\,
{\rm Mpc}$ (L80 and L120) to test the effect of the missing large-scale power,
or with different resolution to check the convergence as the particle
masses are reduced. The cosmological model is flat $\Lambda$CDM with the
following parameters, using standard notation: $\Omega_{\rm 0m}=0.3$,
$\Lambda =0.7$, $\Omega_{\rm 0b}=0.05$, $H_0=70 \, {\rm km}\,
{\rm s}^{-1} {\rm Mpc}^{-1}$, $n_{\rm s}=1$ and $\sigma_8=0.8778$.
The initial conditions are generated using the software
{\small{CAMB}}{\footnote{http://camb.info/readme.html}} and the Zel'dovich
approximation at the initial redshift of $z=49$. The particle
mesh grid used to calculate the long range forces is chosen to have the
same number of cells as the number of gas particles, $512^3$, while the
gravitational softening is 4 kpc$/h$ in comoving units for the $60\, h^{-1}\,
{\rm Mpc}$ box and scales proportionally to the initial particle
separation for the other simulations. The hydrodynamical processes are
followed according to the prescription of \cite{Katz96}. Star formation
is also included in the model with a simplified prescription that allows to
instantaneously convert into a star particle any gas particle of overdensity
larger than 1000 and temperature colder than 10$^5$ K.
This has been demonstrated to have negligible impact on
the \lya forest transmission power spectrum \citep{VHS04}.

  Most of the SPH simulations in Table \ref{tab:sims} are based on the
same cosmological model as the fiducial one. The exceptions are the
simulations S0.76 and S0.64, where the amplitude of the initial power
spectrum is varied, and the Planck and Lagrange simulations. The
Planck simulation uses a model that is consistent with the most recent
CMB measurements from the Planck mission (REFERENCE), with the following
parameters: $\Omega_{\rm 0m}=0.3175$, $\Lambda =0.6825$,
$\Omega_{\rm 0b}=0.049$, $H_0=67.11 \, {\rm km}\, {\rm s}^{-1}\,
{\rm Mpc}^{-1}$, $n_{\rm s}=0.9624$ and $\sigma_8=0.8338$.
The Lagrange simulation is run for the same cosmological model as the
Euler simulation that is run with the Eulerian code, described below.

  Our main goal in analyzing these simulations is to understand how the
non-linear power spectrum varies several physical parameters, such as
the amplitude $\sigma_8$ or the density-temperature relation. These
comparison are limited by the intrinsic random variations in the
measured power due to the sample variance in simulations of limited box
size. To reduce this intrinsic sample variance, we have chosen to
generate the initial conditions of all the {\sc{GADGET-II}} simulations
by setting the amplitude of every Fourier mode to the exact rms
amplitude predicted by the power spectrum, instead of generating it
following the Rayleigh distribution. The mode phases are still generated
randomly. Only for the fiducial simulation, we have generated several
realizations with different random seeds, including some cases where the
Rayleigh distribution for the amplitudes is included, which will be
analyzed in \S \ref{subs:check-size} to check for any effect that this
can have on our results. This implies that the power spectrum of the
initial conditions for our simulations is exactly equal to the value
predicted by the cosmological model at each Fourier mode, without any
variations due to sample variance. The \lya transmission power spectrum
that is obtained, however, has random variations caused by non-linear
Fourier mode couplings.

  The impact of different thermal histories on the \Lya forest is
explored by modifying the Ultra Violet (UV) background photo-heating
rate in the simulations, as in \cite{Bolton08}. A power-law
temperature-density relation, $T=T_{0}(1+\delta)^{\gamma-1}$, arises in
the low density IGM as a natural consequence of the interplay between
photo-heating and adiabatic cooling \citep{Hui97}. Two different values
for the temperature at mean density, $T_{0}$, and three different values
for the power-law index of the temperature-density relation, $\gamma$,
are considered to examine the impact of the temperature-density relation
on the \lya power spectrum; the most recent observational constraints
\citep{Boera14} favor values for these parameters close to those of
our fiducial model.
The different thermal histories are constructed by modifying the
fiducial simulation He II photo-heating rate according to
$\epsilon_{HeII}=\alpha\times\epsilon_{fid,HeII}^\nu$,
changing the parameters $\alpha$ and $\nu$ \citep{Bolton08}. 
The fiducial thermal history appears to be in overall good agreement
with recent determinations based on line profile fitting \citep{bolton14}.
The thermal histories for the SPH runs are built
to guarantee that the $\rho-T$ relation is approximately constant with
redshift in the range $z=2.2-3$, so the quoted values of $\gamma$ and
$T_0$ approximate reasonably well the density-temperature relation in
the whole redshift range investigated here.

\subsubsection{Eulerian simulation}
\label{subsubs:simCen}

  In addition to the SPH simulations based on a Lagrangian approach, a
simulation based on an Eulerian code is used,
\cite[described in][]{Cen90,Cen92,Cen93,Cen02}. The simulation used
here is on a 50 Mpc/h box with $2048^3$ cells, and uses the
cosmological model with parameters $\Omega_{\rm 0m}= 0.28$,
$\Omega_{\rm 0b}= 0.04$, $\Lambda = 0.72$,
$H_0 = 70\, \rm{km\, s^{-1}\, Mpc^{-1}}$, $n_s=0.96$, and
$\sigma_8 = 0.82$. This simulation has been run with standard initial
conditions, with the amplitudes of every Fourier mode following the
Rayleigh distribution. The $\rho-T$ relation is determined by
following the photoionization heating that is derived from a model
of the ionizing background that is computed as the simulation is run.
In the density range that is important for the \lya forest over the
redshift range $2.2 < z < 3$ that we analyze in this paper, this
$\rho-T$ relation is well approximated by the values of $T_0$ and
$\gamma$ in Table \ref{tab:sims}.

  To compare this simulation with an equivalent one run with the SPH
method, we have run the simulation designated "Lagrange" in Table
\ref{tab:sims}. The Lagrange simulation is run for the same
cosmological model and box size as the Euler one, and using HeII
heating parameters to approximately mimick the $\rho$-T relation in
the Euler one. This is further discussed in \S \ref{subs:chek-matvscen}.

\subsection{Extracting the Ly$\alpha$ power spectrum from the simulations}
\label{subs:pows}

\begin{figure}[htbp!]
\centering
\includegraphics[width=0.99\textwidth]{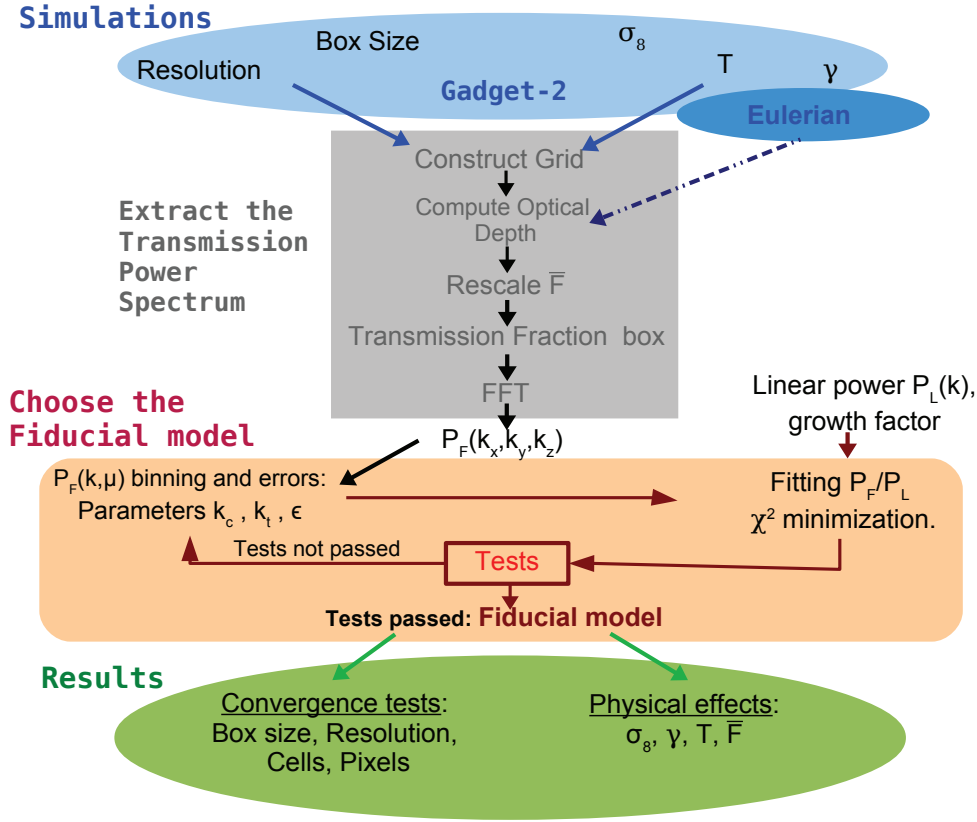}
\caption{\label{fig:diagram} Flow chart of the analysis method of the
power spectrum from simulations followed in this work. A set of
{\sc GADGET-II} simulations
are used with different resolution and box sizes, and different physical
properties: mean transmission $\bar F$, power spectrum amplitude
$\sigma_8$, mean temperature $T_0$, and temperature-density relation
slope $\gamma$. The output of the simulations at various redshifts is
represented in a real space grid (the Eulerian simulation output
directly provides this grid), and then the \Lya optical depth is
computed in redshift space. These optical depths are rescaled to keep
the mean transmission fraction fixed, and then the Fast Fourier
Transform is performed and
the square moduli provide estimates for the power spectrum
$P_F(k_x, k_y, k_z)$.
The power spectrum is then averaged in bins of ($k$, $\mu$), errors
are estimated, and fits to our proposed equation (\ref{eq:Dapm7}) are done.
At first, tests of convergence with parameters related to the binning of
Fourier modes and the errors are carried out. After these tests are
performed, a fiducial model is chosen and further convergence tests are
done for the simulation box size, resolution, and grid cells. Finally,
the dependence of the results on physical effects is examined. }
\end{figure}

  We now start discussing the full procedure for processing the
simulation outputs to obtain an estimate of the \lya transmission power
spectrum, and for analyzing fits to this power spectrum. This procedure
is summarized in the diagram in figure \ref{fig:diagram}, and discussed
in detail in the rest of this section.

%______________________figure______________________%

  For the {\sc GADGET-II} simulations, the SPH formalism for computing
the hydrodynamic variables of gas density, temperature and velocity on a
Cartesian grid, and then extracting mock \Lya spectra, is followed as
described in the Appendix A4 of \cite{Theuns98}. For the Euler
simulation, the Cartesian grid that the simulation is run on is used
directly to obtain the \Lya spectra, as in \cite{MCOR96}. Each
simulation grid in real space is used to generate three distinct boxes
of \Lya spectra, taking each of the three axes as the LOS. For
each of the three axes, the spectra for the entire simulated box are
computed, resulting in $N_c^2$ \Lya spectra.

  Apart from the parameters of each simulation, an additional parameter
is necessary to compute the \Lya spectra: the intensity of the ionizing
background, which can be altered to adjust the mean transmission
$\bar F(z)$ to a certain value. The mean transmission fraction is fixed
to the value given by the expression
\begin{equation}
\bar F(z)= \exp \left[-0.0023 (1+z)^{3.65} \right] ~,
\label{eq:meanF}
\end{equation}
which was found to adequately fit the observational data of
high-resolution spectra by \cite{Kim07}, after subtracting the
estimated metal contribution. We note that more recent
determinations give comparable values of $\bar F$, but there are
substantial uncertainties in this determination
\citep{Becker13,Boera14}.

  The computed \Lya spectra are modified to adjust this value of the
mean transmission by using the approximation that the optical depth
varies at each pixel as the inverse of the intensity of the ionizing
background, and that the gas temperature is not affected by this
background intensity. This assumes that collisional ionization can be
neglected and that the atomic fraction is much smaller than unity, which
is generally an excellent approximation (except in high density regions
where the optical depth is very large in any case, and therefore does
not affect the computed \Lya spectra). The use of this approximation
avoids having to recompute the \Lya spectra every time that the mean
transmission is adjusted to the required value in the expression above.
The assumption that the temperature and hydrodynamic evolution of the
IGM is not affected by the intensity of the ionizing background is not
as accurate because cooling by line excitation is neglected, but in any
case, here we are interested in examining the dependence of the
predictions for $P_F$ on an assumed, fixed $\rho$-T relation, and
separately on $\bar F$.

  We will generally present results at the redshifts $z=2.2$, $2.4$,
$2.6$, $2.8$, and $3$, where equation \ref{eq:meanF} implies rescaling the
\lya spectra to mean transmission values of $\bar{F} = 0.8517$, $0.8185$,
$0.7813$, $0.7404$, and $0.6960$, respectively. We shall use these values
except for a few cases discussed in section \ref{subs:anZ}, where we
examine the variation of $P_F$ under changes in $\bar F$.

 A Fast Fourier Transform is applied to the entire box of \Lya    
spectra, for each of the three cases taking each axis as the LOS.
Usually, the grid is cubic with $N_c^3$ cells, except in the P1024 model in
Table \ref{tab:sims} where the number of pixels is $2N_c$. For the
latter case, we first average the value of $F=\exp(-\tau)$ in every
two pixels to obtain a cubic grid, and then compute the Fourier
transform. The routine {\sc fft} from the {\it scipy} package in Python is
used for this computation{\footnote{http://docs.scipy.org/doc/numpy/reference/routines.fft.html}}.
This results in $N_c^3/2$ 
independent Fourier modes for each of the three axes chosen as the
LOS, each one
with a modulus and a phase. The moduli are used to obtain the estimate
of the \Lya power spectrum $P_F(k,\mu;z)$.

%______________________subsection______________________%
\subsection{Power spectrum estimation, Fourier space binning, and errorbars}
\label{subs:bins}

  We discuss in this subsection the procedure for evaluating the power
spectrum in Fourier space bins, using the Fourier modes of the \lya
transmission field of the simulations. The wavenumbers that are
available in a simulation of box size $L$ go from the minimum value,
$k_1\equiv 2\pi/L$, to the maximum value equal to the Nyquist frequency,
$k_1 N_c/2$. Several parameters are involved
in our choice of binning in Fourier space and the computation of error
bars of $P_F$ used to obtain fits, which we enumerate here:

\begin{enumerate}

\item The cutoff scale $k_c$. This is the maximum scale at which we
consider that $P_F$ can be reliably predicted from a simulation and
measured from the observations. We use only modes with $k<k_c$ to fit
the results of $P_F$, where $k_c$ is less than the Nyquist value.
For our fiducial simulation, the comoving cell size is
$L/N_c=0.12 \hmpc$, corresponding to a velocity width $\sim 12 \kms$
and a Nyquist value $\pi N_c(1+z)/(HL)\sim 0.25\, {\rm s}\, {\rm km}^{-1}$.
Most of the modes in a simulation have wavenumbers near the Nyquist
value, which are affected by the absorption tails of
high column density systems arising from highly non-linear collapsed
structures that may not be correctly modelled in the
simulations. Moreover, in practice the observed power at this large $k$
is highly sensitive to the presence of narrow metal lines, which
are not included in the simulations. Following M03, we consider
that any comparison of theoretical and observed power spectra is not
reliable for wavenumbers above $k(1+z)/H\sim 0.1\, {\rm s}/{\rm km}$, which
is approximately a fixed comoving scale in our examined redshift range
We therefore choose, for our fiducial simulation,
$k_c=100\, k_1 = 10.47\, h/{\rm Mpc}$, and for all simulations we keep
the physical value $k_c = 10.47\, h/{\rm Mpc}$ fixed. All the modes with
$k>k_c$ are discarded. This leaves, for the fiducial model,
$4\times 10^6$ independent Fourier modes with $k<k_c$ to be used in our
analysis.
%This means that our power
%spectrum measurement depends only on a small fraction of all the modes
%obtained from the box ($10^6$ out of $\sim 10^8$ for the fiducial
%simulation), after the ones with highest wavenumber are discarded.

\item The transition scale $k_t$. For effectively computing a $\chi^2$
function to fit the estimated power spectrum from a simulation to an
analytic model, having more than $10^6$ values from independent Fourier
modes is still a very large number, and the vast majority of these
Fourier modes are at high $k$. To reduce this number, we define bins in
the $(k,\mu)$ variables to average the estimated power spectrum within
each bin when $k$ is larger than a transition scale $k_t$. For $k<k_t$,
different modes obtained from the simulation are averaged only when they
have exactly the same values of $(k,\mu)$, and compared to the power
spectrum values from an analytic model at exactly the same $(k,\mu)$ to
compute the $\chi^2$ function.

  We choose $k_t= 1\, h/{\rm Mpc}$ for the fiducial simulation with box
size $L=60 \hmpc$, which is close to the geometric average of $k_c$ and
$k_1$. This results in roughly the same number of bins at $k> k_t$, as
different values of $(k,\mu)$ at $k < k_t$, which optimizes the
efficiency and accuracy of the calculation. For different box sizes, we
keep fixed the value of $k_t L=60$, so that the number of Fourier modes
evaluated without binning remains roughly constant. For example, for our
largest simulation with $L=120 \hmpc$, we use $k_t=0.5\, h/{\rm Mpc}$.

\item Number of bins in $(k,\mu)$. The range of $k$ from $k_t$ to $k_c$
is divided into 16 bins that are equally spaced in $\log k$, and $\mu$
is divided also into 16 linearly spaced, equal bins from 0 to 1. This
gives a total of 256 bins for $k>k_t$ for which the power spectrum is
estimated, for all simulations. Simulations with larger box size, with
a smaller value of $k_t$, have therefore a larger bin size in $\log k$.
The number of different values of $(k,\mu)$ that are obtained at $k<k_t$
turns out to be 296 for our choice of $k_tL=60$, therefore a total of
552 fitting points are used for all simulations when fitting
$P_F(k,\mu)$ to an analytic model.

\item Errors $\sigma_P(k,\mu)$. Evaluating the $\chi^2$ function requires
assigning an error to each Fourier mode estimated from a simulation. We
assume that the Fourier modes are independent, and therefore that our
covariance matrix is diagonal. We keep a count of the number of Fourier
modes, $n_F(k,\mu)$, contributing to each of the 552 values of
$(k,\mu)$ (including all three projections we use of the simulation box
in redshift space, so $n_F$ is always a multiple of 3). We then compute
the error of each power spectrum evaluation according to:
\begin{equation} \label{eq:pferr}
 \sigma_P(k,\mu)= P_F(k,\mu)\left[ 1/ \sqrt{n_F(k,\mu)} + \epsilon \right] \, ,
\end{equation}
where $P_F$ is the estimated value of the transmission power spectrum.
For $\epsilon=0$, this is the expected error owing to the Poisson
variance associated with the number of independent modes available in
the simulation. The constant $\epsilon$ is included to avoid an
excessive weight to the overall fit from the modes with the highest
values of $k$, following the procedure of M03. The number of
modes below a certain value of $k$ grows as $k^3$, so if a reasonable
value of $\epsilon$ is not included, any analytic fit will need to be
very highly accurate for all the high-k modes before the low-k modes are
of any importance in determining the minimum of the $\chi^2$ function.
Unfortunately, there is no clear objective way to decide the value of
$\epsilon$ that one should choose to obtain a fit that is adequately
weighting the results of a simulation over the broad range of $k$ that
is being probed, and the results of the fits depend on $\epsilon$. In
our work we have chosen a constant value of $\epsilon=0.05$ for all the
simulations and analyses. The optimal value of $\epsilon$ is discussed
in Appendix \ref{subs:epsilon}.

\end{enumerate} 

  It is worth going through some examples of the number of independent
modes available for the power spectrum estimate for the smallest
wavenumbers, starting at $k_1=(2\pi)/L$. For the direction parallel to
the LOS (with $k_x=k_y=0$ and $k_z=k_1$, choosing the z-axis as the
LOS), with $\mu=1$, only one independent Fourier mode is obtained (the
mode with $k_z=-k_1$ is not independent because of the condition that
the \Lya transmission field is a real function). Two independent modes
perpendicular to the LOS, with $\mu=0$, are available, for
$k_x=k_1$, $k_y=0$, and $k_x=0$, $k_y=k_1$. The next smallest modes
have $k=\sqrt{2} k_1$, with two independent modes for $\mu=0$
($k_x=k_1$, $k_y=k_1$, and $k_x=k_1$, $k_y=-k_1$), and four
independent modes for $\mu=1/\sqrt{2}$ (with either $k_x$ or $k_y$
being equal to +/- $k_z$). In general, eight independent modes are
available for any values of $k_x$, $k_y$ and $k_z$ when they are all
different from zero and $k_x\neq k_y$ (owing to the symmetry under two
independent sign changes and under the exchange of $k_x$ for $k_y$),
which are used to estimate the \Lya power for the values
$k=(k_x^2+k_y^2+k_z^2)^{1/2}$, $\mu=k_z/k$. In this case, the estimate
of the power at this $(k,\mu)$ will come from $n_F=24$ values, because
eight independent modes are obtained for each of the three axes chosen
as the LOS. For some modes, this number is further increased
whenever several combinations of $(k_x,k_y,k_z)$ yield the same values
of $(k,\mu)$ (e.g., for $k_x=3$, $k_y=4$ and $k_x=5$, $k_y=0$). There
are 296 different values of $(k,\mu)$ that are obtained from a cubic box
with $kL < 60$.

  For the bins at $k>k_t$, the average values of $(k,\mu)$ of the
contributing Fourier modes in each bin are stored, in addition to the
mean power spectrum value. These average values are usually very close
to the central values of the bin (because the number of modes included
in each bin is large), but they are not exactly equal. The model of the
power spectrum to be fitted is then evaluated at these average values
instead of the bin center.
 
%______________________subsection______________________%
\subsection{Parameterized fitting function for the \Lya power spectrum}
\label{subs:fiteq}

  The \lya transmission power spectrum obtained from the simulations
will be fitted to the following analytic model:
\begin{equation}
\label{eq:P3Deq}
P_F(k,\mu) =b^2_{F\delta}\, (1+\beta \mu^2)^2\ P_{L}(k)\, D(k,\mu)~.
\end{equation} 
The first terms on the right hand side are derived from the linear
perturbation theory of \cite{kaiser87}, as explained in
\S \ref{sec:biasf}, where
$P_{L}(k)$ is the mass density fluctuation linear power spectrum,
$b_{F\delta}$ is the density bias factor of the \lya transmission, and
$\beta$ the redshift distortion parameter. The function $D(k,\mu)$ is the
deviation from linear theory due to non-linear evolution, so we expect
$D$ to approach unity in the limit of small $k$.
%Moreover, the
%non-linear deviation $D-1$ should be proportional to the square
%amplitude of the density fluctuations, or $D-1 \propto k^3 P_{L}(k)$,
%in the limit of small $k$, as expected from perturbation theory.

  We shall use two different fitting models for $D(k,\mu)$ in this
paper, although we have tested many others before deciding on a formula
that provides good fits. First, the expression used by M03,
which we designate $D_{0}$, with a total of 8 free parameters,
\begin{equation}
 D_0(k,\mu)= \exp \left[ \left(\frac{k}{k_{nl}} \right)^{a_{nl}} - 
 \left(\frac{k}{k_p} \right)^{a_p}  -  
 \left( \frac{k \cdot \mu}{ k_{v0} (1+k/k_{v1})^{a_{v1}} }\right)^{a_{v0}}
   \right].
\label{eq:DvPm10}
\end{equation} 
Second, the expression we shall use in most of our fits, $D_1$, is a new
one that has only 6 free parameters, and actually only 5 are used in
most of our fits. We make the ansatz that the
non-linear correction should behave as $D-1 \propto k^3 P_{L}(k)$ in the
limit of small $k$, because perturbation theory predicts that the
second order terms of $P_F$ should depend on integrals of products of
four linear perturbations. First, we define the Fourier amplitude of
linear density fluctuations as
\begin{equation}
\label{eq:D2k}
  \Delta^2(k) = {1\over 2\pi^2} k^3 P_L(k) ~.
\end{equation}
The fitting formula $D_1$ that we use is
\begin{equation}
 D_{1}(k,\mu)= \exp \left\{ \left[ q_1 \Delta^2(k) + q_2 \Delta^4(k) \right]
 \left[ 1 - \left(\frac{k}{k_v} \right)^{a_v} \mu^{b_v} \right] -
  \left(\frac{k}{k_p} \right)^{2}   \right\} .
\label{eq:Dapm7}
\end{equation} 
These equations are to be understood as simple fitting formulae that
have been found to provide useful fits to the numerically obtained
power spectra from the simulations by experience. However, some
physical motivation for the various terms can be provided as follows:
\begin{itemize}

\item Non-linear enhancement: The power spectrum is increased on scales
near the onset of non-linearity, because non-linear collapse of
structure tends to enhance the power relative to the linear prediction.
For the $D_1$ formula, and ignoring for now the $\mu$ dependence, this
term is $q_1 \Delta^2(k)$, with the optional
addition of the higher order term $q_2 \Delta^4(k)$ that may be
included to improve the fit. The constants $q_1$ and $q_2$ are
dimensionless and control the importance of this non-linear enhancement.
In the $D_0$ formula this term is $(k/k_{nl})^{a_{nl}}$.
The scale $k_{nl}$ at which
non-linear effects start being important should roughly obey
$\Delta^2(k_{nl})\sim 1$, so the dimensionless constants $q_1$ and $q_2$
in the $D_1$ formula are expected to be of order unity.
The local slope of the power spectrum near
this non-linear scale at $z\simeq 2.5$ is $n_{eff}\simeq -2.3$, implying
that $\Delta^2 \sim k^{0.7}$ near this scale, which agrees with the
typical value found for $a_{nl}$ in M03.
In the limit of small $k$, the difference $D-1$ is
proportional to $\Delta^2(k)$ in our new formula, so an extrapolation
to $k$ values smaller than those probed by our simulations can give
a reasonable prediction, while the formula $D_0$ of M03
is expected to overpredict $D-1$ for small $k$.

\item Jeans smoothing: The gas pressure suppresses the power below the
Jeans scale. Small-scale power is present in regions of high density,
where the Jeans scale is reduced in the highly non-linear regime, but
this should not greatly affect the \Lya forest which is mostly sensitive
to moderately overdense structures. This power reduction is modelled by
the isotropic term with the scale $k_p$ playing a role that is
reminiscent of a Jeans scale, although this cannot be taken literally
because our formulae are simply a fit to a highly non-linear numerical
result. The $D_0$ formula has a free power-law $a_p$ for this term, but
we have found that good fits are obtained by fixing $a_p=2$, and
therefore we fix this in the $D_1$ formula.

\item Line-of-sight broadening: Finally, non-linear peculiar velocities
and thermal broadening cause a smoothing of the correlation along the
LOS, and therefore a suppression of power that increases with
$\mu$. This suppression was found in M03 to be well matched by
a power-law dependence on $\mu$ inside the exponential, but several
parameters had to be added to fit the $k$-dependence of this non-linear
anisotropic term, the third one in the $D_0$ formula. We find that
by multiplying this term by the same non-linear enhancement depending
on $\Delta^2(k)$, a simple power-law dependence on $k$ also provides
a good fit.
\end{itemize}

  Therefore, our new formula $D_1$ has two advantages over $D_0$: it
has the correct behavior for $D_1 - 1$ in the limit of small $k$, and
it reduces the number of non-linear parameters from eight to six, and
in fact to five in most of the simulations analyzed in this paper where
we will set $q_2=0$, still providing sufficiently good fits.

  To calculate the fit of the values of $P_F(k,\mu)$ extracted from the
simulations to equation (\ref{eq:P3Deq}), we use the Montecarlo Markov
Chain method \cite[hereafter, MCMC; see, e.g.,][for a review on the Metropolis
algorithm that we use]{Bhanot88} to minimize the $\chi^2$ function,
which we compute as
\begin{equation}
 \chi^2 = \sum {(P_s - P_m)^2 \over (\sigma_P\, P_m/P_s)^2 } =
   \sum {(P_s^2/P_m - P_s)^2 \over \sigma_P^2 } ~,
\end{equation}
where $P_s$ is the power spectrum measured from the simulation, $P_m$
is the power spectrum of the model being fitted, and the sums are over
all the bins in $(k,\mu)$. The errors $\sigma_P$ are computed using
equation (\ref{eq:pferr}) with $P_F=P_s$, so that they do not depend
on the model, and are the ones shown in our figures. However, the
$\chi^2$ function is obtained with the errors computed from the fitted
model, which are $\sigma_P\, P_m/P_s$.

%______________________Section______________________%

\section{Results: the L120 Simulation}
\label{subs:fid}

\begin{figure}[!htbp]
\centering  

\hspace*{-0.02\textwidth}        
   \includegraphics[width=0.49\textwidth]{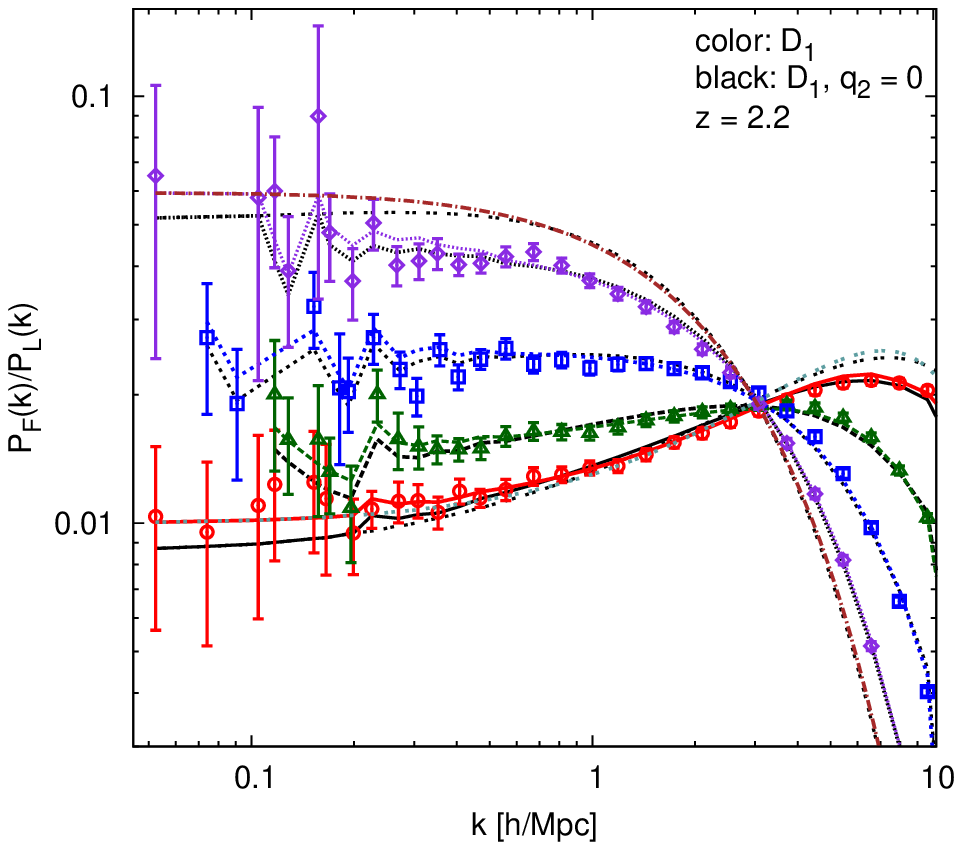}
\hspace*{0.01\textwidth}                
    \includegraphics[width=0.49\textwidth]{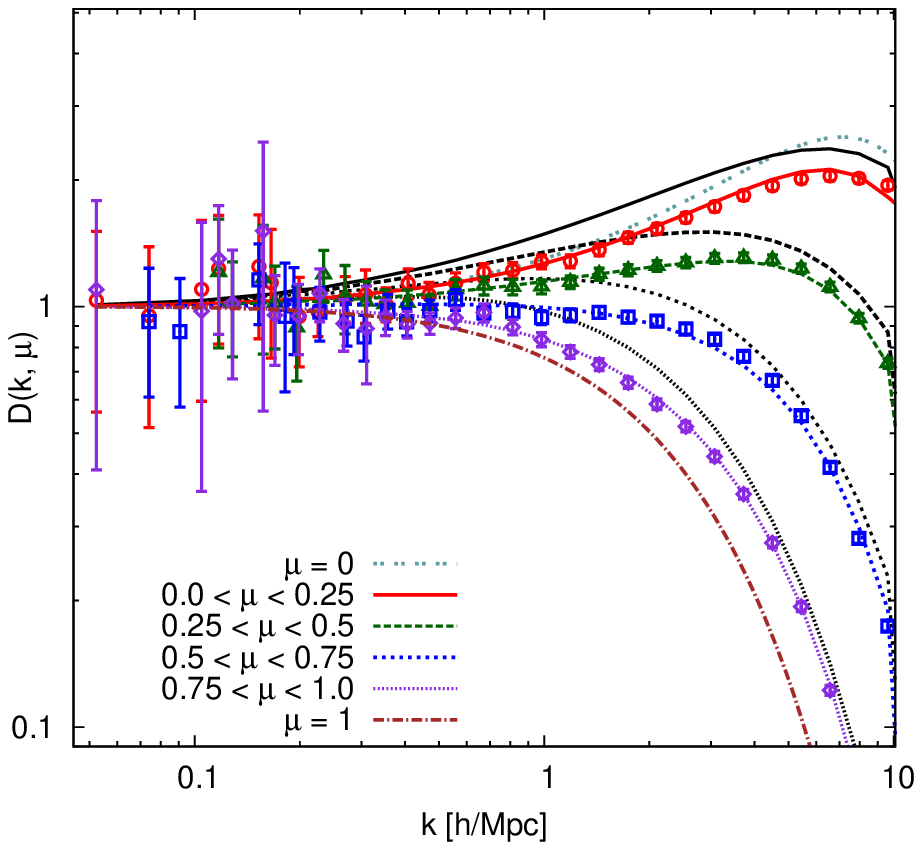}

\caption{\label{fig:reffig} Power spectrum for the $120\, {\rm Mpc}/h$
box at $z=2.2$, averaged over the indicated 4 bins in $\mu$, and with
bins in $\log k$ as described in the text. Points with error bars in the
left panel are the results from the simulation, and colored lines are
the fit to equation (\ref{eq:Dapm7}) with 6 free
parameters (in addition to the two bias factors).
Black curves are the fit to 5 free non-linear parameters, when setting
$q_2=0$ in $D_1$.
The left panel shows the ratio of the transmission power spectrum to the
linear one, and the right panel shows the non-linear term $D(k,\mu)$.
Points in the right panel are shown only for the fit with $q_2$ as free
parameter, to avoid cluttering.
Brown dashed curves are the fitted model computed at $\mu=0$ and
$\mu=1$, shown to indicate the difference with the averaged bins
$0 < \mu < 0.25$ and $0.75 < \mu < 1$; black dashed curves (shown only
in the left panel) are the same for the fit with $q_2=0$.
}
\end{figure}

  This section presents the results of the non-linear \lya transmission
power spectrum for the L120 simulation, the largest SPH simulation we
have analyzed, with $768^3$ particles (see Table \ref{tab:sims}). We
start using the fitting formula $D_1$ in equation (\ref{eq:Dapm7}),
which will be used in all our results for other simulations. Figure
\ref{fig:reffig} shows the estimated ratio $P_F(k,\mu)/P_L(k)$ at
$z=2.2$ as colored points with error bars, and the resulting fits as
curves, in the left panel. The non-linear term $D(k,\mu)$ is shown in
the right panel. The colored curves are the fit to the 6 free parameters
in $D_1(k,\mu)$ plus the two linear bias parameters, and the black
curves show the fit that is obtained when the $q_2$ parameter,
multiplying the second-order term in $\Delta^2(k)$, is set to zero.

  The structure of these two panels will be the same for the various
models analyzed in the rest of this paper. A set of four curves and
points are shown for each fit, corresponding to the four intervals of
$\mu$ indicated in the figure. An additional two curves show in this
case the fit model at $\mu=0$ and $\mu=1$.
%In this case the points are for a single
%simulation but the two different fits mentioned above are shown.
The points are obtained by averaging the modes within the bins in $k$
that are plotted and these four bins in $\mu$. Even though the fit is
performed with 16 bins in $\mu$ at $k>k_t$, as described in \S
\ref{subs:bins}, the results are then further averaged into 4 bins for
the purpose of display only. The fit also uses, as described above, 16
bins in $\log k$ within $k_t < k < k_c$, which are directly plotted, and
individual mode values for $k<k_t$. These individual mode values are
also averaged, for display purposes, into 16 bins in $\log k$ between
$k_1=2\pi/L$ and $k_t$, allowing for easy visualization of the results
in plots that are similar to those in M03. Note that, for small
$k$, some of the bins in $\log k$ and $\mu$ do not include
any of the actual values of $(k,\mu)$ from the simulation modes, and in
this case they are absent from the plot.

  The way in which the values and errors of the power spectrum estimates
in the original bins used for the fit are averaged into the bins used to
make the figures is as follows: each original bin is assigned a weight
$w(k,\mu)=1/\sigma_P^2(k,\mu)$, using the errors in equation
(\ref{eq:pferr}). The values of the power spectrum and the mean coordinates
of the coarse bins for the plot are obtained by averaging $P_F/P_L$,
$\log k$ and $\mu$ with these weights, and the new error in the coarse
bin is set to the inverse square root of the sum of the weights $w(k,\mu)$,
assuming Gaussian independent errors. The error bars therefore indicate
the effective weight that each plotted point, as the average of several
points used in the actual fit, is given to obtain this fit.

%______________________figure______________________%

%______________________figure______________________%
\begin{figure}[!htbp]
\centering

\hspace*{-0.02\textwidth}          
   \includegraphics[width=0.49\textwidth]{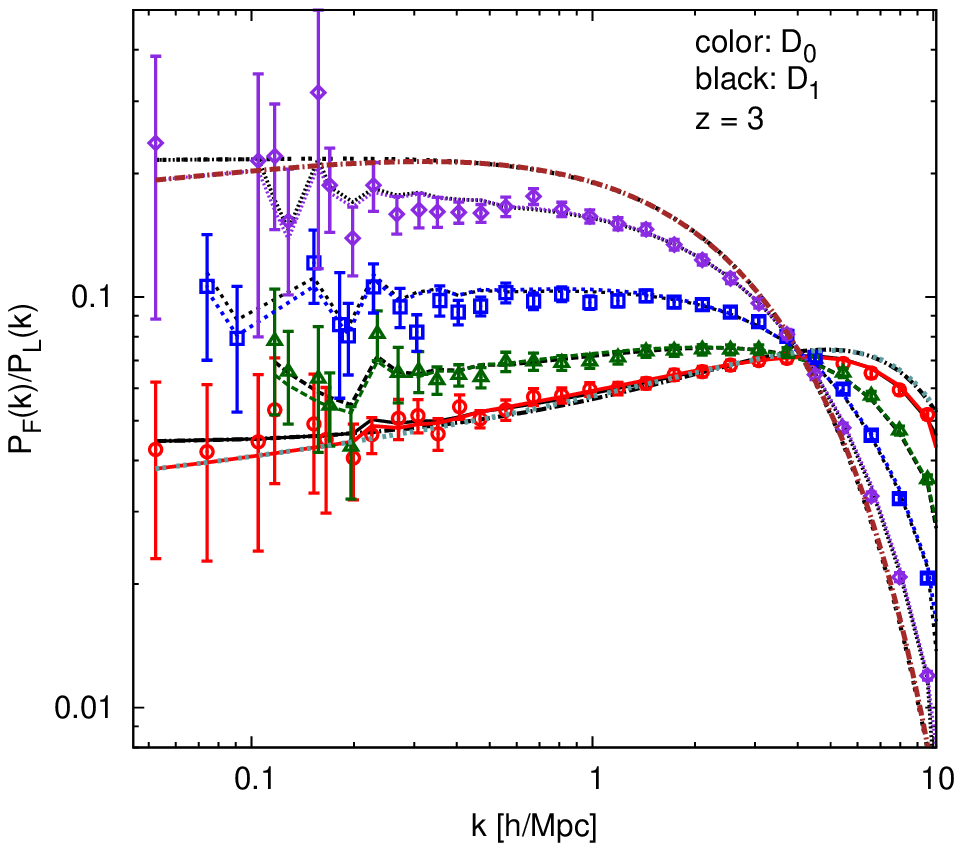}
\hspace*{0.01\textwidth}                
    \includegraphics[width=0.49\textwidth]{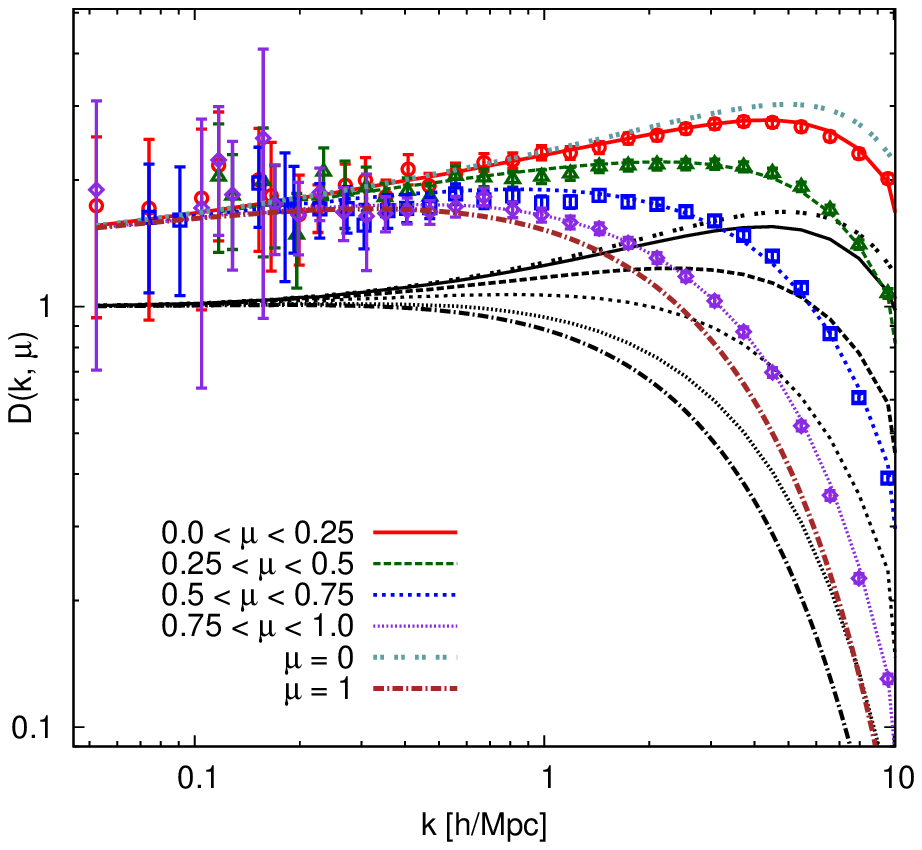}

\caption{\label{fig:mc10}Same as figure \ref{fig:reffig}, but comparing
the fit using the $D_0$ formula (colored curves and points) with the one
using $D_1$ with free $q_2$ (black curves; black points omitted in right
panel), and at $z=3$.}
\end{figure}

  The curves are the result of the model fit, when the model is computed
on the same values of $(k,\mu)$ of all the bins used for the fit, and
then averaged in the same way as the simulation points for display
purposes. This averaging is the reason for the discontinuities in these
curves in the left panel, which are particularly apparent at low $k$.
The four curves for the four values of $\mu$ start from the left side at
different values of $k$, depending on the smallest $k$ value for which
there exists a mode having $\mu$ within each bin. The smallest
wavenumber, with $k=k_1$, exists only for $\mu=0$ and $\mu=1$. The model
predictions are shown for the same values of $(k,\mu)$ as the points,
and are plotted as a continuous line only to guide the eye. Finally, the
cyan and brown dash-dot lines (black for the $q_2=0$ case) are the model
predictions for $\mu=0$ and $\mu=1$ as a function of $k$ (this time, not
averaging over any bins), shown to indicate the difference with the
results in the averaged bins of smallest and largest $\mu$.

  When the points indicating simulation results and the curves showing
the model in the left panel are divided by the expression
$[b_{F\delta} (1+\beta\mu^2)]^2$ in each of the computational bins,
using the bias factors obtained in each fit, the non-linear term
$D(k,\mu)$ is obtained, plotted in the right panel. The
colored points are now valid only for the free $q_2$ fit; points for the
$q_2=0$ fit are omitted to avoid excessive cluttering.
As before, the points and model predictions are averages of the bins
used for computing the fit over the coarser bins used to make the plot.
The averaging of the values and errors is done using the same weights
as for the ratio $P_F/P_L$.

  Figure \ref{fig:reffig} shows that a very good fit is obtained by
varying only the 6 parameters in equation (\ref{eq:Dapm7}), plus the two
linear bias factors (colored curves). A value of $\chi^2 = 296.7$ is
%$\chi^2 = 296.805$
obtained, for $552-8=544$ degrees of freedom at redshift 2.2.
When fixing $q_2=0$ at $z = 2.2$, the value of $\chi^2$ increases to
$349.8$. This value of $\chi^2$ does not
reflect a real ``goodness of fit'' because we have added the parameter
$\epsilon$ in equation (\ref{eq:pferr}) to reduce the weight of the
modes at high $k$, and because the initial conditions were generated
without including the Rayleigh distribution in the amplitude of each
Fourier mode, so the scatter of the values of $P_F/P_L$ arises only from
non-linear coupling. The value of $\chi^2$ is therefore substantially
less than the number of degrees of freedom. Nevertheless, its variation
with the model fit can still indicate if the improvement of a fit is
significant. The fit with $q_2=0$ is already quite good, but it improves
substantially by including $q_2$ as free parameter. In particular, the
low-$k$ points are better matched for free $q_2$, increasing the
model power at small $k$ by $\sim 15\%$, implying an increase of
$\sim 7\%$ in the density bias factor. The difference between the two
fits for $D(k,\mu)$ in the right panel goes up to nearly $20\%$ and
extends to higher $k$, because of degeneracy of the non-linear
parameters with the linear bias factors.

  Results at the higher redshift $z=3$ are shown in Figure
\ref{fig:mc10}, this time comparing the fitting formula $D_1$ with all 6
free parameters (black curves), with the result for $D_0$ with 8
non-linear parameters (colored curves). As before, black points are
omitted in the right panel. The fits are fairly close to each
other for $P_F/P_L$, with values of $\chi^2=252.0$
%$\chi^2=251.969$
for $D_1$, and $\chi^2=262.3$
%$\chi^2=262.301$
for $D_0$, again with a
difference at low $k$ where $D_1$ better matches the simulation points.
In this case, the fit with $D_1$ fixing $q_2=0$ is much closer to being
optimal than at $z=2.2$, with $\chi^2=254.6$.
The fit with the $D_1$ formula is clearly better than with $D_0$, with
two fewer parameters; we have found this to be true also at $z=2.2$.
Moreover, we see in the right panel
that the formula $D_0$ converges very slowly toward unity at small $k$.
The reason is that the value of the $a_{nl}$ parameter in equation
(\ref{eq:DvPm10}) obtained for this fit at $z=3$ is small,
$a_{nl}\simeq 0.2$, and there is a strong degeneracy with the value of
the linear bias factors. This formula therefore easily leads to
unphysical parameter values when the linear bias factors are not
determined independently from the fit to the power spectrum of a
simulation of limited box size. The values of the parameters
%and $\chi^2$
are listed in the tables in Appendix \ref{subs:tables}.

\subsection{Results for the linear bias factors}

  In the limit of small $k$, the ratio $P_F/P_L$ approaches
$b_{F\delta}^2(1+\beta\mu^2)^2$. This is indeed the behavior shown by
our results in the left panel of Figures \ref{fig:reffig} and
\ref{fig:mc10}. The
size of the box limits the number of Fourier modes available in the
simulation at low $k$ and hence the accuracy to which the linear
bias factors can be measured from our fit. In addition, degeneracies
between these bias factors and the non-linear parameters are present
when fitting the simulation results for $P_F/P_L$.

  The values of the bias factors and the redshift distortion parameter
are shown in Figure \ref{fig:zplot_L120_1,2} at the 5 redshift outputs
that we will use in most of our models, and for the three cases we have
considered: the $D_1$ formula with free $q_2$, fixing $q_2=0$ in $D_1$,
and the $D_0$ formula. The two physical bias factors, $b_{\tau\delta}$
and $b_{\tau\eta}$, are obtained from the transmission bias factors
$b_{F\delta}$ and $b_{F\eta}$ derived from our fits of
$P_F(k,\mu)/P_L(k)$ through equation (\ref{eq:btau}), and are shown in
the center and right panels. Errorbars are derived from the Monte Carlo
Markov Chain computed for the fits, and they generally overestimate the
purely statistical errors for a fixed fitting function because our
$\chi^2$ function is too low due to the $\epsilon$ parameter and the
absence of Rayleigh-distributed amplitudes in the simulation. However,
they are dependent on the fit model and they do not include the
systematic errors that are investigated in \S \ref{sec:tests}.

  The results of the bias factors for $D_0$ differ substantially from
the ones derived with $D_1$, and they are less reliable because the
function $D(k,\mu)$ converges too slowly towards unity at small $k$, as
seen in Figure \ref{fig:mc10}. This slow convergence is unphysical
because non-linear effects are actually very small on the largest modes
of the L120 simulation box. This problem for $D_0$ is worse at high
redshift and, together with strong degeneracies with the non-linear
parameters, is the cause of the strange redshift dependence of the bias
factors. For this reason, we focus in the rest of this paper on results
obtained with $D_1$.

%______________________figure______________________%
\begin{figure}[!htbp]
\centering
    \includegraphics[width=1.0\textwidth]{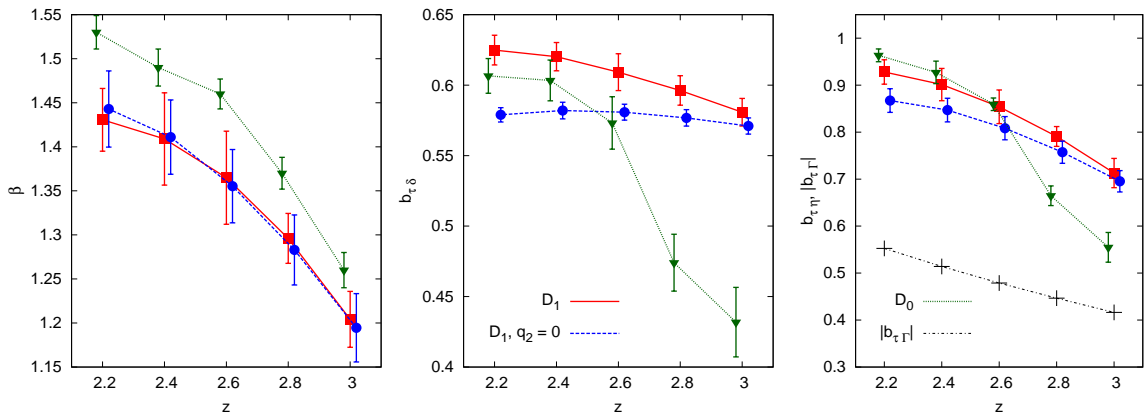}
\caption{\label{fig:zplot_L120_1,2} Results of the fitted values for the
two bias factors $b_{\tau\delta}$ and $b_{\tau\eta}$, and the redshift
distortion parameter $\beta$ as a function of redshift, for the 5
redshifts outputs of the L120 simulation. The error bars are 1$\sigma$,
as returned from the MCMC fitting. To avoid superposition of errorbars,
green points are shifted slightly left and blue points slightly right,
but they are all at our 5 standard redshift outputs. Results are shown
for the fitting formula $D_1$ with all 6 free non-linear parameters (red
squares), for $D_1$ setting $q_2=0$ (blue circles), and for the formula
$D_0$ with 8 free parameters (green triangles). The differences among
the fitting models are caused by degeneracies with the non-linear
parameters, which are more severe for $D_0$. In the right panel,
the absolute value of the radiation bias parameter $b_{\tau\Gamma}$ is
also shown as crosses.} \end{figure}

  The redshift distortion parameter $\beta$, shown in the left panel,
is predicted to have a value near $1.4$ at the most commonly observed
redshift in BOSS, $z\simeq 2.3$, and to decline with redshift. There is
little variation of the value of $\beta$ when fixing $q_2$ to zero. The
value we predict is slightly smaller than that of M03, who
found $\beta=1.58$ at $z=2.25$. Our result for $\beta$ for the $D_0$
formula is closer to that of M03, but this is not for the
same reason since an independent method was used by M03 to
determine the bias factors. The dependence of $\beta$ on the fitting
method and physical parameters will be further discussed in \S
\ref{sec:get}.

 The physical bias parameters have a relatively weak
dependence on redshift. Previous results reported in terms of the
transmission bias factors indicated a very rapid evolution with
redshift, due mostly to the change in the mean transmission $\bar F$.
The predicted value of $b_{\tau\delta}\simeq 0.6$, nearly constant with
redshift, has the physical meaning that the \lya effective optical depth
fluctuates on large scales by $\sim$ 60\% of $\delta$, the fluctuation
in the mass density field. As mentioned above, the density
bias factor drops by $\sim 8\%$ at $z=2.2$ when fixing $q_2=0$,
corresponding to the $\sim 15\%$ variation of power at low $k$ seen in
Figure \ref{fig:reffig}. This is an indication of the uncertainty due to
the degeneracy with non-linear parameters and the use of different
fitting formulae.

  The bias factor of the peculiar velocity gradient is computed from
equation (\ref{eq:rdp}). For the logarithmic derivative of the growth
factor, we use the values for the cosmological model of our fiducial
simulation: $f(\Omega)= (0.9875$, $0.9895$, $0.9911$, $0.9924$,
$0.9935)$ at $z= (2.2$, $2.4$, $2.6$, $2.8$, $3)$. The value of
$b_{\tau\eta}$ is below unity and decreases with
redshift. This means that the \lya forest behaves differently from a
large-scale structure survey of objects with a selection function that
is independent of the peculiar velocity gradient along the LOS, $\eta$:
the effective optical depth fluctuates only by
$\sim 70\%$ of the fluctuation in $\eta$ at $z=3$. We shall return in
\S \ref{sec:disc} to the physical reason why $b_{\tau\eta}$ is less
than unity and decreases with redshift, which is a general
characteristic in the results of all our simulations.
% As mentioned in \S
%\ref{sec:biasf}, a model where the \lya forest is made of a series of
%clouds that are not systematically aligned with the surrounding
%large-scale structure should always have $b_{\tau\eta}=1$. The lower
%value we find for $b_{\tau\eta}$ is explained if the filamentary
%structures giving rise to the \lya forest have internal velocity
%gradients that follow that of the large-scale structure: for example,
%when the absorption is stretched out in a region of negative $\eta$
%(faster expansion along the LOS), the gas in the filaments
%can also have a faster internal velocity gradient, broadening the
%absorption lines in velocity and reducing the degree of saturation.
%As the redshift is reduced, saturated absorption lines become more rare
%and correspond to increasingly virialized objects, and therefore
%$b_{\tau\eta}$ increases closer to unity.

  Finally, we compute also the radiation bias factor $b_{\tau\Gamma}$,
defined at the end of \S \ref{sec:biasf}, using the probability
distribution of $F$ in the simulation L120 at each redshift output,
and equation (\ref{eq:bfgam}). The results are shown as crosses in
the right panel of figure \ref{fig:zplot_L120_1,2}. The values are
a factor of $\sim 2$ smaller than $b_{\tau\eta}$. This shows that the
approximation proposed by \cite{seljak12}, in which $b_{\tau\eta}=
b_{\tau\Gamma}$, fails in an important way. Following on our discussion
at the end of \S \ref{sec:biasf}, we believe the reason is that
non-linear evolution of the small scale fluctuations in the \lya forest
substantially modifies the value of $b_{\tau\eta}$.

  As far as the linear power spectrum is concerned, the main goal of
numerical simulations of the \lya forest should be to accurately predict
the value of $\beta(z)$ and $b_{\tau\eta}(z)$, and to examine the model
dependence of these functions, to compare to observational
determinations. Our results in this section already show the main
difficulty involved in this goal: the numerical fits we obtain depend
on the fitting formula that is used, and on the value of the $\epsilon$
parameter we have chosen in equation (\ref{eq:pferr}) to increase the
importance of the low-$k$ modes. When setting $q_2=0$, the fit to the
low-$k$ points is worse and both $b_{\tau\delta}$ and $b_{\tau\eta}$ go
down. However, if $q_2$ is set free then the $q_1$ parameter determining
the limit of the non-linear correction at low $k$ becomes highly
degenerate. Determining accurate values of the bias parameters with a
reliable non-linear correction can only be done with large numbers of
simulations on large boxes, to have better statistics for the power at
low-$k$. Since we have only one simulation with $L=120 \hmpc$ and a few
on smaller boxes, the results in this paper on linear bias factors are
of limited accuracy, as exemplified by the difference in our two
different fits with $D_1$.
%not accurate to better than $\sim 10\%$.
  
%______________________figure______________________%
\begin{figure}[!htbp]
\centering
    \includegraphics[width=1.0\textwidth]{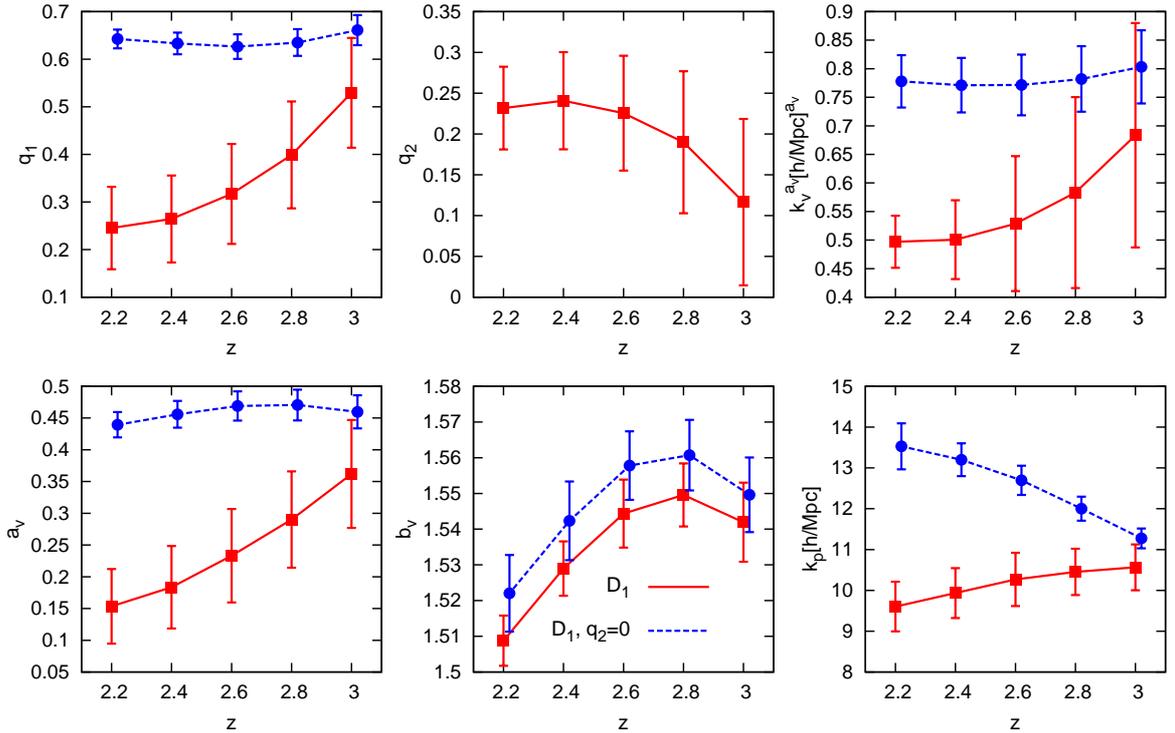}
\caption{\label{fig:zplot_L120_3-7} Results of the non-linear parameters of
$D_1$ as a function of redshift, for the 5 redshifts outputs of the L120
simulation. The error bars represent the 1$\sigma$ contour as obtained
from the MCMC fitting. Solid red lines are for all 6 parameters left free,
and dashed blue ones are for fixing $q_2=0$.}
\end{figure}

\subsection{The non-linear part of the power spectrum}

  The values of the non-linear parameters are shown for the fitting
model $D_1$ in Figure \ref{fig:zplot_L120_3-7}, for all 6 parameters
being free (red solid line), and for $q_2=0$ (blue dashed line). We
plot $k_v^{a_v}$ instead of $k_v$ to reduce the amount of degeneracy
among parameters. We
comment here on the principal features of the results on these
parameters and what this reflects on the shape of the non-linear
function $D(k,\mu)$.

  The scale $k_p$ is generally near $10\, h/{\rm Mpc}$, or
$k_p/H\sim 0.1 \, {\rm s/km}$. This is a characteristic value for the
Jeans scale of the photoionized gas in the IGM, as discussed in
\S \ref{subs:fiteq}. This parameter is used to fit the declining
power with $k$ that occurs above this scale even for $\mu=0$,
clearly seen in Figures \ref{fig:reffig} and \ref{fig:mc10}. The
value of $q_1$, controlling the amplitude of the non-linear
power enhancement, is $q_1\simeq 0.6$ when no second-order term is
included, but a substantial degeneracy occurs with the parameters
$q_2$, $k_v$, $k_p$ and $a_v$ when $q_2$ is included in the fit,
causing large changes of $q_1$ that vary with redshift.

  Non-linear effects imply a change of the sign of the quadrupole
of $P_F(k,\mu)$ as $k$ increases. At small $k$, the power is largest
at $\mu=1$ due to the Kaiser effect, and at large $k$ the effect of
velocity dispersion takes over and the power is largest at $\mu=0$.
The quadrupole is zero at a point where curves of different $\mu$
cross each other. This point reflects a characteristic scale at which
non-linearity changes the sign of the power spectrum anisotropy, and
shifts to the left (larger scales) as the amplitude of the mass power
spectrum increases from $z=3$ to $z=2.2$.

%______________________figure______________________%
\begin{figure}[!htbp]
\centering

\hspace*{-0.02\textwidth}                                                           
   \includegraphics[width=0.49\textwidth]{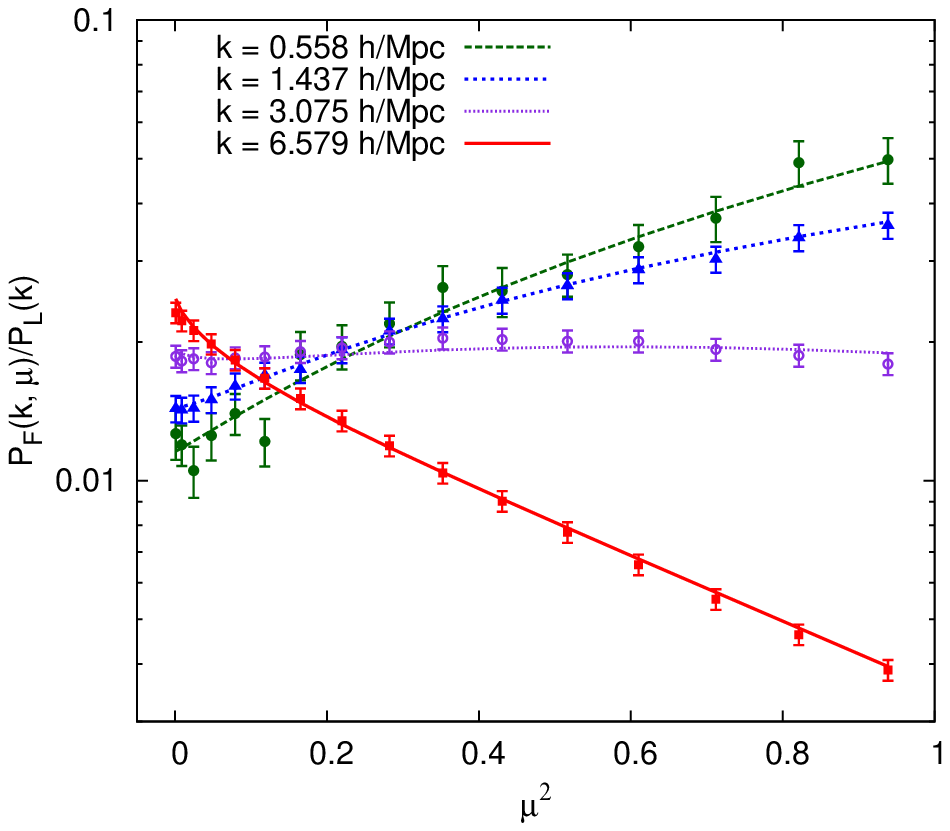}
\hspace*{0.01\textwidth}                
    \includegraphics[width=0.49\textwidth]{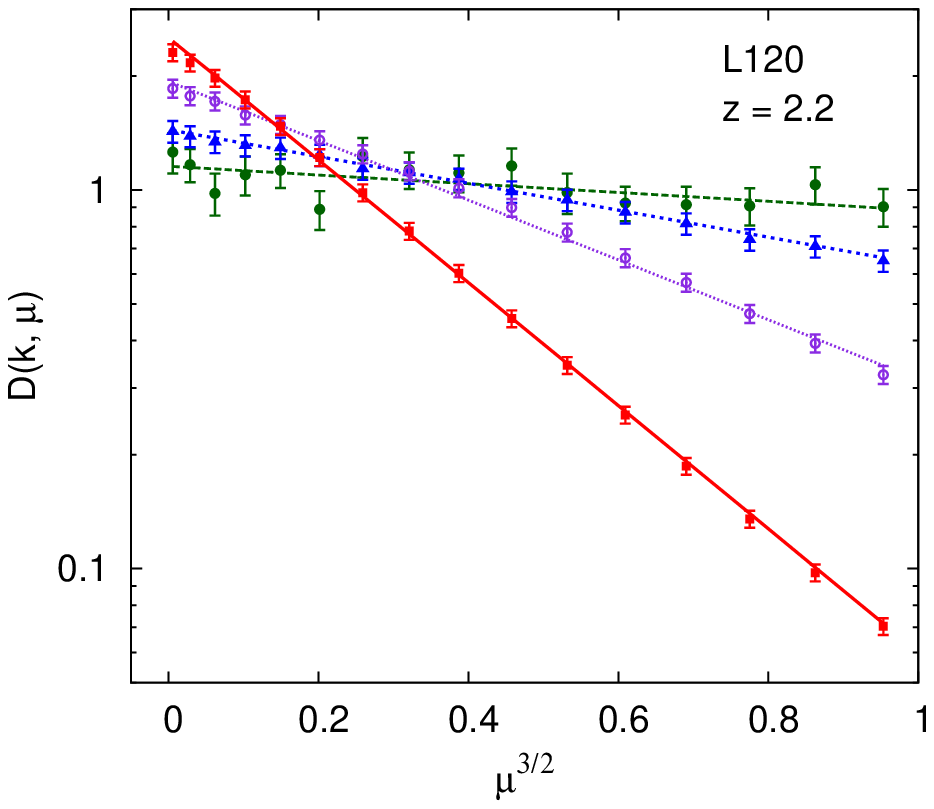}

\caption{\label{fig:pmu} Left panel: $P_F(k,\mu)/P_L(k)$ as a function
of $\mu^2$, for the bins centered at $k= (0.56, 1.44, 3.08, 6.58)\,
% $k= (0.55, 1.42, 3.04, 6.51)\,
h/{\rm Mpc}$ (or bins number 1, 6, 10 and 14 above $k_t$). Right panel:
same points for $D(k,\mu)$ as a function of $\mu^{1.5}$. }
\end{figure}

  The dependence of the non-linear correction on $\mu$ is very well
represented by the power-law $\mu^{b_v}$ inside the exponential in
equations (\ref{eq:DvPm10}) and (\ref{eq:Dapm7}), where $b_v$ is close
to $1.5$ at all redshifts and is subject to little degeneracy with other
parameters. To see in greater detail the dependence of the power on
$\mu$, Figure \ref{fig:pmu} shows $P_F/P_L$ as a function of $\mu^2$ in
the left panel, and $D(k,\mu)$ as a function of $\mu^{1.5}$ in the right
panel, for a set of four selected bins in $\log k$, corresponding to
bins number 3, 7, 11 and 16 in Figure \ref{fig:reffig} starting from the
right edge. The four values of $k$ of these bins are indicated in the
figure, in units of $h/{\rm Mpc}$. This figure shows the actual value of
$P_F$ that we use in our fits for the 16 bins in $\mu$ in our analysis,
instead of the averages over coarse bins as in previous figures. The
results are shown at $z=2.2$ and for the $D_1$ fit including $q_2$ as
free parameter. We see in the left panel that at the scale where the
quadrupole changes sign ($k= 3.04h/{\rm Mpc}$), the power is indeed
almost independent of $\mu$. The right panel shows that the
$\mu$-dependence of the non-linear correction to the power is
surprisingly well fit by a simple power-law at all values of $k$.

  It will be interesting to express our non-linear parameters in terms
of the characteristic scale where the anisotropy changes sign. We
define the wavenumber $k_{na}$ (the subindex is for the scale of
{\it non-linear anisotropy}) as the one that obeys
\begin{equation}
 P_F(k_{na},\mu=0) = P_F(k_{na},\mu=1) ~.
\end{equation}
The $\mu$ dependence of the transmission power spectrum for our fitting
formula $D_1$ is
\begin{equation}
\label{eq:fpfmu}
 P_F(k,\mu) \propto (1+\beta\mu^2)^2 \,
    \exp \left[ - \beta_a(k) \mu^{b_v} \right] ~,
\end{equation}
where, if we use for simplicity the formula $D_1$ with $q_2=0$, we have
\begin{equation}
 \beta_a(k)= q_1 \Delta^2(k)\, \left( k\over k_v \right)^{a_v} ~.
\end{equation}
The value of $k_{na}$ can be computed from our non-linear parameters
from equation
\begin{equation}
\label{eq:betaa}
 \beta_a(k_{na})= 2\log(1+\beta) ~.
\end{equation}
As it turns out, the function of $\mu$ in equation (\ref{eq:fpfmu})
happens to be nearly constant in the range $0 < \mu < 1$ when
$b_v\simeq 1.55$ and $\beta\simeq 1.4$, if $\beta_a$ obeys equation
(\ref{eq:betaa}). As we shall see, these values of $b_v$ and $\beta$
do not change much for all the other models we examine in this paper.
This is the reason that all the curves for $P_F/P_L$ with different
$\mu$ cross each other nearly at the same wavenumber $k_{na}$ for all
our models.

  We compute the value of $k_{na}$ for all our models with $q_2=0$, by
iteratively solving the equation
\begin{equation}
 k_{na} = \left[ (2\pi)^2 \, k_v^{a_v} \log(1+\beta) \over
   q_1\, P_L(k_{na}) \right]^{1/(3+a_v)} ~.
\end{equation}
The parameter values of our fits are given in the tables of Appendix
\ref{subs:tables}.
The parameter $k_{na}$ is particularly useful because its correspondence
with the crossing point of the curves means that it has very little
degeneracy with all other parameters, and therefore its error is small.

  We have not been able to think of an analytical explanation for the
surprisingly well matched $\mu$-dependence of $\log D(k,\mu)$ to the
power-law $\mu^{b_v}$, with $b_v\simeq 1.55$. We note that $b_v < 2$
implies a singular second derivative of $P_F(k,\mu)$ at $\mu=0$.

  Finally, we also note that once $q_2$ is introduced as a free
parameter, the power-law index $a_v$ becomes very small, particularly
at low redshift, and a large degeneracy is introduced among the
parameters $k_v$, $a_v$, $q_1$ and $q_2$. The improvement in the fit
obtained by including $q_2$ is modest, and this improvement is further
reduced for simulations in smaller boxes, where less information is
available on the $k$-dependence (and the parameter degeneracy
increases). We therefore will fix $q_2=0$ for our fits to the
results of most of our simulations.
% This suggests that the
% additional freedom for the $k$-dependence of the non-linear correction
% that is included with $q_2$ nearly removes the need to have a
% $k$-dependent factor multiplying the $\mu^{b_v}$ term; in other words,
% the $k$-dependence of the isotropic and anisotropic parts of the
% non-linear terms become almost the same.

%% file: tables/sims.tex
%_________________________________________TABLE______________________________________________%

\begin{table}[h]
\centering

	\begin{tabular}{|c|cccc|ccc|}
	\hline
	Name &  Box size & Particles & $N_c$ &  Pixels & $\sigma_8$ & $ \gamma$ & $\log(T_0)$ \\
	\hline
	\hline
	Fiducial & 60 Mpc/h & $512^3$ & $512^3$ & $512$ &  0.8778 & 1.6 & 4.3  \\
	% \hline%\vspace{2 mm}
	P1024 & 60 Mpc/h & $512^3$ & $512^3$& $1024$ & 0.8778 & 1.6 & 4.3  \\
	C256 & 60 Mpc/h & $256^3$ & $256^3$ & $256$ & 0.8778 & 1.6 & 4.3  \\
	R384 & 60 Mpc/h & $384^3$ &  $512^3$ & $512$ & 0.8778 & 1.6 & 4.3 \\
	R384C & 60 Mpc/h & $384^3$ &  $256^3$ & $256$ & 0.8778 & 1.6 & 4.3 \\
	R640 & 60 Mpc/h & $640^3$ &  $512^3$ & $512$ & 0.8778 & 1.6 & 4.3  \\
	R640C & 60 Mpc/h & $640^3$ &  $256^3$ & $256$ & 0.8778 & 1.6 & 4.3 \\
	% \vspace{1 mm}
	L80 & 80 Mpc/h & $512^3$ &  $512^3$ & $512$ &  0.8778 & 1.6 & 4.3  \\
	L120 & 120 Mpc/h & $768^3$ &  $512^3$ & $512$ &  0.8778 & 1.6 & 4.3  \\
	\hline% \vspace{3 mm}
	Euler &  50 Mpc/h & --- & $2048^3$ & $2048$& 0.82 & 1.54 & 4.03  \\
	Lagrange &  50 Mpc/h &  $512^3$ & $512^3$ & $512$& 0.82 & 1.58 & 4.10  \\
	Planck &  60 Mpc/h &  $512^3$ & $512^3$ & $512$& 0.8338 & 1.4 & 4.2  \\
	\hline% \vspace{3 mm}
	G1.3 & 60 Mpc/h & $512^3$ &  $512^3$ & $512$ &  0.8778 & 1.3 & 4.3  \\
	G1.0 & 60 Mpc/h & $512^3$ &  $512^3$ & $512$ &  0.8778 & 1.0 & 4.3  \\
	G1T4 & 60 Mpc/h & $512^3$ & $512^3$ & $512$ & 0.8778 & 1.0 & 4.0  \\
	S0.76 & 60 Mpc/h & $512^3$ &  $512^3$ & $512$ &  0.7581 & 1.6 & 4.3  \\
	S0.64 & 60 Mpc/h & $512^3$ &  $512^3$ & $512$ &  0.6396 & 1.6 & 4.3  \\

	\hline

	\end{tabular}
%\end{sidewaystable}
\label{tab:sims}
\caption{List of simulations and analysis variations used in this paper.
The first column lists the name we give to the simulation/analysis, the
second the simulation box size, and the
third indicates the number of particles used in the simulation (both for
dark matter and gas, except for the simulation named Euler, which uses
a fixed Eulerian grid instead of gas particles). The
fourth column
gives the number of cells, $N_c$, used to represent the hydrodynamic variables
in the spatial grid that is computed to obtain the \lya forest spectra,
and the fifth column is the number of pixels on the line of sight
direction used to compute the \lya spectra. The last three columns give
the power spectrum amplitude and the $(T_0,\gamma)$ parameters of the
temperature-density relation in the simulation, where $T_0$ is expressed
in kelvin.}
% before $^\dagger$ For the Eulerian simulation  $\Omega_{m0}$ = 0.28, $\Omega_{b0}$ = 0.046, $\Omega_{\Lambda}$   = 0.72; while for the rest $\Omega_{m0}$ = 0.3  $\Omega_{b0}$ = 0.05. $\Omega_{\Lambda}$ = 0.7
%$^\ast$ this will be the reference model with that configuration and trough all the work it will be compared with all the rest of the simulations analysed. $H_0$ =70km/(s Mpc), $n_s$ = 0.96 for all. }

\end{table}

%_________________________________________ end TABLE ______________________________________________%

%% file: tests.tex
\section{Convergence Tests: Resolution, Box Size and Numerical Method}
\label{sec:tests}

This section addresses the degree to which our fit results from the
various simulations we analyze have converged when box size, resolution,
grid size and numerical method are
%and the binning to represent the \Lya transmission power spectrum,
varied. Variations of the power spectrum with physical parameters of
the \Lya forest will be discussed in \S \ref{sec:get}.
% The more technical tests for the dependence on the grid size, binning and
%fitting parameters are described in the Appendix.

  We remark here one important difference between our work and that of
M03. We do not use any splicing technique (see also
\citep{borde}) to combine results from simulations of different
resolution or box size, a method introduced by M03 to attempt
to better reach a convergent solution. In this work we simply fit both
the linear bias factors and the non-linear parameters of the analytic
formulae of equations (\ref{eq:DvPm10}) and (\ref{eq:Dapm7}) to the
power spectrum of each simulation. Our results can probably be improved
by using this splicing technique, in particular for the linear bias
factors, but we did not have enough simulations in this work to do this
for all the models we analyze, and we left this for future studies. The
emphasis of our analysis is more focused on the variations of the bias
parameters and the non-linear form of the \lya power spectrum with
respect to a reference model (our fiducial simulation), rather than
aiming for a highly accurate convergence of the results, for which a
larger number of simulations on large boxes are required.
%We find in this section
%that our results are generally reliable at a level better than
%$\sim 10\%$ in relative accuracy for the power spectrum.

  Before we start, we also note that all our SPH simulations generally
use the same random initial conditions for the phases of Fourier modes
of fixed $kL$.
Moreover, as described in \S \ref{subsubs:simmatt}, the amplitudes of
the modes are always equal to the variance predicted by the power
spectrum, instead of being generated randomly with the Rayleigh
distribution. This minimizes the random sampling variance due to the
finite box size and allows for a more direct comparison of different
models. In general, however, there is no guarantee that the average
$P_F$ obtained by suppressing the Rayleigh distribution of amplitudes
in the initial conditions is the same as the correct $P_F$ that can be
derived only by including the Rayleigh distribution, and this will
need to be tested in future studies.
As we shall see in this section,
% and in Appendix \ref{subs:chek-Errors},
the variations of the low-$k$ power spectrum from
simulations of the same model but different random initial conditions
are similar to the differences introduced by resolution and box size,
and the overall accuracy to which we can measure the bias factors from
our simulations is not better than $\sim 10\%$.

%______________________subsection______________________%

\subsection{Resolution}
\label{subs:checkres}

  We start by checking the dependence of our results on the resolution.
There are three different quantities in relation to the resolution of
the simulation and the analysis that is done for computing the power
spectrum that must be tested for convergence: the number of particles
in the simulation, the number of cells in the grid to compute the
hydrodynamic variables, and the number of pixels in the simulated \Lya
spectra.

\begin{figure}[!htbp]
\centering
\includegraphics[scale=0.99]{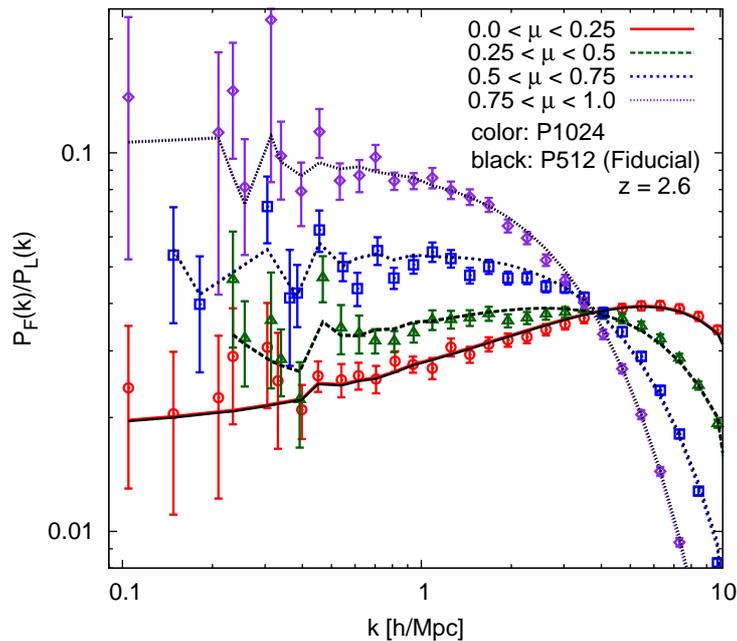}
\caption{ \label{fig:mcd_Pix} Comparison of the power spectrum result
obtained for our fiducial simulation and model analysis (black curves;
box size $L=60 \hmpc$, $512^3$ particles and grid cells, and 512
spectral pixels), with the case of doubling the spectral pixels to 1024,
at redshift $z = 2.6$ (colored curves). Black points are omitted to avoid
cluttering.}
\end{figure}

  We compare first results for the power spectrum from the same
simulation and spatial grid, but varying the pixel resolution of the
computed \lya spectra. Results for $P_F/P_L$ are shown in figure
\ref{fig:mcd_Pix}, where black curves are for our fiducial model and
colored curves and points are for the P1024 model. The fiducial model is
the same as the one used in \S \ref{subs:fid}, but with the box size
$L=60 \hmpc$, and the P1024 model is exactly the same but with the
number of pixels in the \lya spectra doubled to 1024 (Table
\ref{tab:sims}). Black points are omitted to avoid cluttering, but they
follow the black curves similarly to the colored ones. The P1024 model
is computed by calculating the optical depth in 1024 pixels along the
LOS, and then averaging $\exp(-\tau)$ for every two pixels to obtain a
cubic grid of $512^3$ values of the transmission fraction, on which a
Fast Fourier Transform is done in the same way as for all our other
models. Both models are fitted with our $D_1$ formula with $q_2=0$.

  The results are shown at $z=2.6$, and are very similar at other
redshifts. The increase in the pixel resolution practically does not
affect the transmission at the level that is discerned in figure
\ref{fig:mcd_Pix} (the variation is less than 2\%).
Clearly, the spectral pixel resolution is sufficient for our purpose,
and any errors it is causing are much smaller than those due to the
grid size and number of SPH particles in the simulation.
%but needs to be increased for studies aiming at percent level accuracy.
%The
%bias factor $b_{\delta\tau}$, shown in figure \ref{fig:Pix-1,2},
%increases by only $\sim$ 1\% when the pixel resolution is doubled, and
%the redshift distortion parameter $\beta$ is also minimally affected.

%______________________figure______________________%
\begin{figure}[!htbp]
\centering
\includegraphics[scale=0.99]{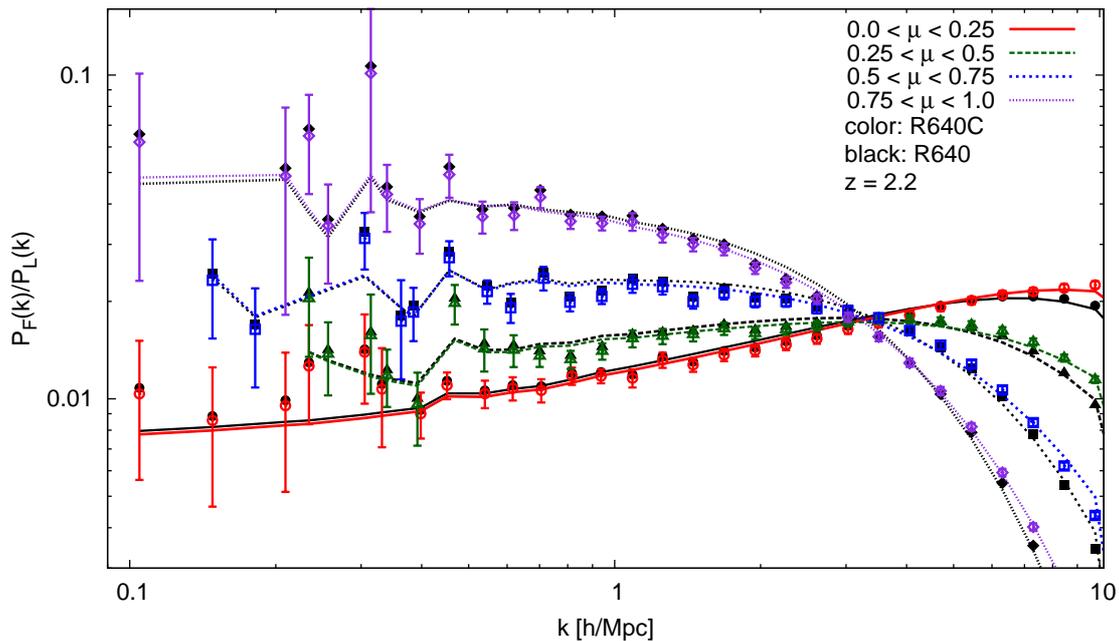}
\caption{\label{fig:mcd-cells26} Comparison of $P_F/P_L$ at $z=2.2$,
%(left panel) and $z=3$ (right panel)
for the R640 simulation with $640^3$ particles analyzed with a grid of
$512^3$ cells (black curves and points), with the same simulation
analyzed with a coarser grid of $256^3$ cells (R640C, colored curves and
points). Errorbars for the black points are suppressed to avoid cluttering,
but they are identical to the colored ones.}
\end{figure}

  Next, we compare two analyses of a simulation using spatial grids of
different resolution. For that purpose, we take our highest physical
resolution simulation, R640 (equal to the fiducial one but with $640^3$
particles instead of $512^3$, although still analyzed on a $512^3$ cells
grid; see Table \ref{tab:sims}), and compare it to R640C, the same
simulation analyzed on a grid of only half the sampling, with $256^3$
cells. The results for $P_F/P_L$ are shown in figure
\ref{fig:mcd-cells26}, at redshift $z=2.2$. Larger differences can be
appreciated in this figure. First of all, a coarse grid increases the
power on small scales, $k > 5\, h/{\rm Mpc}$, probably due to the noise
introduced by a grid cell of $0.23 \hmpc$ in R640C. This also generates
a roughly constant decrease of the power on large scales in R640C
compared to R640. In fact, for $k < 0.5 \hmpc$, the black points are
systematically higher than the colored ones by $\sim 7\%$ at all values
of $\mu$. The change in the small-scale power alters non-linear
couplings to produce an impact on the large-scale power that becomes
constant on very large scales, therefore altering the bias factors.
Results at other redshifts are qualitatively similar.

  Figure \ref{fig:mcd-cells26} also exemplifies the uncertainty in our
fits for recovering the low-$k$ power: the colored curves move above the
black ones at high $\mu$ and low $k$, which is opposite to the
simulation results for $P_F/P_L$ indicated by the points. This is again
reflecting the uncertainty due to the fitting formula that is chosen.
The fitted curves can better reflect the behavior of the simulation
points when the $q_2$ parameter is left free, but the reliability of the
derived bias factors is still subject to a similar error because the
small box size does not allow one to obtain the bias factors more
accurately.

%%______________________figu______________________%
%\begin{figure}[!htbp]
%\centering
%\includegraphics[width=1.0\textwidth]{fig8bias_Pix_Cell.eps}
%\caption{\label{fig:Pix-1,2} Bias factor and redshift distortion
%parameter of the numerical fits to $D_1$ (with $q_2=0$) as a function of
%redshift, showing the differences between the fiducial model when the
%spectral resolution is increased to 1024 pixels (P1024), when the
%resolution of the simulation is increased to $640^3$ particles (R640),
%and when the grid resolution is decreased from $512^3$ to $256^3$
%(R640C).}
%\end{figure}
\begin{figure}[!htbp]
\centering
\includegraphics[width=1.0\textwidth]{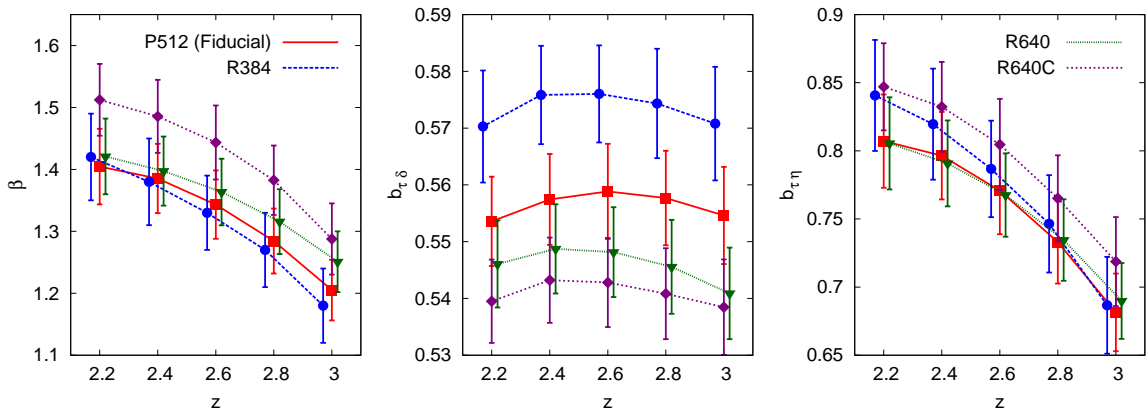}
\caption{\label{fig:res1,2} Redshift distortion parameter and bias factors
fitted for three simulations of increasing physical resolution: R384,
fiducial, and R640. These three models all have a $512^3$ grid, and we
compare them also to R640C, with a coarser grid of $256^3$ and the same
particle number as R640. Blue circles for R384 are shifted slightly left,
and green triangles for R640 slightly right, to avoid superposition of
errorbars.}
\end{figure}

  The bias factors derived from the $q_2=0$ fits are shown in figure
\ref{fig:res1,2}. The errors are the formal ones derived from the
$\chi^2$ fits, but the more important errors on the biases are of a
systematic nature arising from the limited size of our boxes, and the
need to fit the curves of $P_F/P_L$ at small scales where the
simulations provide much more information than at large scales. The
results are fairly similar for all the models shown in this figure.
The largest difference is the higher value of $\beta$ derived for
R640C, but as discussed above, this does not reflect the true difference
of power at low $k$ between R640C and R640 shown by the colored and
black points in figure \ref{fig:mcd-cells26}, and is an artifact of our
fit with $q_2=0$. The real systematic errors in the bias factors are
better reflected by the difference with the fits with free $q_2$.

  We conclude that the errors introduced by the limited resolution of
the grid size are substantial, and that future work should
% investigate
%the best grid size to use for a given number of particles, and 
study the
convergence with both particles and grid size for safely obtaining
results for $P_F$ to accuracies better than 10\%. We note that our L80
and L120 simulations are analyzed also with a $512^3$ grid, which for a
larger box size implies a lower physical resolution, with an expected
impact on the power spectrum and bias factors following that seen in
figures \ref{fig:mcd-cells26} and \ref{fig:res1,2}.

%______________________figu______________________%
%\begin{figure}[!htbp]
%\centering
%\includegraphics[scale=1.09]{zevolCell1,2,_new.eps}
%\caption{\label{fig:cells1,2} Redshift distortion parameter and bias
%factor obtained for the fits of the R640 simulation with the standard
%$512^3$ grid (solid red curves), and for a smaller $256^3$ grid
%(R640C, dashed blue curves).}
%\end{figure}

%______________________figu______________________%
%\begin{figure}[!htbp]
%\centering
%\includegraphics[scale=1.09]{zevolCell3,4,5,6,7,_new.eps}
%\caption{\label{fig:cells3-8} Non-linear parameters of the fit to $D_1$
%in equation (\ref{eq:Dapm7}), for the R640 simulation analyzed with a
%$512^3$ grid, and the same simulation analyzed with a smaller $256^3$
%grid (R640C). The relatively large differences in $k_p$, $k_v$ and $a_v$
%reflect an impact of coarse grid resolution on the small scale power.}
%\end{figure}

%______________________figu______________________%
\begin{figure}[!htbp]
\centering
\hspace*{-0.02\textwidth}
   \includegraphics[width=0.49\textwidth]{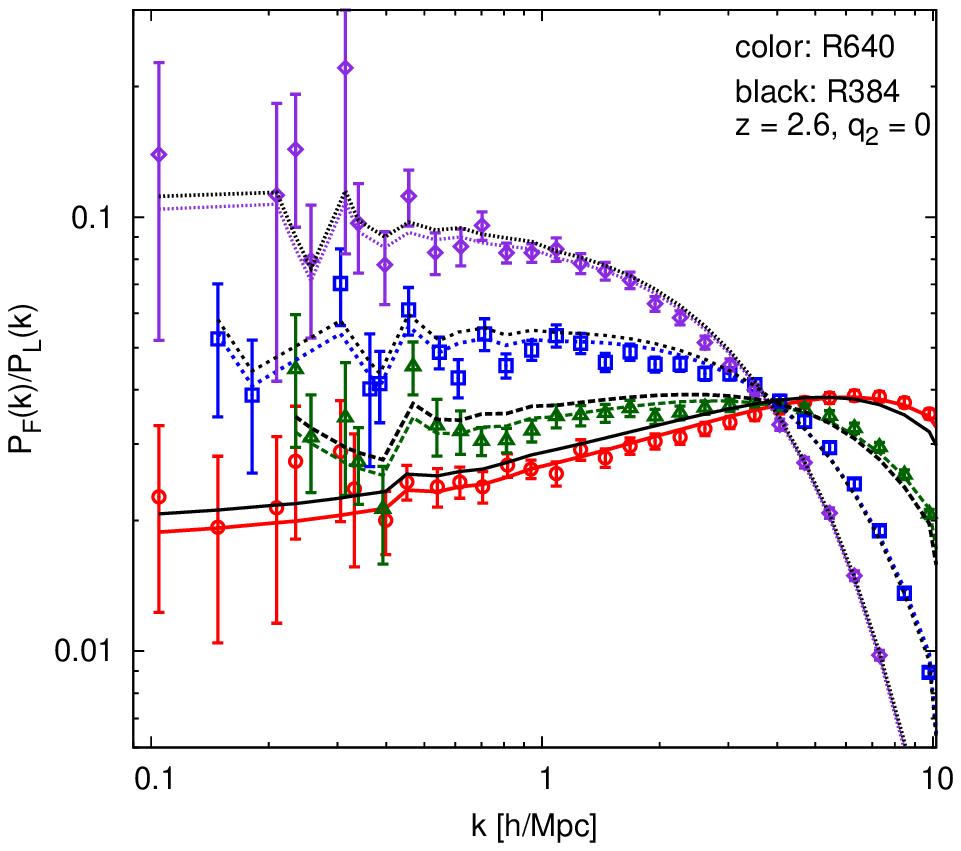}
\hspace*{0.01\textwidth}
    \includegraphics[width=0.49\textwidth]{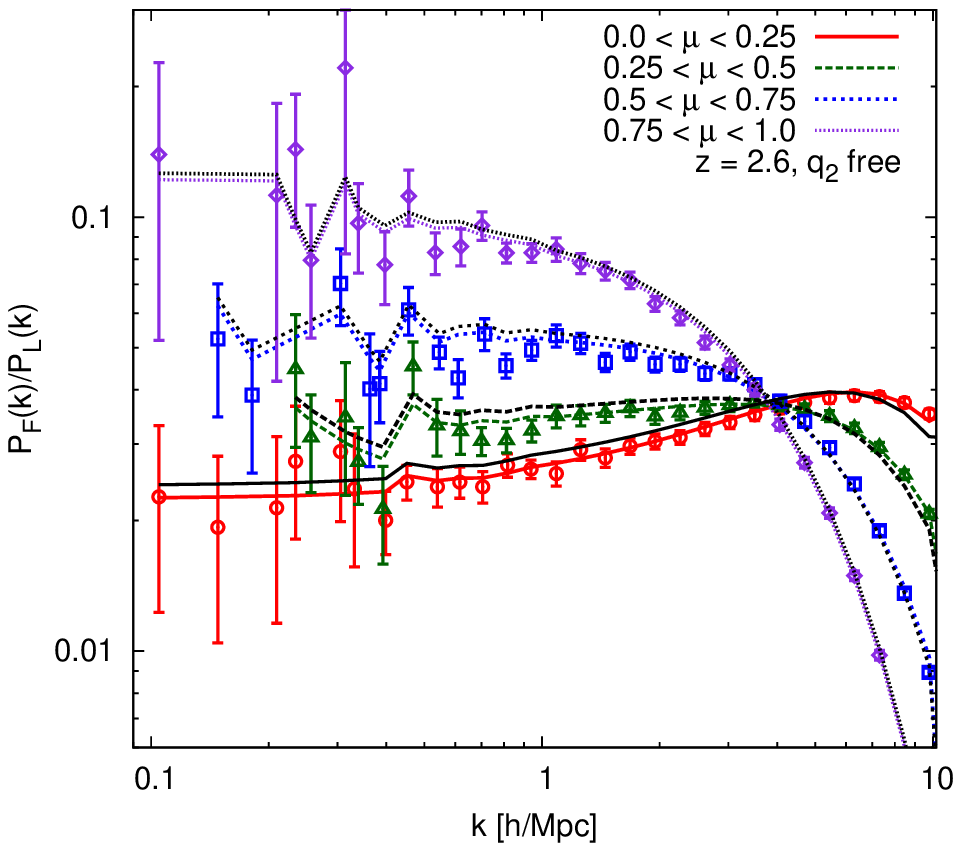}
\caption{\label{fig:mcdres} Comparing the power spectrum of simulations
R640 (colored curves and points) and R384 (black curves, points
omitted), at $z=2.6$, with higher and lower resolution compared to the
fiducial model, respectively. {\it Left panel}: fits with $q_2=0$.
{\it Right panel:} fits with free $q_2$. }
\end{figure}

  Finally, we compare simulations of different resolution, varying the
number of SPH particles. The simulations R640 and R384, which are the
same as the fiducial one but varying the number of particles to $640^3$
and $384^3$, respectively, produce results for $P_F/P_L$ that are shown
in figure \ref{fig:mcdres} at $z=2.6$. In general, the changes due to
the physical resolution are comparable to those caused by the resolution
of the spatial grid but of opposite sign: decreasing the particle
resolution decreases the power at high $k$, although in this case this
is clear only for low $\mu$. In addition, the power at low $k$ generally
drops with increasing resolution. In this case, we show in the left
panel the two fits with $q_2=0$, and in the right panel the two fits
with free $q_2$. The fit at low $k$ follows the points better for free
$q_2$, as expected because of the extra degree of freedom, and the
difference between the curves also better reflects the differences in
the points for R640 and R384. However, as discussed before, it is not
clear from the present simulations if the fits with free $q_2$ (which
have a lower value of $q_1$ and therefore converge faster to the linear
solution) are a more reliable result for the bias factors and the
non-linear correction $D(k,\mu)$ at low $k$.

  The derived bias factors for $q_2=0$ are also shown for both models in
figure \ref{fig:res1,2}. The bias factors for free $q_2$ are not shown,
but are similar to those of the fiducial model shown below in figure
\ref{fig:boxes1,2}. In general the density bias factor decreases by
$\sim 5\%$ when the particle resolution is doubled at fixed grid size,
and increases by a smaller amount when the grid resolution is doubled.
We note here that the changes at low $k$ induced by the
small-scale resolution effects are due to non-linear couplings that
depend on the fact that we always keep a fixed value of $\bar F$; the
result would in general be different if instead we kept a fixed value
of the ionizing background intensity.

  We therefore conclude that the limited resolution and grid size in our
simulations affect the values of the bias parameters by $\sim$ 5\%, and
unfortunately our results do not show yet a clear convergence with the
resolution. Increasing the number of particles or the number of grid
cells modifies the results in opposite directions.
% indicating that an
%optimal ratio of particles to grid cells needs to be maintained: a grid
%that is too coarse induces a loss of resolution, while an excessively
%fine grid may cause shot noise from individual SPH particles, which
%affects the predicted $P_F$.
Higher resolution simulations will be
needed to better understand the conditions for convergence and to obtain
more accurate results for the non-linear power $P_F(k,\mu)$ and the bias
factors. Nevertheless, our derived values of the bias factors and $\beta$
at any redshifts vary by less than $10\%$ in all of our simulations, and
the predicted non-linear shape is also not subject to greater relative
changes arising from resolution, so we believe that our results are
reliable within this level of uncertainty.

  Our results on the sensitivity to the simulation resolution can be
compared to those obtained in \cite{Lukic15}: their figure 12 shows
the quantity $\Delta_F^2=k^3P_F/(2\pi^2)$ (equal to what we plot below
in figure \ref{fig:sigpf}). In agreement with our results in figure
\ref{fig:mcdres}, lower resolution decreases the power at high $k$
and increases it at low $k$, and these changes are more pronounced at
low $\mu$. However, the sensitivity to resolution seems to be weaker
in our SPH simulations. Our R640 and R384 simulations have an
interparticle separation that is about twice the cell size of the
$L10_N128$ and $L10_N256$ simulations of \cite{Lukic15}, and yet the
differences between the two simulations are $\sim 40\%$ in
\cite{Lukic15}, and only $\sim 10\%$ between R640 and R384 in our
case. The likely explanation for this is that the value of $\sim F$
is not held fixed when comparing simulations of different resolution
in \cite{Lukic15}, and that our results shown at $z=2.6$ are mostly
sensitive to gas densities above the mean density of the universe,
where the resolution of particle-based codes improves over that of
Eulerian codes. At higher redshift, however, the \lya forest probes
lower overdensities and the resolution effects in our simulations
probably become more severe.

%______________________subsection______________________%

\subsection{Simulation Variance and Box Size}
\label{subs:check-size}

%______________________figu______________________%
\begin{figure}[!htbp]
\centering
\includegraphics[scale=0.99]{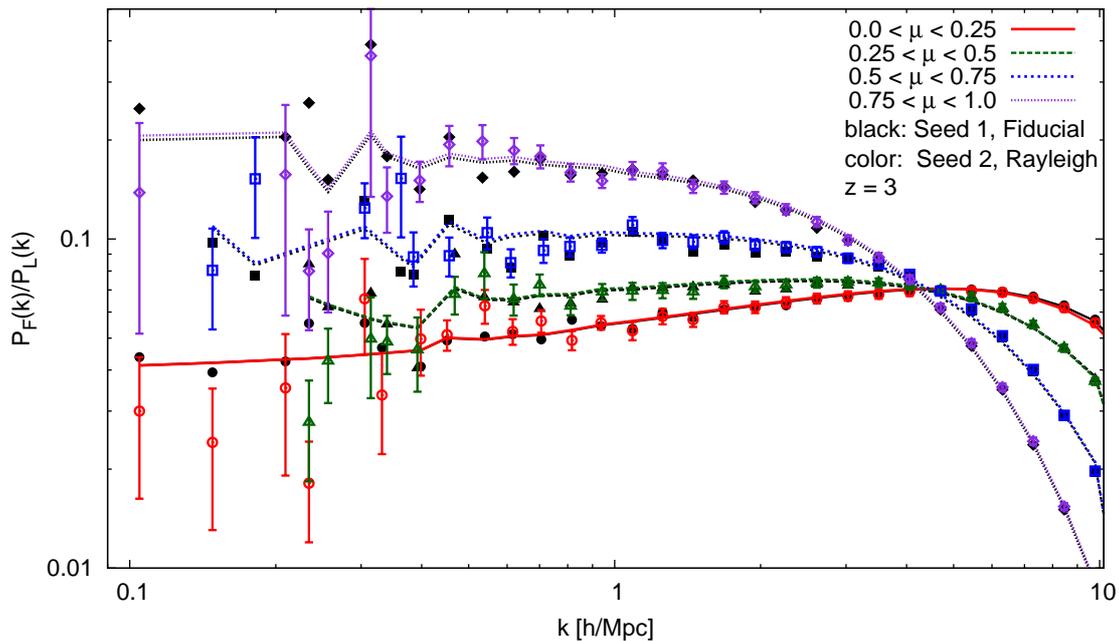}
\caption{\label{figseedP} Comparison of $P_F/P_L$ for two different
random initial conditions for the fiducial simulation at z=3, with a
60 Mpc$/h$ box size. Seed 1 is our usual fiducial simulation (black
curves and points with errorbars omitted), with all Fourier mode square
amplitudes fixed to the mean value predicted by the matter power
spectrum of the model. Seed 2 is for different initial conditions with
the Rayleigh distribution of amplitudes (colored curves and points with
errorbars).}
\end{figure}

  The limited box size of any cosmological simulation implies the
presence of a statistical error and a systematic error on any quantities
that are measured from the simulation, such as the power spectrum of any
tracer. The statistical error is due to the shot noise arising from the
finite number of structures of a given scale that are formed within the
simulation volume; if a number $N_{sim}$ of simulations with the same
box size are performed with different random initial conditions, this
statistical error is reduced as $N_{sim}^{-1/2}$. The systematic error,
which is not reduced by averaging over several simulations, is due to
the discretization of Fourier modes, which need to have cartesian
components on the box axes that are multiples of $k_1=2\pi/L$, and the
absence of any modes below the wavenumber $k_1$. We therefore start by
looking at the statistical variations of the power spectrum in the
fiducial model, with $L=60 \hmpc$, by examining three simulations run
with three different random seeds for the initial conditions: Seed 1 is
for our fiducial model shown in previous figures, Seed 2 is for
independent initial conditions where the Fourier mode amplitudes are
generated with the correct Rayleigh distribution, and Seed 3 for
independent initial conditions varying only the Fourier mode phases of
Seed 1, but keeping the square amplitudes fixed to the mean value
determined by the matter power spectrum, as in Seed 1 (see \S
\ref{subsubs:simmatt}).

%______________________figu______________________%
\begin{figure}[!htbp]
\centering
\includegraphics[width=1.0\textwidth]{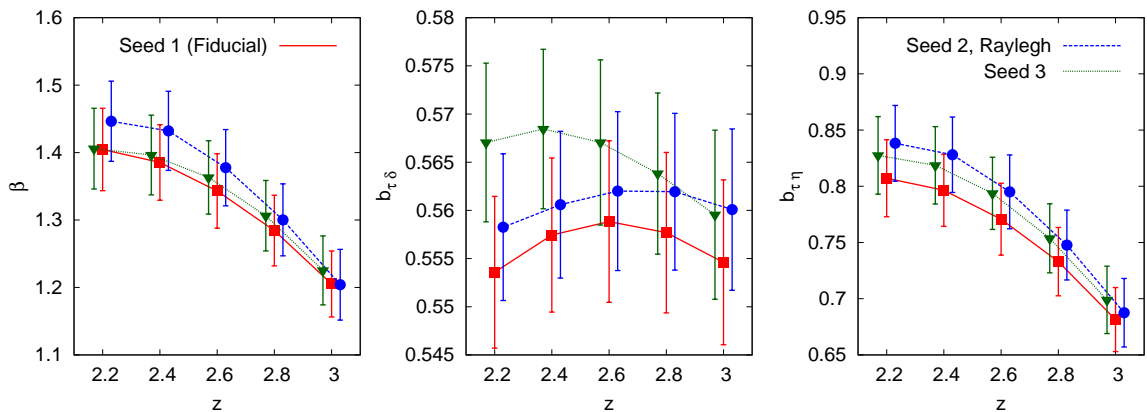}
\caption{\label{fig:seedb} Comparison of the redshift distortion parameter
and linear bias factors for three different random initial conditions for
the fiducial simulation, obtained from the fits with $q_2=0$.}
\end{figure}

  Figure \ref{figseedP} shows $P_F/P_L$ at $z=3$ for the first two
seeds as points, and the fits with $q_2=0$ as curves. Seed 2, with the
Rayleigh distribution, is shown as colored curves and points, and our
usual fiducial simulation (Seed 1) is the black curves and points. The
difference in the two fits is very small, the largest difference occurs
at low $k$ and high $\mu$ and is only $\sim 3\%$. This is despite the
large random differences seen at low $k$ due to the introduction of
random amplitudes for seed 2. The fit does not have much freedom to vary
owing to the large number of modes at small scales, resulting in very
small variations for the high-$k$ points between the two simulations, so
the large scatter at low $k$ gives small variations in the derived bias
factors. Despite this, we see that the fits are generally good for both
seeds. At $z=3$, the $q_2=0$ fits are in fact better at low-$k$ than the
cases at lower redshift shown in previous figures, and are also closer
to the free $q_2$ fits. Note that the even though the errorbars are
shown around the measured simulation points, they need to be computed
for the predicted $P_F$ of the fitted model, so points that are below
the model are not as statistically discrepant as they look in the figure
(e.g., fourth red circle and first green triangle starting from the
left).

  The derived linear bias factors are plotted in Figure \ref{fig:seedb}
at all redshifts, for all three seeds, for the $q_2=0$ fits. Typically,
the two bias factors can have random variations of up to $\sim 4\%$ when
varying the initial conditions, and this is not strongly affected by
including the Rayleigh distribution of Fourier mode amplitudes. We have
tested that these conclusions remain valid for an additional two
independent seeds, not shown here. This uncertainty due to the random
initial conditions is therefore comparable to the one due to the
resolution of the simulations. These tests also show that the
difference between including or not the Rayleigh distribution of mode
amplitudes in the derived average $P_F$ is small enough for the purpose
of the present study.
%We also find that if the fit is done with the free $q_2$
%parameter in our $D_1$ formula, the scatter in the bias factors with
%different seeds is increased because of the degeneracy with the
%non-linear parameters.
%We consider that the scatter that we obtain
%fixing $q_2=0$ is more representative of the actual uncertainty due to
%the sample variance of the initial conditions, which is relatively
%small compared to uncertainties arising from the resolution.

%______________________figu______________________%
\begin{figure}[!htbp]
\centering
\includegraphics[scale=0.99]{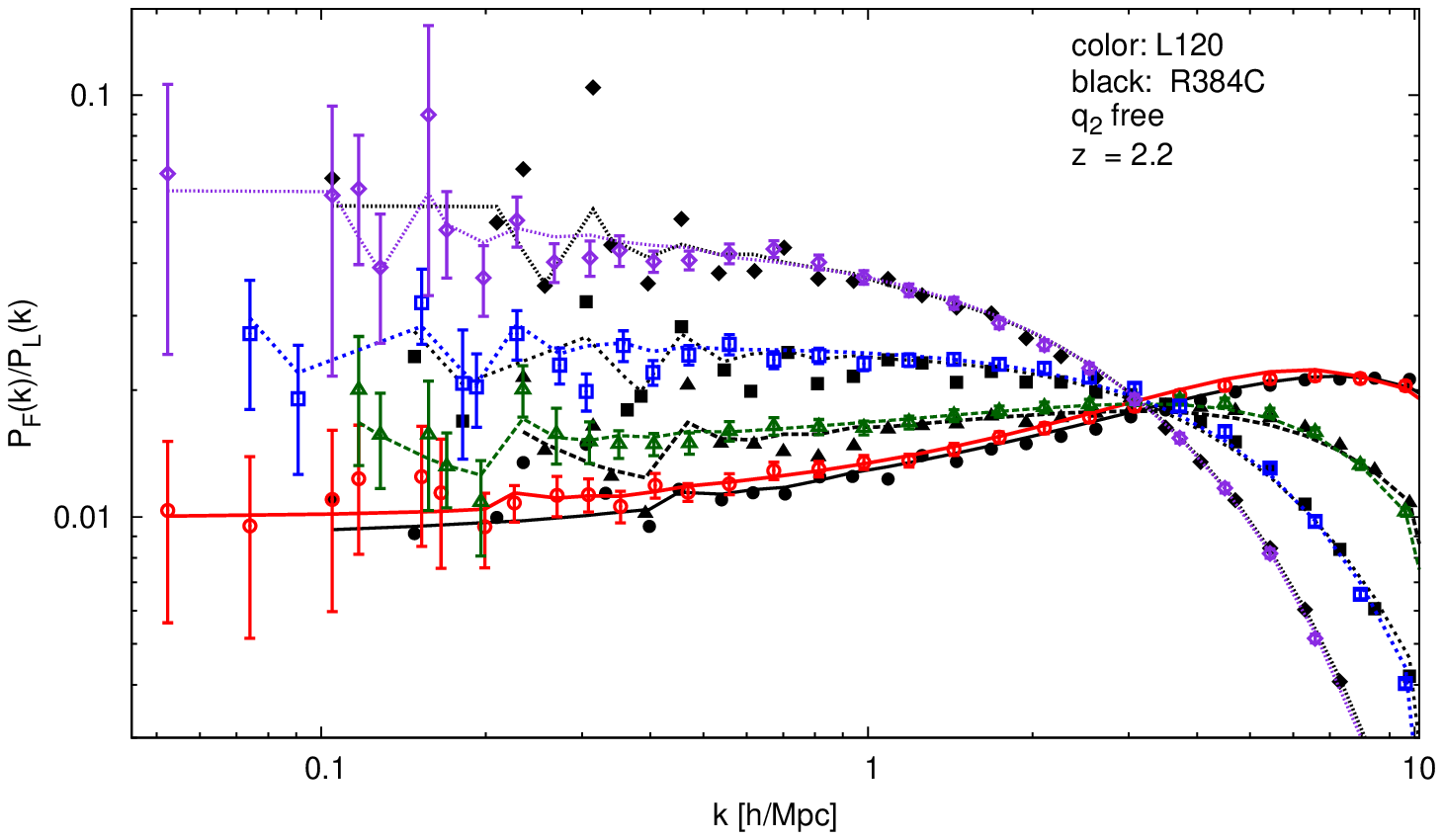}
\includegraphics[scale=0.99]{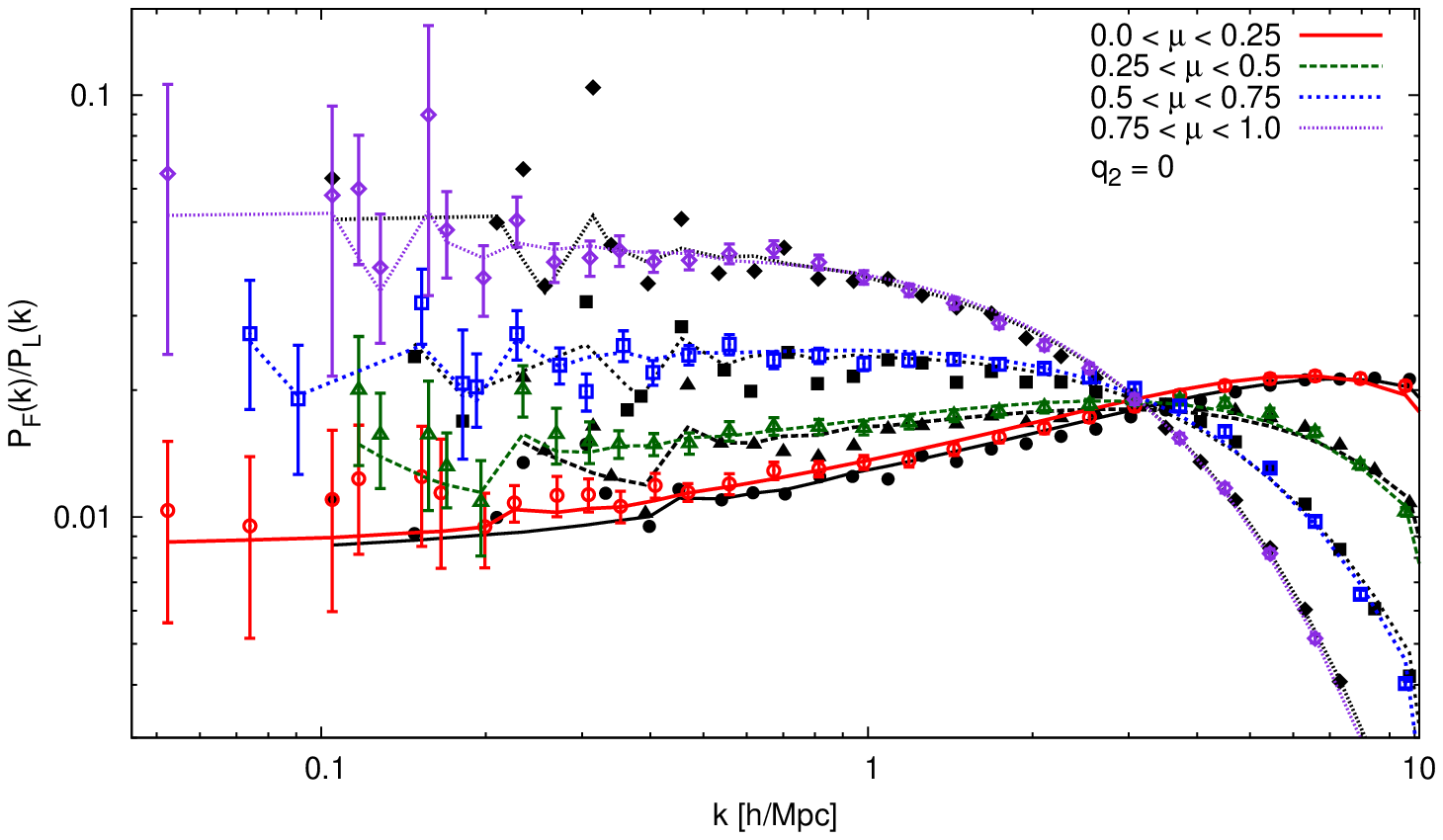}
\caption{\label{fig:boxesMcd} Comparison of $P_F/P_L$ from simulations
L120, with box size of 120 Mpc$/h$ (colored points and curves), and
R384C, on a 60 Mpc$/h$ box (black curves and points with omitted
errorbars), at $z=2.2$. The two simulations have equivalent physical
resolution, with $768^3$ and $384^3$ particles, and $512^3$ and $256^3$
grid cells, respectively, at $z = 2.2$, so differences should be due to
the effects of box size and the random initial conditions displaced by
a factor of 2 in $k$ only. Upper panel shows the fit with $q_2$ as a
free parameter, and bottom panel is for $q_2=0$.}
\end{figure}

  Next, we compare two simulations of the same model that differ only
in the box size, but have the same physical resolution in both particles
and grid cells: L120 and R384C. The R384C simulation has box size
$L=60 \hmpc$, half of that of L120, and also half the number of
particles and grid cells along a LOS, so that the initial
interparticle spacing and grid cell size are the same. The ratio
$P_F/P_L$ is shown in figure \ref{fig:boxesMcd} for these two models at
$z=2.2$. The upper panel shows the fits with free $q_2$, and the lower
panel the fits with $q_2=0$.
Two types of differences induced by the finite box size are seen in this
plot. First, at low $k$, the accuracy in fitting bias factors is
increased as the box size is increased and more points are available to
determine the linear limit
of the power spectrum. This is seen more clearly for the lowest $\mu$
curve, which is raised by $\sim 10\%$ from R384C to L120 in the free
$q_2$ fit. Second, the points for the R384C simulation are generally
lower than for L120 even at high $k$, with the largest difference
ocurring also for the low $\mu$ curve, where it reaches $\sim 5\%$.
The main reason for this difference at high $k$ is that we are always
computing the \lya transmission power after fixing $\bar F$, and the
value by which we need to multiply the optical depth to achieve a fixed
$\bar F$ changes in the various models we show.
The large-scale power that is missing in the R384C simulation results
in larger voids with lower densities, and for a fixed intensity of the
ionizing background, this would result in a higher $\bar F$. The
intensity of the ionizing background therefore needs to be lowered to
have the same $\bar F$ in the two simulations, and this induces the
change in power at high $k$. At the highest values of $k$ near
$10 \hmpc$, the power measured in the two simulations shown by the
colored and black points is very close, but as we have seen previously
this power is quite sensitive to the particle and grid resolution,
which are the same in these two simulations.

%______________________figu______________________%
\begin{figure}[!htbp]
\centering
\includegraphics[width=1.0\textwidth]{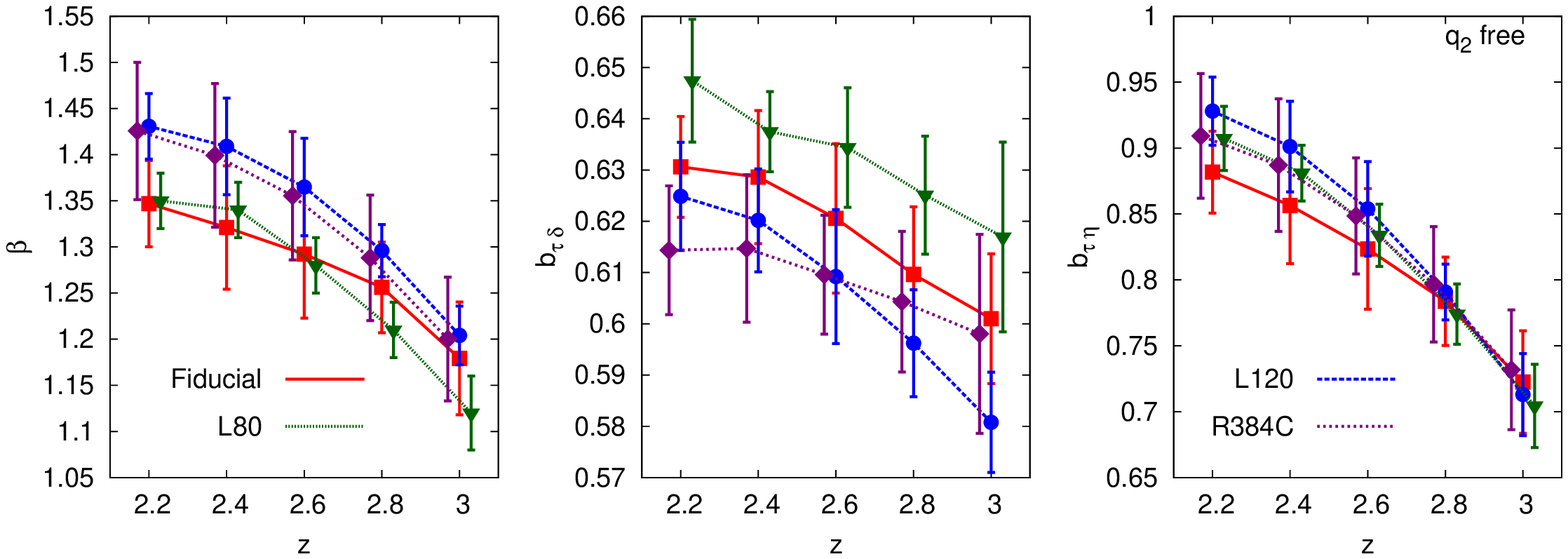}
\caption{\label{fig:boxes1,2} Comparison of the bias and $\beta$
parameters for different box sizes. Notice that while the R384C and L120
simulations have equivalent physical resolution, L80 has the same
particle density but a different physical grid size. The results shown
here are for free $q_2$ in the $D_1$ formula.}
\end{figure}

  Linear bias factors are shown in figure \ref{fig:boxes1,2} for the
fits with free $q_2$ (blue circles for L120 and purple rhombi for
R384C), where a third model of intermediate box size, L80, has been
added (green triangles). The results for the fiducial model, on a
$60 \hmpc$ box, are also shown as red squares (these were shown only for
the $q_2=0$ fit in previous figures). In general, the bias factors are
higher by $\sim 10\%$ for the free $q_2$ fit compared to the $q_2=0$
fit. While the bias factors are nearly the same for L120 and R384C,
the L80 simulation has a larger value of $b_{\tau\delta}$ by $\sim 5\%$
at all redshifts, while $b_{\tau\eta}$ is nearly the same, and therefore
$\beta$ is lower for L80. This may seem strange because L80 is a
simulation of intermediate box size. The L80 simulation also has the
same particle resolution as the others, but its grid size is smaller
(with the same number $512^3$ of cells as the larger box of L120). As
seen previously in figure \ref{fig:res1,2}, increased grid resolution
raises the value of $b_{\tau\delta}$ and lowers $\beta$, and correcting
for this brings the green lines (for L80) a bit closer to the blue and
purple lines (for L120 and R384C) in figure \ref{fig:boxes1,2}. However,
most of the difference remains, and we believe this can be assigned to
the effects of random initial conditions. Even though these simulations
have the same Rayleigh-suppressed initial conditions on modes of fixed
$kL$, the power of a simulation at fixed $k$ still has random
variations, and therefore we can expect some of the variations in bias
factors analogous to the ones between seeds 1 and 3 seen in figure
\ref{fig:seedb}.

  We conclude that the systematic effects of the box size are comparable
to the random variations due to different initial conditions seen in
figure \ref{fig:seedb}, and also to the systematics due to limited
resolution in our simulations. If anything, the error resulting from
using different fits to the whole shape of $P_F/P_L$, which we have
illustrated by including or not including the additional free parameter
$q_2$ in this paper, is a bit larger: both $b_{\tau\delta}$ and
$b_{\tau\eta}$ are systematically larger by $\sim 10\%$ for free $q_2$
than for fixed $q_2=0$ at low redshift in most of our models, with a
smaller difference at high redshift. The better fit that is obtained
for free $q_2$ to the low-$k$ points suggests this to be more reliable,
but this fit also implies a rather small value of the $q_1$ parameter
(see figure \ref{fig:zplot_L120_3-7}), and therefore a surprisingly fast
convergence of the non-linear factor $D_1$ to unity at low $k$. This
remains to be tested with more large box simulations.

%{\bf MV this plot describing the non-linear parameters I would put in the appendix and I will add here just one-two sentences: The non linear parameters seen in \hyperref[fig:boxes3-8]{ figure \ref*{fig:boxes3-8}} show large discrepancies, and even for two parameters, $k_p, \, b_v$ L80 does not behave as expected, their values not falling in between L60 and L120 but lower for the two cases. This  behaviour might have to do with the method used to select the fiducial model (explained in the appendix), or the different physical resolutions used when comparing L80 with the others, nevertheless the effects are minor and on the fitting parameters, but show how complex this study is. }

%______________________figu______________________%
%\begin{figure}[!htbp]
%\centering
%\includegraphics[scale=1.09]{zevolLbox3,4,5,6,7,8,_new.eps}
%\caption{\label{fig:boxes3-8} {\bf MV: I would put this in the appendix} Comparing the nonlinear parameters for different box sizes. Notice that while L60 and L120 have equivalent physical resolution, i.e. L60 has  $384^3$ particles and $256^3$ cells, L120 has $768^3$ par and $512^3$ cells L80 has both different physical and simulation resolution since it has $512^3$ particles and $512^3$ cells.  }
%\end{figure}

%______________________subsection______________________%
\subsection{Comparing Lagrangian and Eulerian simulations}
\label{subs:chek-matvscen}

  As a final test, we analyze here a simulation that is directly run on
an Eulerian grid instead of using gas particles. This is the simulation
labeled "Euler" in table \ref{tab:sims} and described in section
\ref{subsubs:simCen}. As a comparison, we have run an SPH simulation
designed to match the physical parameters of the Euler simulation, so
that a comparison can be made. This simulation is labeled "Lagrange"
in table \ref{tab:sims}: it has the same box size of $L=50 \hmpc$, and
resolution of $512^3$ both in number of particles and grid cells. The
model mass fluctuation power spectrum is the same, and we have also
attempted to match the two simulations to have the same
density-temperature relation, which depends on the model of the ionizing
background and the heating and reionization history. Unfortunately, it
is not possible to do this match exactly owing to the different methods
for computing heating in the two simulations. In addition, the initial
conditions were not the same in the two simulations, (in this case both
of them include the Rayleigh distribution of Fourier mode amplitudes),
so this inevitably introduces some difference in the results.

%______________________figu______________________%
\begin{figure}[!htbp]
\centering
\includegraphics[scale=0.59]{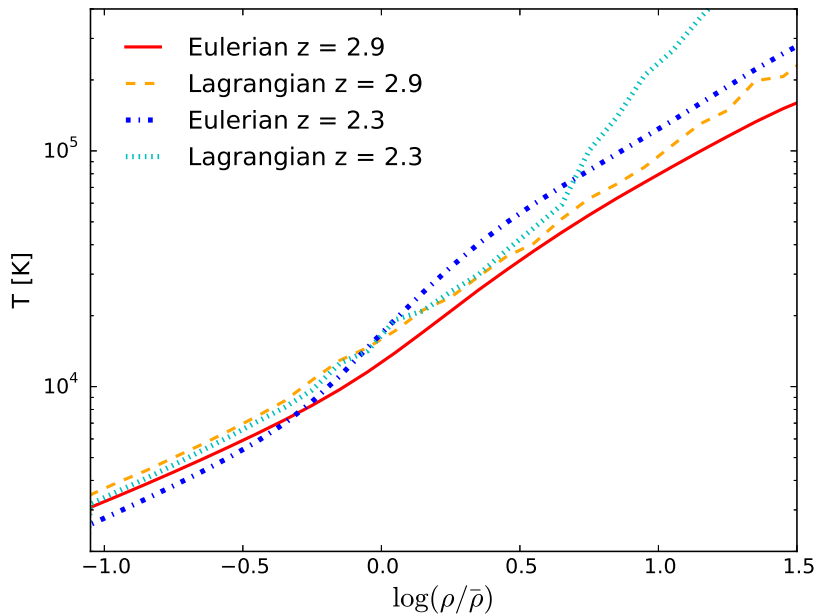}
\caption{\label{fig:pfeul} Mean temperature as a function of density for
the Eulerian and Lagrangian simulations, at $z=2.3$ and $z=2.9$. }
\end{figure}

  The values of the mean transmission for the Eulerian simulation were
determined by the model of the ionizing background evolution used in
that simulation, and are equal to the following values:
$\bar{F}=0.8042$, $0.7484$ and $0.6725$ at the three redshift outputs
of $z=2.3$, $2.6$ and $2.9$, respectively. We choose to rescale the
optical depths in the Lagrange simulation to match these same values of
the mean transmission, which are different from our standard ones used
in previous figures. The redshift outputs for the simulation Lagrange
are our standard set of 5 values, for which we set the mean transmission
to the same value as the Eulerian one at $z=2.6$, and to values
following a power-law dependence of the same form as in equation
\ref{eq:meanF}, separately chosen to match the mean transmission of the
Euler simulation for $z<2.6$ and for $z>2.6$. This results in the
following values: $\bar F = (0.8212, 0.7864, 0.7484, 0.6989, 0.6452)$
at $z = (2.2, 2.4, 2.6, 2.8, 3.0)$.

%______________________figu______________________%
\begin{figure}[!htbp]
\centering
\includegraphics[scale=0.99]{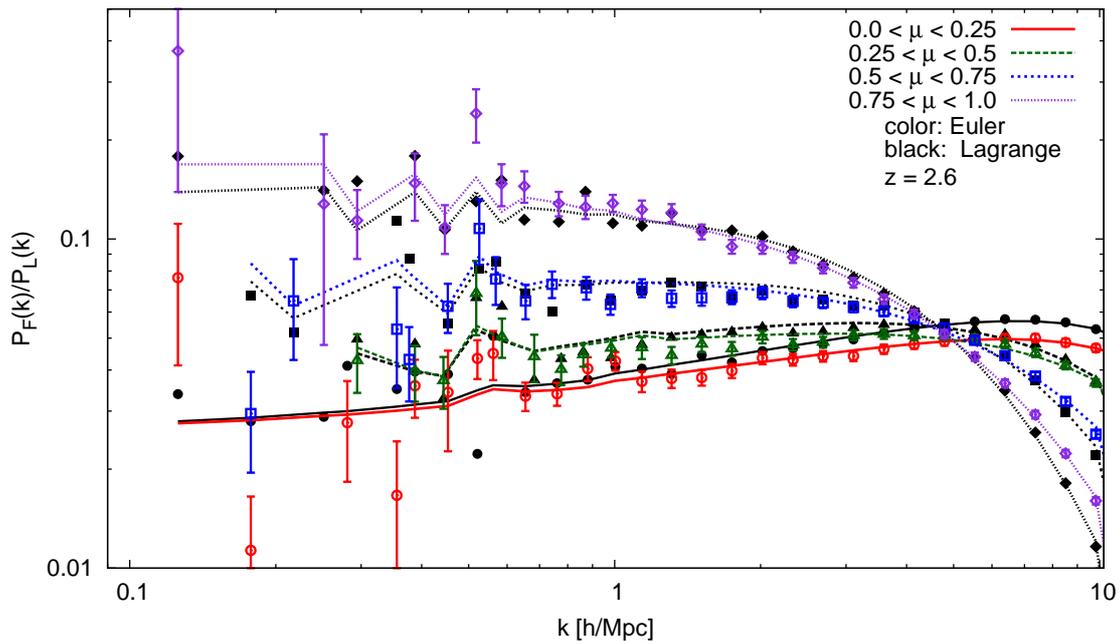}
\caption{\label{fig:pfeul} Comparison of $P_F/P_L$ for the Euler
and Lagrange simulations, using an Eulerian code and the {\sc GADGET-II}
code, of the same cosmological model, at $z=2.6$.}
\end{figure}

%______________________figu______________________%
\begin{figure}[!htbp]
\centering
\includegraphics[width=1.0\textwidth]{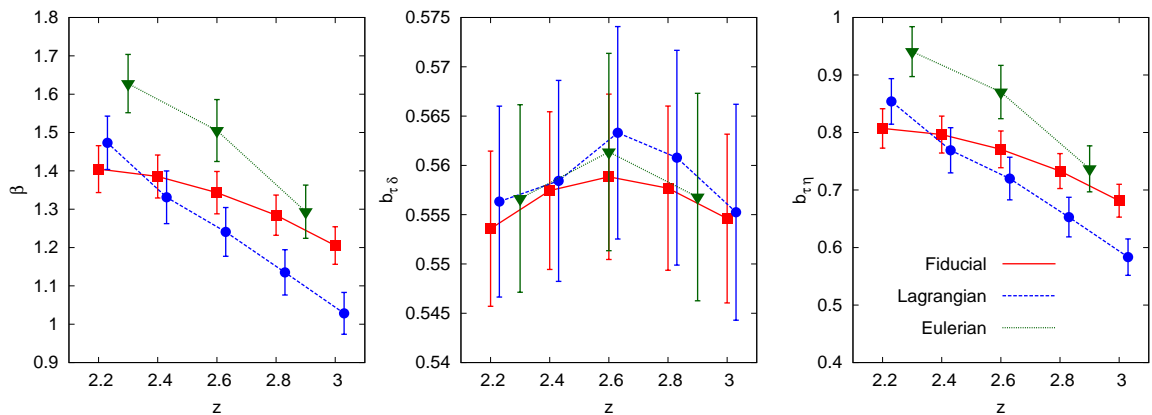}
\caption{\label{fig:euler1,2} Redshift evolution of $\beta$ and the bias
factors for the Euler and Lagrange simulations, from the fits with
$q_2=0$.}
\end{figure}

  The full density-temperature relation of the Euler simulation is shown
in figure \ref{fig:pfeul}, at redshifts $z=2.3$ and $z=2.9$. The
temperature that is shown is the average in all the cells that have the
density in each bin of width $\Delta\log\rho = 0.1$. The relation is not
exactly a power-law, and it steepens above $\rho/\bar\rho \sim 1$ due to
shock-heating of the gas in the sheets and filaments of the \lya forest
as structure formation develops. The two simulations are fairly close to
each other, but in the low-density regime the Eulerian one has lower
temperature by $\sim 0.1$ dex. At densities $\rho/\bar\rho > 1$ the
shock heating in the Eulerian simulation is higher than in the
Lagrangian one, particularly at low redshifts, and at
$\rho/\bar\rho > 5$ the Lagrange simulation shows more heating at low
redshift. This seems to be affected by the collapse and sudden heating
of a few clusters in the simulation and is therefore affected by random
variance depending on initial conditions. We have obtained a linear
regression fit to the $\log T - \log \rho$ relation of the Euler
simulation in the low-density regime
$-1.5 < \log(\rho/\bar\rho) < -0.3$, in which shock heating is not
important, with the result indicated in Table \ref{tab:sims} for
$z=2.6$, which has a weak variation with redshift. The difficulty in
obtaining a more precise match of the density-temperature relation in
the Eulerian and Lagrangian simulations illustrates the problems that
are encountered when testing that numerical simulations using different
codes and different methods make the same predictions for the \lya
forest power spectrum.
More precise tests, that start from exactly the same initial conditions
and force the same thermal histories, should be performed in order to
investigate the differences between the two frameworks more completely
(see, e.g., \cite{regan07}, where a comparison between SPH and
Eulerian codes in terms of the 1D transmission power spectrum is
presented and where differences of order $<5\%$ are found).

  The ratio $P_F/P_L$ is shown in figure \ref{fig:pfeul}, for the Euler
simulation as colored curves and points, and the Lagrange one for black
curves and points with omitted errorbars, at $z=2.6$. The curves are
fits to the $D_1$ formula with $q_2=0$. Important differences between
the two simulations are apparent: first, at low $k$, the high-$\mu$
curves are higher for the Euler simulation, implying a higher predicted
value of $\beta$. At high $k$, the curve at lowest $\mu$ is lower for
Euler than Lagrange, whereas the high-$\mu$ curves are also at lower
power except at $k> 5 \hmpc$, where the power decreases more slowly for
Eulerian simulation. The differences are clear and at the level of
$\sim 10\%$, and they may be due to several effects having to do with
the different resolution and methods of the simulations. It is useful in
particular to compare with figure \ref{fig:mcdres}, which shows that
poor resolution produces a sharper decline of the power at the highest
$k$, suggesting that the poorer resolution of the Lagrange simulation
compared to the Euler one at mean density can explain this difference
in figure \ref{fig:pfeul}.
Nevertheless, the fact that the two simulations agree within 10\%
on the predicted power (except at low $k$ and high $\mu$ where the
difference grows to 20\%, but is sensitive to random variance from the
different initial conditions) is reassuring.

  The bias factors are shown in figure \ref{fig:euler1,2}. The redshift
distortion factor (left panel) predicted by the Euler simulation is
substantially higher than for the other simulations analyzed so far,
corresponding to the higher power of the colored curve at low $k$ and
high $\mu$ in figure \ref{fig:pfeul}, whereas the variations in
$b_{\tau\delta}$ are very small, but as discussed above this is
sensitive to the variance in the different initial conditions. The
differences in the predictions between the two methods will need to be
understood in more detail before theoretical predictions can be made
from these hydrodynamic cosmological simulations at a high level
of accuracy. However, the main features of the evolution of these bias
factors are common to all our results: $b_{\tau\delta}$ is roughly
constant with redshift and near 60\%, and $b_{\tau\eta}$ is slightly
below unity and declines with redshift, implying also a decline of
$\beta$ with redshift.

%______________________figu______________________%
%\begin{figure}[!htbp]
%\centering
%\includegraphics[scale=1.09]{fig18par_Eul.eps}
%\caption{\label{fig:euler3-8} Differences in the measured non-linear parameters for Euler and Lagrange simulations. {\bf MV: probably would move this figure in appendix}.   }
%\end{figure}

%\input{./tables/eulertab}
%\clearpage 

%% file: physical.tex
%________________________SECTION_______________________________%

\section{Dependence of the \lya Power Spectrum on the Physical Model}
\label{sec:get}

  We analyze in this section the dependence of the \lya transmission
power spectrum on the physical characteristics of the IGM and the
cosmological model. As described in \S \ref{sec:intro}, the \lya power
spectrum depends mainly on the amplitude and shape of the initial matter
power spectrum, the density-temperature relation, and the value of the
mean transmission at each redshift. Other properties of the cosmological
model should not affect $P_F(k,\mu)$, except for rescalings: for example,
changes in $H_0$, $\Omega_b$ and $\Omega_m$ simply imply a rescaling of
the Jeans scale and any physical scale in the power spectrum in terms of
angular and redshift separations, which are the observed coordinates of
spectral pixels. The details of the gas cooling time, which depend on
$\Omega_b$ and the ionizing background spectrum, can be thought of as
being incorporated in the density-temperature relation and its history,
which determine the impact of the Jeans scale and the thermal broadening
on the \lya power spectrum.

  The dependence of the \lya power spectrum with the most relevant
physical parameters are described below, by examining the following
cases:
\begin{itemize}
\item Dependence on the amplitude of the power spectrum, characterized
by the parameter $\sigma_8$, which we vary for our fiducial model (\S
\ref{subs:ansig}).
\item Dependence on the power spectrum slope on the range of scales
examined by our simulations, which we illustrate by comparing the
fiducial and Planck models (\S \ref{subs:pslope}).
\item Dependence on $\bar F$ at a fixed redshift, and redshift
evolution (\S \ref{subs:anZ}).
\item Dependence on the density-temperature relation, parameterised as
$T=T_0 (\rho/\bar\rho)^{\gamma-1}$ (\S \ref{subs:angam}).
\end{itemize}
Throughout this section, we use only fits with the formula $D_1$ fixing
$q_2=0$.
 
%______________________subsection______________________%
\subsection{Power spectrum amplitude}
\label{subs:ansig}

%______________________figu______________________%
\begin{figure} [!htbp]
\centering
\hspace*{-0.02\textwidth}
   \includegraphics[width=0.49\textwidth]{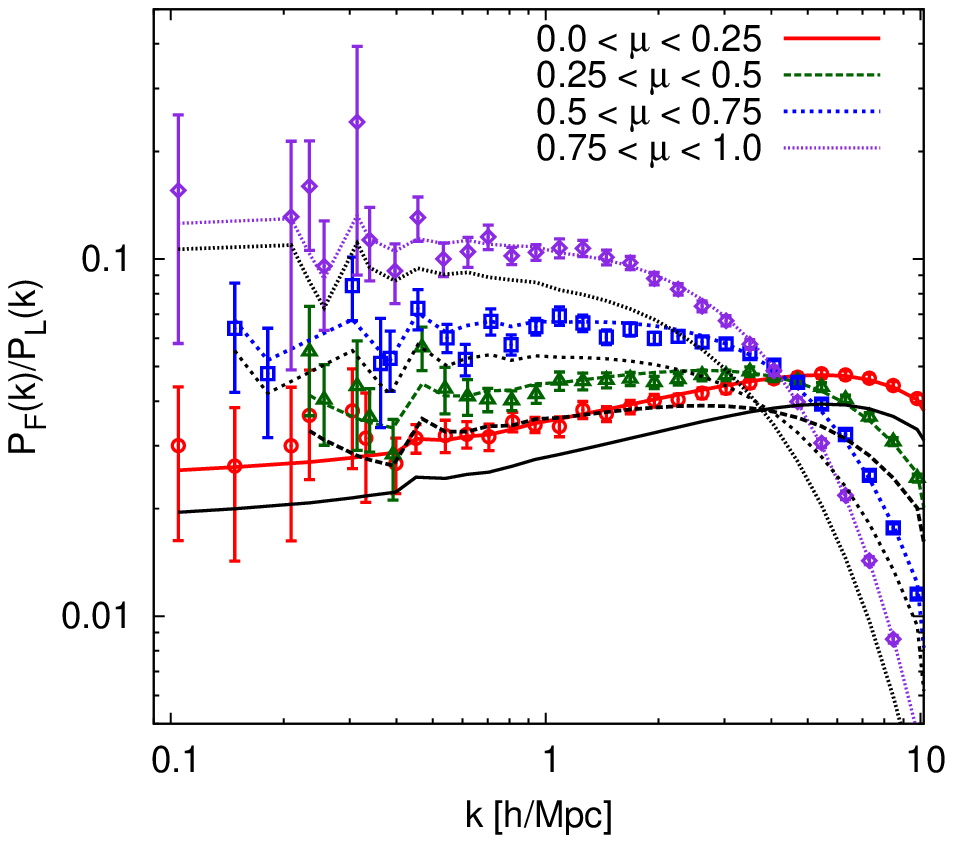}
\hspace*{0.01\textwidth}
    \includegraphics[width=0.49\textwidth]{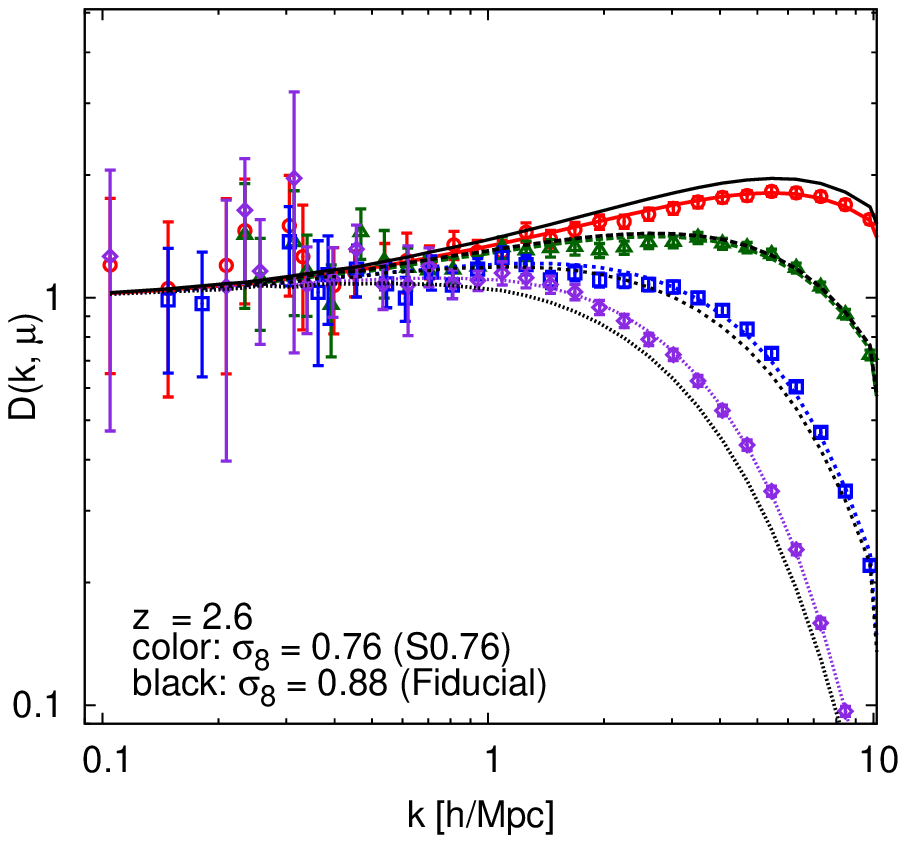}
\caption{\label{fig:sigMcd} Comparison of $P_F/P_L$ (left panel) for
models with different power spectrum amplitude, S0.76 with
$\sigma_8=0.7581$ (colored curves and points), and the fiducial model
with $\sigma_8=0.8778$ (black curves; points omitted), at z=2.6. The
non-linear term $D(k,\mu)$ is shown in the right panel. As expected,
$D-1$ increases with the amplitude, and the characteristic scale at
which non-linear anisotropy starts dominating, where curves with
different $\mu$ cross each other in the left panel, increases ($k_{na}$
decreases) with the amplitude.}
\end{figure} 

  The ratio $P_F/P_L$ is compared in figure \ref{fig:sigMcd} (left
panel) for the model S0.76 and the fiducial model. The only difference
between these two models is in the amplitude of their linear power
spectrum, which is lower by a factor $(0.7581/0.8778)^2=0.746$ in S0.76.
The colored curves are generally higher than the black ones by a factor
that is slightly smaller than $0.746$, indicating that $P_F$ increases
slowly with $\sigma_8$, so $P_F/P_L$ decreases although not as fast as
$\sigma_8^{-2}$. As the power spectrum amplitude increases, we expect
that non-linear effects become important at increasingly long scales.
This is confirmed by the fact that the characteristic wavenumber
$k_{na}$ at which the curves with different $\mu$ cross each other
decreases with $\sigma_8$. This crossing point represents the
characteristic scale where the power spectrum quadrupole changes sign,
switching from the linear Kaiser effect to the non-linear velocity
dispersion effects. In the right panel, the non-linear correction
$D(k,\mu)$ is shown, which generally increases its departure from unity
as $\sigma_8$ is increased. Black points are omitted in these plots to
avoid excessive cluttering, but they are similarly well fitted by the
black curves as for the colored ones.

%______________________figu______________________%
\begin{figure} [!htbp]
\centering
\includegraphics[scale=0.99]{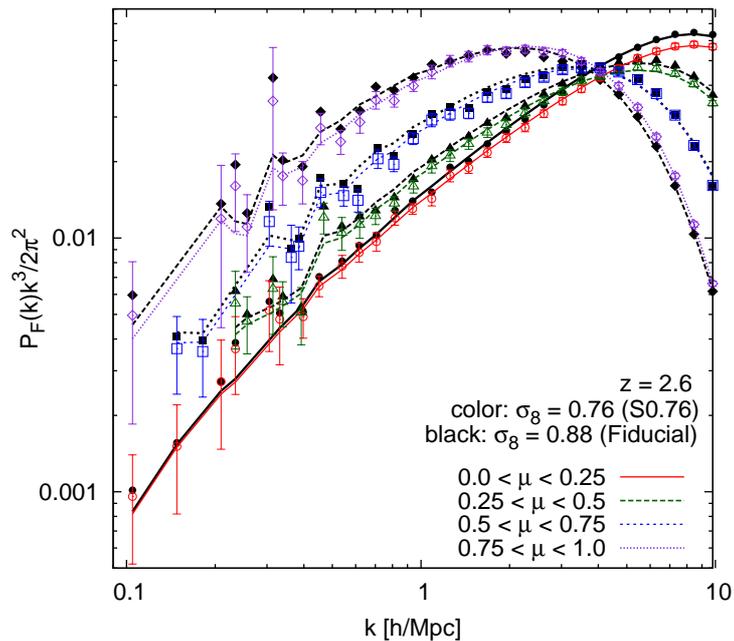}
\caption{ \label{fig:sigpf} Amplitude of transmission fluctuations in
Fourier space, defined as $P_F k^3/(2\pi^2)$. The power $P_F$ increases
slowly with the amplitude of the linear power spectrum parameterized by
$\sigma_8$, except at high $k$ and $\mu$ where it decreases with
$\sigma_8$. Black points, for the fiducial model, have their errorbars
omitted but they are equal to the colored ones.}
\end{figure}

  The direct dependence of $P_F$ on the amplitude is better visualized
in figure \ref{fig:sigpf}, where we plot the quantity
$P_F(k)k^3/(2\pi^2)$ for the same models as in figure \ref{fig:sigMcd}.
Analogously to the amplitude $\Delta^2(k)$ in equation (\ref{eq:D2k}),
this is the dimensionless Fourier amplitude of transmission
fluctuations. Here the black points are also included (with omitted
errorbars, which are equal to the colored errorbars). At low $k$, $P_F$
increases very little for
$\mu=0$ as $\sigma_8$ increases, but this increase is more pronounced at
high $\mu$, indicating an increase of $\beta$ with $\sigma_8$. At high
$k$, $P_F$ continues to increase with the linear amplitude at low $\mu$,
but it decreases at high $\mu$. This corresponds to the decrease of the
wavenumber $k_{na}$ for the crossing of the curves in figure
\ref{fig:sigMcd} with increasing $\sigma_8$.
 
%______________________figu______________________%
\begin{figure} [!htbp]
\centering
\includegraphics[width=1.0\textwidth]{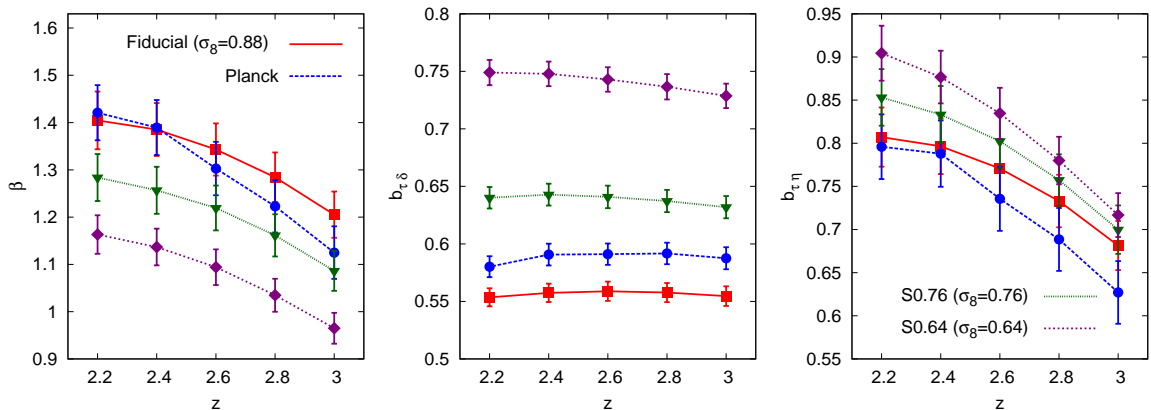}
\caption{\label{fig:sigb} Bias and redshift distortion factors as a
function of redshift for models S0.76 and S0.64, which differ from the
fiducial one only for a lower linear amplitude of the mass power
spectrum, and of the Planck model, which also has a different power
spectrum slope. All values are obtained from the fits with $D_1(k,\mu)$
fixing $q_2=0$.}
\end{figure} 

  The change of bias and redshift distortion factors is shown in figure
\ref{fig:sigb}. The very weak dependence of $P_F$ on the linear
amplitude at low $k$ and low $\mu$ implies that $b_{\tau\delta}$
decreases with the amplitude nearly as fast as $\sigma_8^{-1}$. At high
$\mu$ the power $P_F$ has a substantial increase with $\sigma_8$, so the
bias factor $b_{\tau\eta}$ decreases more slowly with $\sigma_8$,
particularly at high redshift, and as a consequence $\beta$ increases
with $\sigma_8$, as seen in the left panel.

\subsection{Power spectrum slope}
\label{subs:pslope}

%______________________figu______________________%
\begin{figure} [!htbp]
\centering
\hspace*{-0.02\textwidth}
   \includegraphics[width=0.49\textwidth]{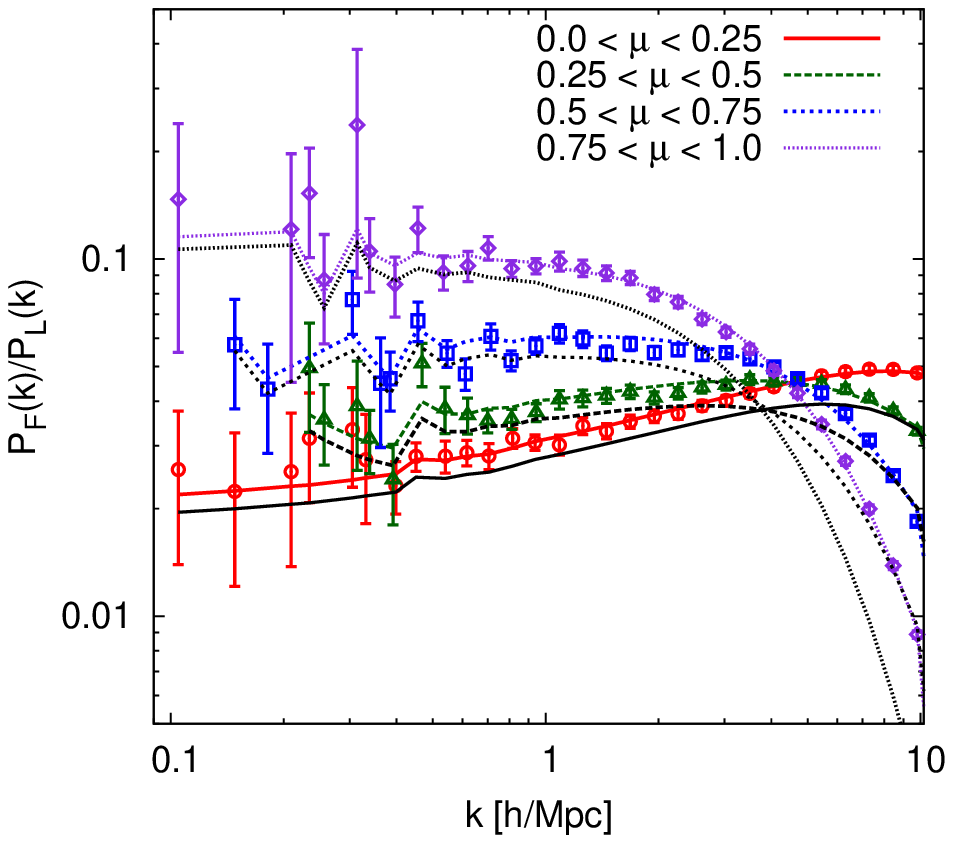}
\hspace*{0.01\textwidth}
    \includegraphics[width=0.49\textwidth]{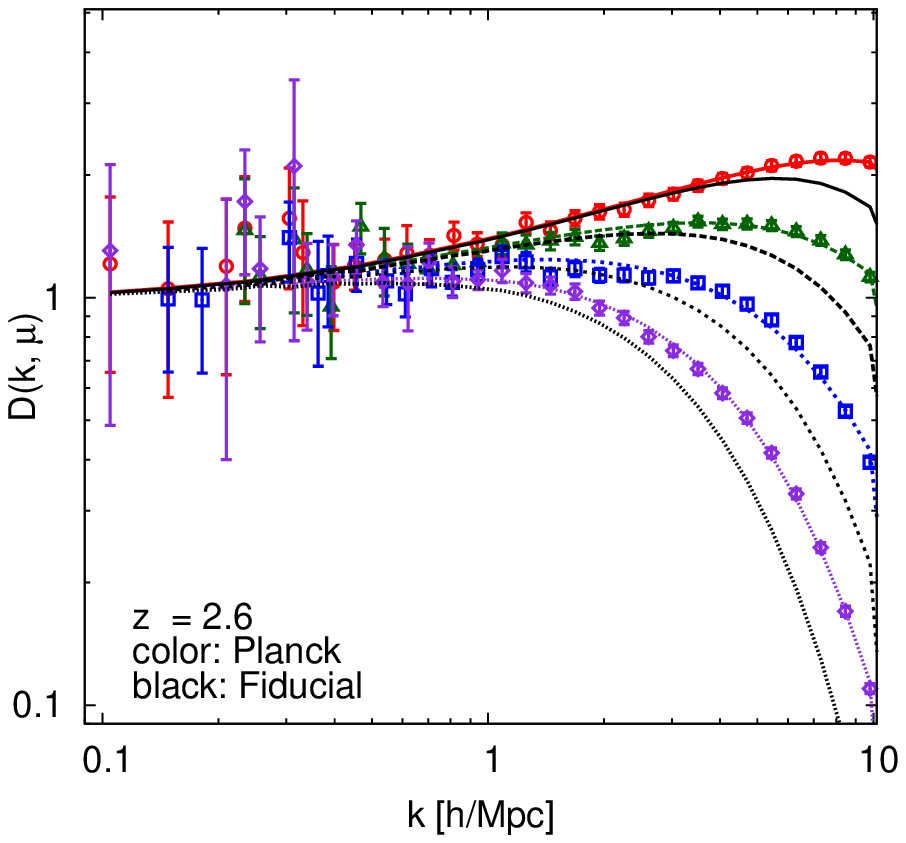}
\caption{\label{fig:PlaMcd} Comparison of $P_F/P_L$ (left panel) for
the fiducial model (with $n_s=1$) and the Planck model (with
$n_s=0.9624$). }
\end{figure} 

\begin{figure} [!htbp]
\centering
\includegraphics[scale=0.99]{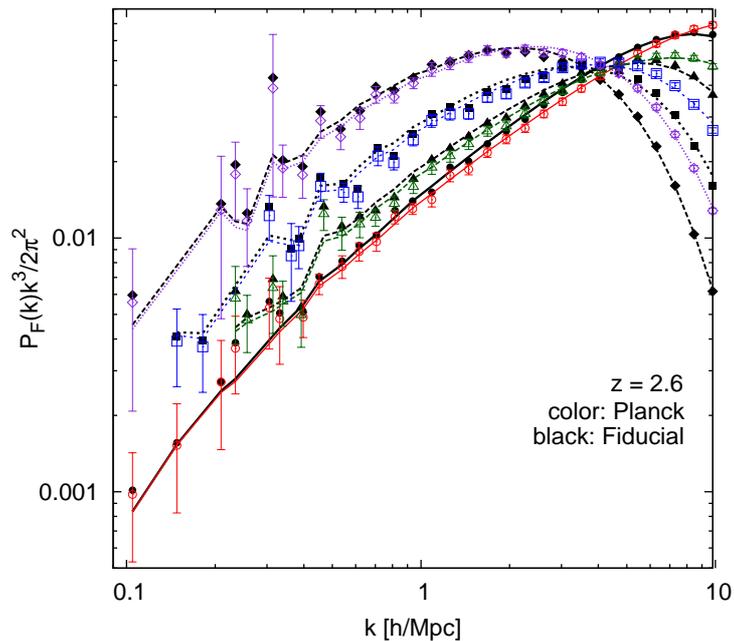}
\caption{ \label{fig:Plapf} Amplitude of transmission fluctuations
for the Planck model (colored curves and points) and the fiducial model
(black curves and points with errorbars omitted). }
\end{figure}

  We now compare the model labeled as "Planck" in table \ref{tab:sims},
which has parameters consistent with the ones measured by the Planck
mission, with the fiducial model. Figure \ref{fig:PlaMcd} shows
$P_F/P_L$ and $D(k,\mu)$ for the Planck model as colored curves and
points, compared to the fiducial model as black curves (with points
omitted). The transmission fluctuation amplitude, $P_F(k)k^3/(2\pi^2)$,
is shown also in figure \ref{fig:Plapf}, in all cases at $z=2.6$.
Interpreting the differences between these two models is in this case
more complicated than in the previous subsection, because now both the
amplitude and slope of the power spectrum are different. In addition the
density-temperature relation also differs slightly for these two models
as specified in table \ref{tab:sims}.

%______________________figu______________________%

  The fiducial and Planck models have slightly different values of
$\Omega_{0m}$, implying a difference in their growth factors. It is
therefore useful to express the mass fluctuation amplitudes on spheres
of radius $8\hmpc$ at $z=2.6$, instead of the present epoch, which have
the following values: $\sigma_8(z=2.6) = (0.3102, 0.2910, 0.2679)$ for
the fiducial, Planck and S0.76 models, respectively. The Planck model
therefore has a linear amplitude that is close to the average of the
fiducial and S0.76 models at the scale of $\sigma_8$. In addition,
comparing figures \ref{fig:sigpf} and \ref{fig:Plapf}, we can see that
the amplitude of $P_F$ of the Planck model is intermediate between that
of the fiducial and S0.76 models at the lowest values of $k$ probed by
our simulations, $k\sim 0.2 h/{\rm Mpc}$. This value of $k$ roughly
corresponds to the scale at which the parameter $\sigma_8$ measures the
amplitude of density fluctuations, and can therefore act as an
approximate pivot scale for comparing amplitudes of models with
different slope. In agreement with this, the Planck model has a bias
factor $b_{\tau\delta}$ intermediate between the values for the fiducial
and S0.76 models (see figure \ref{fig:sigb}). This bias factor has
therefore little sensitivity to the power spectrum slope, and depends
mostly on the amplitude parameter $\sigma_8(z)$. We caution, however,
that this bias factor may also depend on the temperature-density
relation, and is sensitive to the value of $\bar F$. The behavior of
$b_{\tau\eta}$ and $\beta$ shows a steeper redshift evolution of the
Planck model compared to the fiducial one with varying amplitude. This
different evolution may be affected by the lower temperature of the
Planck model compared to the fiducial one, as will be discussed in \S
\ref{subs:angam}.

  The ratio of the linear power spectra for the Planck model to that of
the fiducial model varies approximately as $k^{-0.03}$ over the range of
scales we explore, as caused mostly by the difference in the primordial
spectral index $n_s$ (with a smaller effect arising from the different
value of $\Omega_{\rm 0m}h$). One might expect a similar $k$-dependence
of the ratio of the transmission power $P_F$ for the two models, seen in
figure \ref{fig:Plapf}, but this would generally be misleading. In fact,
figure \ref{fig:sigpf} shows that the ratio of $P_F$ for the S0.76 and
fiducial models has a similar $k$-dependence at low $\mu$, even though
these two models have exactly the same $P_L(k)$ except for the
normalization. This is caused by the increase of the non-linear
correction $D$ with $\sigma_8$ at low $k$ and $\mu$.

  In general, the sensitivity of $P_F$ to $P_L$ is relatively weak on
scales $k \gtrsim 0.3 \hmpc$, when non-linear effects start. This means
that $P_F/P_L$ tends to behave inversely with $P_L$, as found in \S
\ref{subs:ansig} when analyzing the dependence of $P_F$ on the
amplitude, and therefore to increase faster with $k$ for the Planck
model compared to the fiducial one. Of course, on very large scales the
constancy of the bias factors implies that $P_F$ is proportional to
$P_L$ for a given model. At the same time, on very small scales ($k
\gtrsim 3 \hmpc$), the reason for the slower decline of $P_F/P_L$ for
the Planck model compared to the fiducial one is caused mostly by the
lower temperature in the Planck simulation (see \S \ref{subs:angam}).

%______________________figu______________________%
\begin{figure} [!htbp]
\centering
\includegraphics[width=1.0\textwidth]{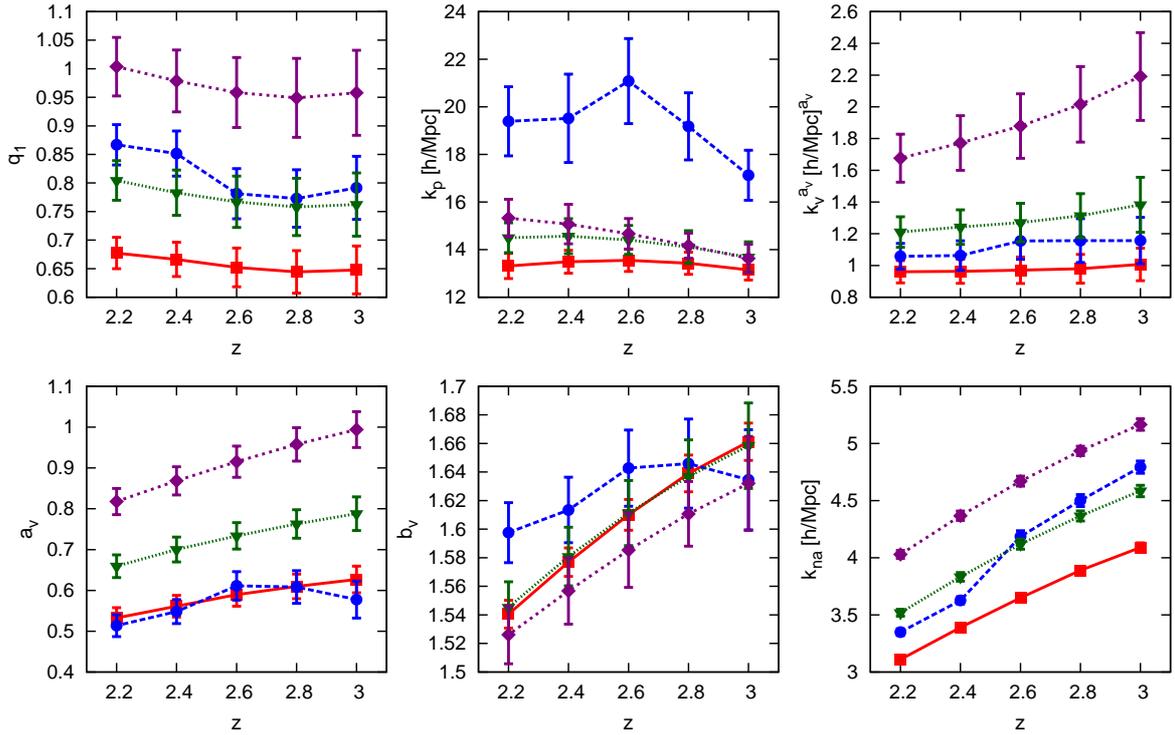}
\caption{\label{fig:sigpar} Parameters of the non-linear function
$D_1(k,\mu)$ for the fiducial model (red squares), for the S0.76
and S0.64 models with a decreasing amplitude of the linear power
spectrum (green triangles and purple rhombi), and for the Planck
model, with an intermediate amplitude between the fiducial and S0.76
models and a lower $n_s$ (blue circles). The sixth panel shows the
parameter $k_{na}$, the wavenumber at which curves of different $\mu$
cross each other in the diagram of $P_F(k,\mu)/P_L(k)$, computed
from equation (\ref{eq:betaa}).}
\end{figure} 

  Finally, figure \ref{fig:sigpar} shows the non-linear parameters of
the function $D_1(k,\mu)$ obtained in our fits for the fiducial, S0.76,
S0.64 and Planck models. The parameter $k_p$ is mostly dependent on the
gas temperature and is higher for the Planck model owing to its lower
temperature, and very similar for the other three models. The
parameters $q_1$, $a_v$ and $k_v^{a_v}$ mostly follow a progression
depending on the amplitude of $P_L$: as the amplitude increases, $q_1$
is decreased because the impact of non-linear terms does not increase as
fast as $\Delta^2(k)$ in equation (\ref{eq:Dapm7}), and $a_v$ and
$k_v^{a_v}$ are also reduced. Variations of $b_v$ are rather small.
The sixth panel in the figure shows $k_{na}(z)$, representing the scale
at which the curves of $P_F/P_L$ at different $\mu$ cross each other.
As expected, the errors of this parameter are small because most of the
degeneracy with other parameters is removed. The variation with redshift
and $\sigma_8$ confirms what we have described: as the power amplitude
increases, $k_{na}$ decreases.

%\pagebreak
%\newpage
 
%______________________subsection______________________%
\subsection{Mean transmission fraction and redshift evolution}
\label{subs:anZ}

%______________________figu______________________%
\begin{figure} [!htbp]
\centering
\hspace*{-0.02\textwidth}
   \includegraphics[width=0.49\textwidth]{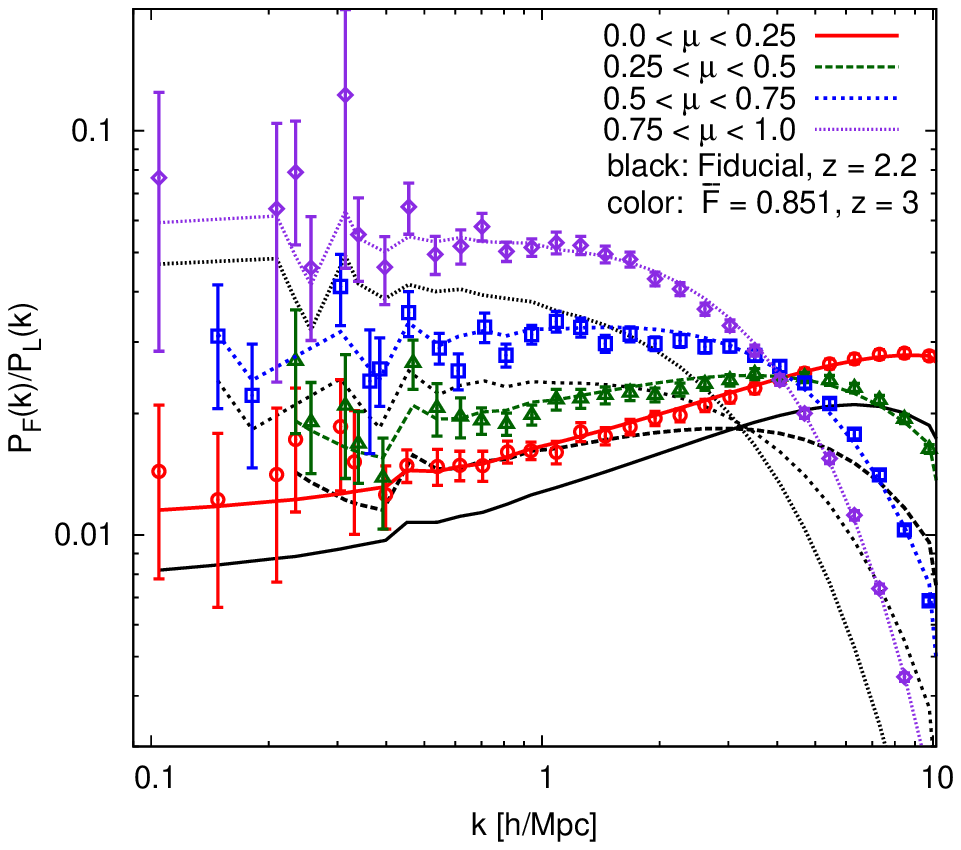}
\hspace*{0.01\textwidth}
    \includegraphics[width=0.49\textwidth]{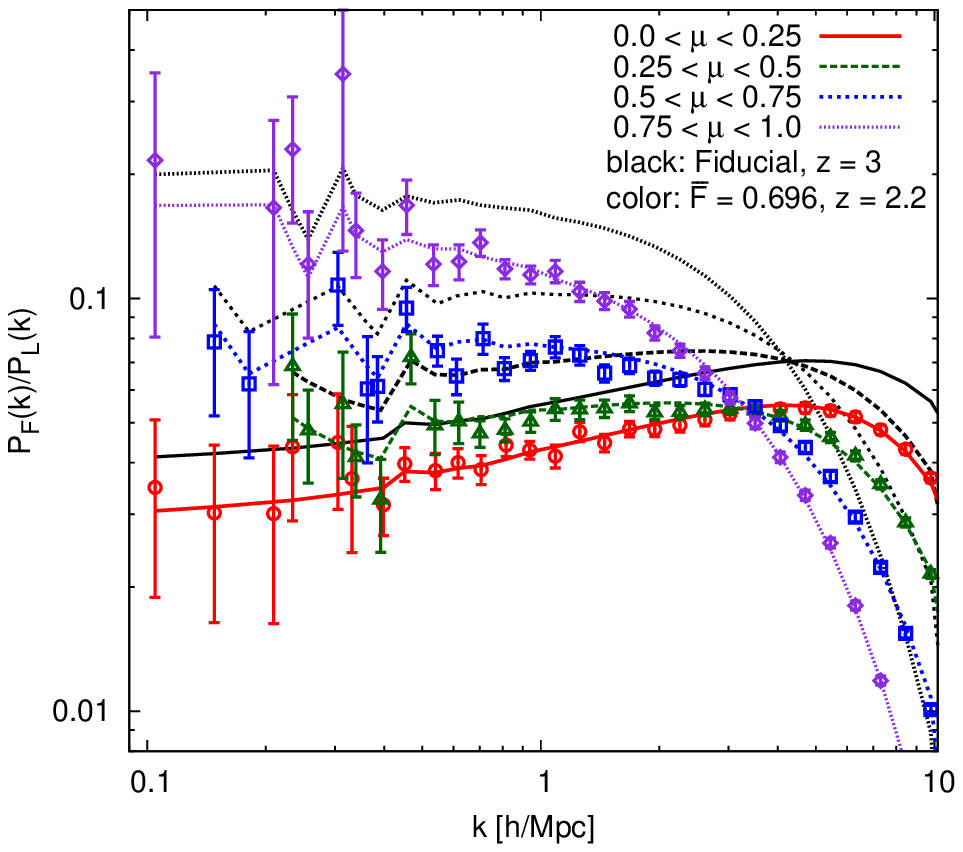}
\caption{\label{fig:PlaMcd} Evolution of $P_F/P_L$ with redshift when
the mean transmission is kept fixed at $\bar F = 0.851$ (left panel)
and $\bar F=0.696$ (right panel). The result of the fiducial model
is shown in both panels at $z=2.2$ and $z=3$, as black curves (with
points omitted) when the true value of $\bar F$ is used, and as colored
curves and points when $\bar F$ is modified to the value at the other
redshift. The observed evolution, from the black curves in the left
panel to those in the right panel, is a result of both the physical
redshift evolution at a fixed $\bar F$, and the variation of $\bar F$
with redshift.}
\end{figure} 

  The observed evolution with redshift of the non-linear power spectrum
is a consequence of two different effects: the physical evolution that
would be observed at fixed $\bar F$, and the change in $P_F$ that occurs
when varying $\bar F$ at a fixed redshift. These two effects are clearly
separated in figure \ref{fig:PlaMcd}, where the change from $z=2.2$ to
$z=3$ for $P_F/P_L$ is shown for the fiducial model at fixed $\bar F$.
In the left panel, we choose the value of $\bar F$ that corresponds to
$z=2.2$, and the standard result for $z=2.2$ for the fiducial model is
shown as the black curves. Colored curves and points show the result at
$z=3$ when using the value of $\bar F$ that corresponds to $z=2.2$.
Similarly, the black curves in the right panel show our standard result
at $z=3$, while the colored curves and points show the $z=2.2$ result
for the value of $\bar F$ that corresponds to $z=3$. The
density-temperature relation of our models is nearly constant with
redshift, so this should have a minimal impact on the variations seen
in this figure.

  The redshift evolution at fixed $\bar F$ arises basically as a
consequence of the varying amplitude of the linear power spectrum. The
amplitude $\sigma_8(z)$ varies by $4/3.2=1.25$ from $z=3$ to $z=2.2$
(neglecting the influence of the cosmological constant at these high
redshifts), so the redshift evolution behaves in the same way as the
variation with $\sigma_8$ in figure \ref{fig:sigMcd}: $P_F/P_L$
decreases nearly as fast as $\sigma_8(z)^{-2}$ at low $k$ and $\mu$,
with an increase of $\beta$ with $\sigma_8(z)$, and the value $k_{na}$
where the quadrupole changes sign decreases (i.e., shifts to larger
scales) with $\sigma_8(z)$. It is interesting to note that this value
of $k_{na}$ is not shifting at all under the large change of $\bar F$
from $0.851$ in the left panel to $0.696$ in the right panel, and
depends on the redshift only despite the clear variations of the shape
of the curves with $\bar F$.  

%______________________figu______________________%
\begin{figure} [!htbp]
\centering
\includegraphics[width=1.0\textwidth]{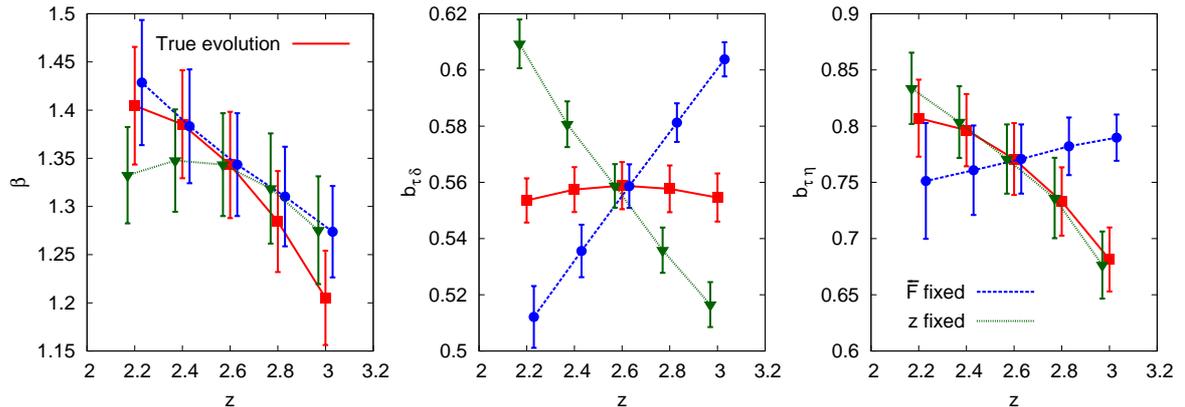}
\caption{\label{fig:zevb} Bias and redshift distortion factors for the
fiducial model. Red squares show our standard result for the redshift
evolution, when $\bar F(z)$ varies with redshift as observed. Blue
circles show the redshift evolution when we fix $\bar F=0.781$, equal to
the value at $z=2.6$. Green triangles show the variations at a
fixed redshift $z=2.6$, varying only $\bar F(z)$ to the value it has at
each of our standard redshift outputs. Blue circles (green triangles)
are shifted slightly right (left) to avoid superposition of errorbars. }
\end{figure} 

  The bias and redshift distortion factors are shown in figure
\ref{fig:zevb} for the fiducial model, for three different cases. The
red squares show their true evolution when $\bar F(z)$ is varied at
the same time as the redshift. The blue circles show the isolated
redshift evolution, when keeping $\bar F=0.781$ fixed to its value at
$z=2.6$, and the green triangles show the result when varying only
$\bar F$ but keeping fixed $z=2.6$ (only the redshift $z$ is shown
in the horizontal axis but the green points are all for $z=2.6$, and
for $\bar F$ equal to the value at the redshift in the axis).

  The redshift evolution of $b_{\tau\delta}$ for fixed $\bar F$ is, as
mentioned in \S \ref{subs:ansig}, equivalent to the variation with the
power spectrum amplitude, increasing with redshift nearly as
$b_{\tau\delta}\propto \sigma_8(z)^{-1}$, in agreement with the blue
points in the middle panel. When varying $\bar F$ at fixed
redshift, $P_F$ increases rapidly as $\bar F$ declines but this is
mostly due to the relation between $b_{F\delta}$ and $b_{\tau\delta}$.
The residual variation left for the optical depth bias on $\bar F$,
shown as the green points, is actually of opposite sign: as $\bar F$
decreases, regions of declining overdensity in the IGM are responsible
for the dominant variations measured in the \lya forest spectra, and
they have a decreasing physical bias factor. The true evolution we can
measure of $b_{\tau\delta}$, shown as the red points, is nearly constant
with redshift due to a fortuitous cancellation of the effects of
declining power spectrum amplitude and declining $\bar F$ with redshift.
This cancellation is, however, not a very precise prediction from our
simulations because it is sensitive to the fit we use to obtain the bias
factors, as can be seen from the redshift evolution of the fiducial
model bias factors for the free $q_2$ fit that is shown in figure
\ref{fig:boxes1,2}.

  The right panel of figure \ref{fig:zevb} shows that the usual decrease
of $b_{\tau\eta}$ with redshift is due to the variation of $\bar F$. In
fact, for fixed $\bar F$, $b_{\tau\eta}$ actually has a slow increase
with redshift. The reason why the redshift distortion factor $\beta$
generally decreases with redshift is therefore not very straightforward.
At fixed $\bar F$, the evolution of $\beta$ has a relatively simple
explanation: the decreasing power spectrum amplitude with redshift
implies an increase of $b_{\tau\delta}$ with redshift, while
$b_{\tau\eta}$ does not change much at fixed $\bar F$, so $\beta$
decreases with redshift. However, the reason why the ratio
$b_{\tau\eta}/b_{\tau\delta}$ has a redshift evolution that is not
strongly changed by the variation of $\bar F$ is not so simple.

%______________________figu______________________%
\begin{figure} [!htbp]
\centering
\includegraphics[width=1.0\textwidth]{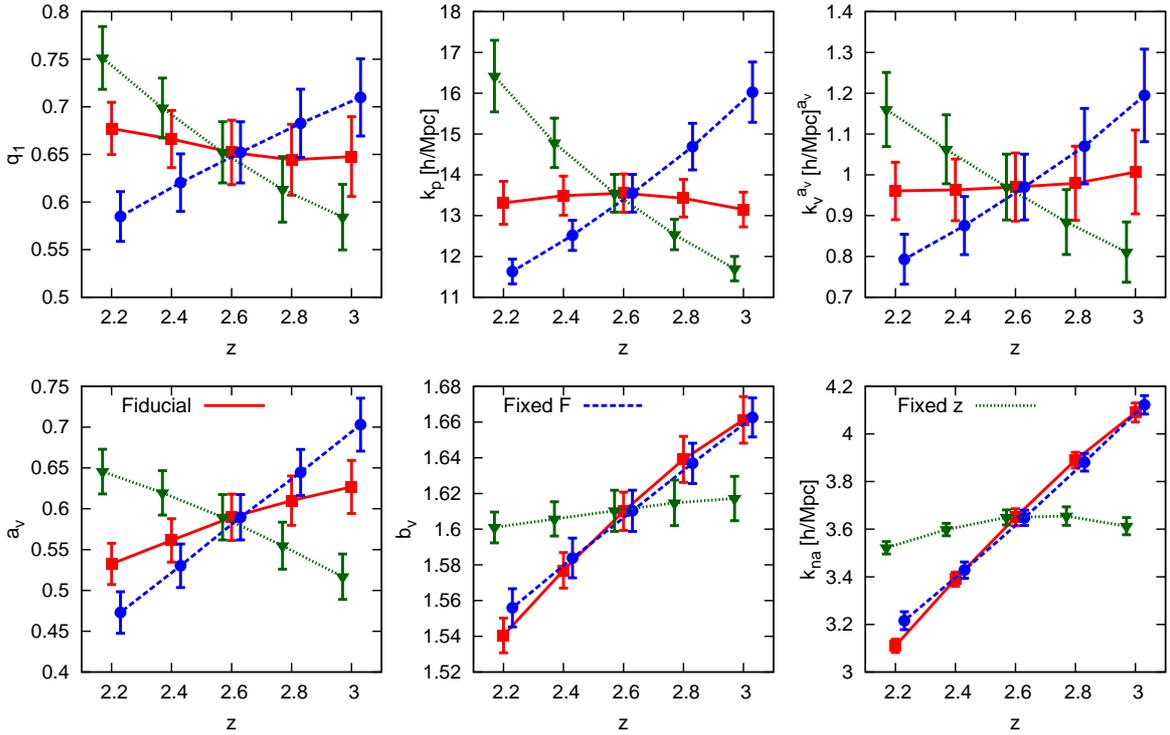}
\caption{\label{fig:flupar} Parameters of the non-linear function
$D_1(k,\mu)$ for the cases of true redshift evolution (red squares),
fixing $\bar F$ to its value at $z=2.6$ (blue circles), and fixing
$z=2.6$ but using the value of $\bar F$ corresponding to each redshift
(green triangles). }
\end{figure} 

  The non-linear parameters of the fits can be seen in figure
\ref{fig:flupar}. Some of the variations with redshift correspond to
the variations we have seen before with the amplitude in figure
\ref{fig:sigpar}: the parameters $q_1$, $k_v^{a_v}$ and $a_v$ increase
with redshift for fixed $\bar F$. However, these parameters also
increase with $\bar F$ and the result for the true evolution is a much
slower variation. It is interesting that the parameter $k_p$ also
varies with redshift when computed at fixed $\bar F$ and that this is
also cancelled by the dependence on $\bar F$, an effect that likely
depends on the slope of the density-temperature relation.

%______________________subsection______________________%
\subsection{Temperature-Density relation}
\label{subs:angam}

%______________________figu______________________%
\begin{figure} [!htbp]
\centering
%\hspace*{-0.02\textwidth}
   \includegraphics[scale=0.99]{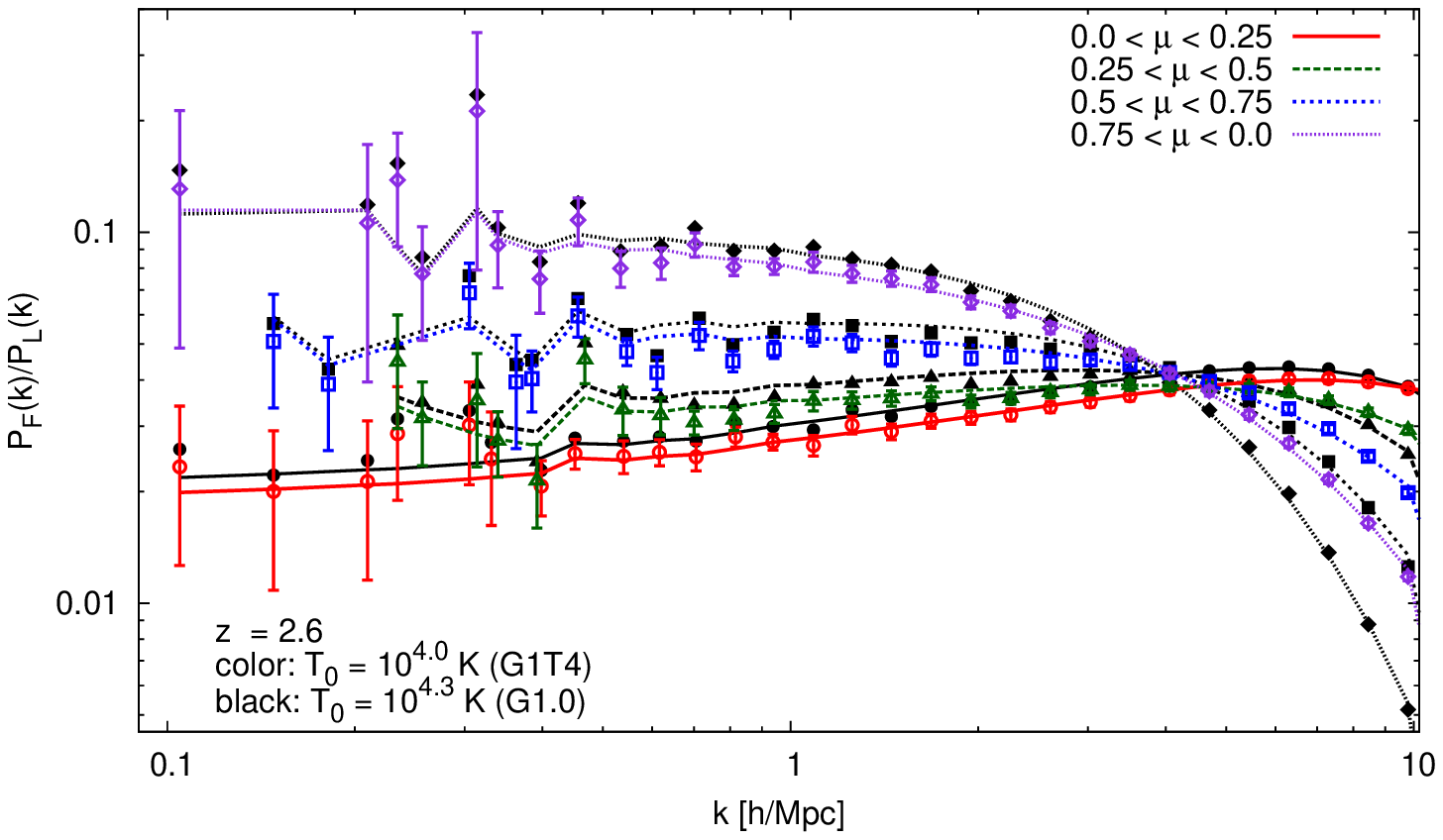}
%\hspace*{0.01\textwidth}
%    \includegraphics[width=0.49\textwidth]{fig28D_Tmp_z26.eps}
\caption{\label{fig:T0Mcd} Dependence of $P_F/P_L$
% (left panel) and the non-linear correction $D(k,\mu)$ (right panel)
on the gas temperature, at $z=2.6$. The two models shown have a
temperature-density relation slope $\gamma=1$, and the gas temperature
is twice higher for the black points compared to the colored ones. }
\end{figure} 

%______________________figu______________________%
\begin{figure} [!htbp]
\centering
%\hspace*{-0.02\textwidth}
   \includegraphics[scale=0.99]{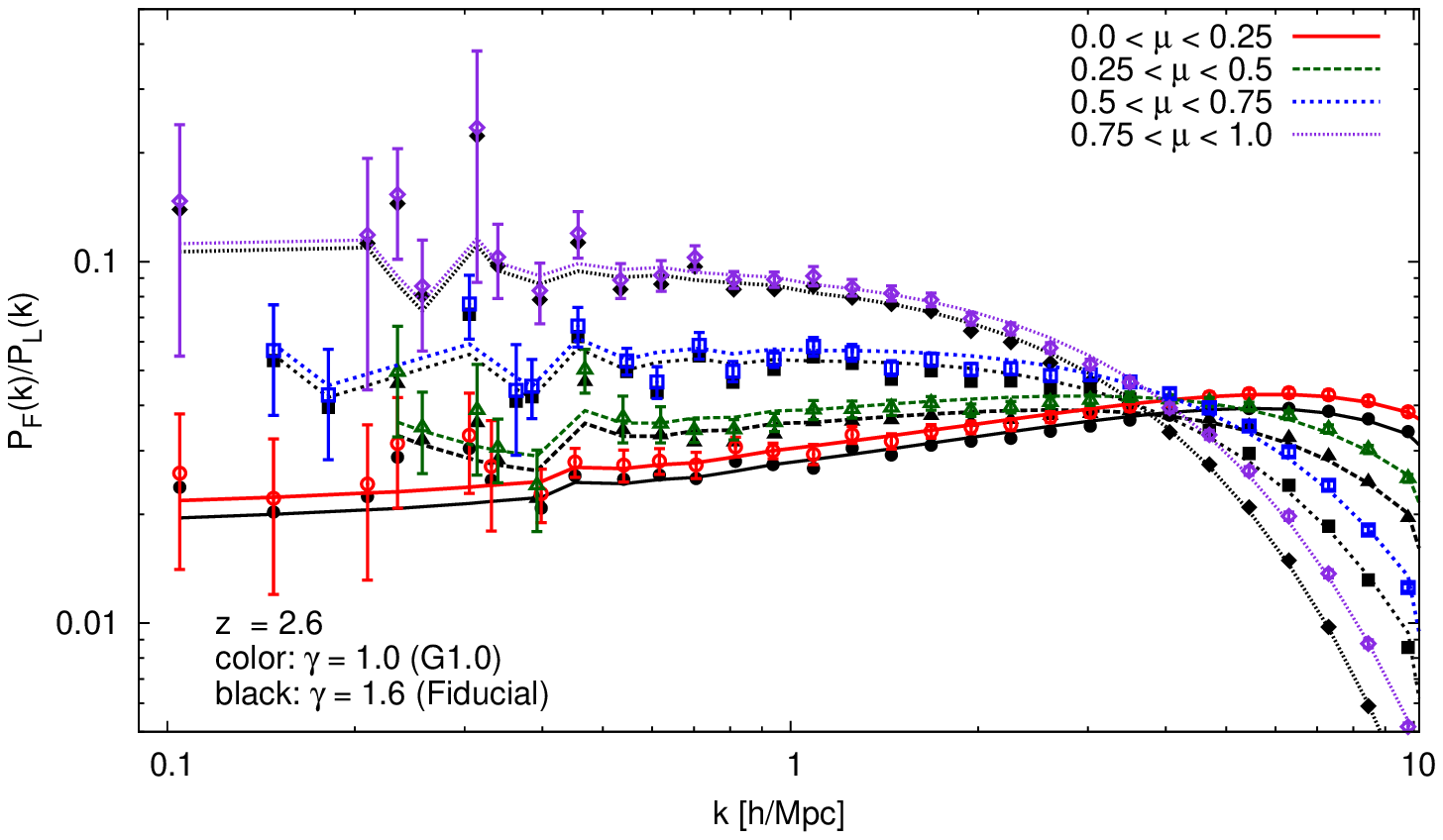}
%\hspace*{0.01\textwidth}
%    \includegraphics[width=0.49\textwidth]{fig29D_Gm_z26.eps}
\caption{\label{fig:GmMcd} Dependence of $P_F/P_L$ on the slope of the
temperature-density relation at $z=2.6$. The black
points are for the fiducial model, and the colored points show the
G1.0 model result, when modifying the slope to $\gamma=1$ (which is the
same model as the black curves in figure \ref{fig:T0Mcd}. }
\end{figure} 

  To finish this section, we investigate the dependence of $P_F/P_L$
on the temperature-density relation. First, we compare in figure
\ref{fig:T0Mcd} two models that have both $\gamma=1$ (a flatter relation
than in our fiducial model, which has $\gamma=1.6$), when we vary the
temperature $T_0$ at the mean density from $10^4$ K in model G1T4, shown
as colored points and curves, to $10^{4.3}$ K in model G1.0, shown as
black curves and points. The results are shown at $z=2.6$ and the fits
are for $q_2=0$. A lower temperature reduces the Jeans length
and, as expected, the damping of the power at small scales is reduced,
as seen clearly in figure \ref{fig:T0Mcd}. This change in the
small-scale power induces a modification of the bias factors on large
scales, which as discussed for previous cases, depends on the
requirement of keeping a fixed value of $\bar F$. Unfortunately, when we
compare the difference of the colored and black points at low $k$ in
figure \ref{fig:T0Mcd} with the difference between the fits, we see that
whereas the points do not indicate any variation of $\beta$ with
temperature (the difference in the points is constant at low $k$ without
an appreciable dependence on $\mu$), the fits show a different behavior:
they reflect the difference between the two models at low $\mu$, but at
high $\mu$ the two curves are nearly the same, implying a higher $\beta$
for the low-temperature models. This results from the need to obtain a
good fit to the well-determined points at high $k$, and the large errors
at low $k$, and therefore the implied small change of $\beta$ with
temperature is not reliable.

  The more reliable modification induced by the gas temperature is the
variation of the scale $k_{na}$: a lower gas temperature implies lower
velocity dispersions and therefore smaller non-linear effects on the
power spectrum anisotropy, so the scale of sign reversal of the power
spectrum quadrupole decreases. Figure \ref{fig:T0Mcd} shows that
$k_{na}$ decreases by $\sim 10\%$ when the temperature increases by a
factor of two. This provides a possible clue to resolve degeneracies
with $\sigma_8$ and $\bar F$ in using the shape of the non-linear power
to constrain the IGM temperature from future measurements (e.g., by
using both the value of $k_{na}$ and the damping rate of the power at
higher $k$).

  The effect of varying the temperature-density relation slope is shown
in figure \ref{fig:GmMcd}, where we compare the G1.0 model (with
$\gamma=1$; colored curves and points) with the fiducial model
($\gamma=1.6$; black curves and points), also at $z=2.6$. The
temperature is kept fixed at the mean density, at $T_0=10^{4.3}$ K.
There is also a reduction of the damping at small scales in the G1.0
model, probably because the relevant value of the temperature is
at a density above the mean when we use $\bar F(z=2.6) = 0.781$, where
the G1.0 model has the lower temperature. This agrees also with the
fact that the value of $k_{na}$ decreases with the slope $\gamma$, in
the same way as it decreases with the temperature as in figure
\ref{fig:T0Mcd}. However, the reduction in slope produces an increase
of the bias factors, contrary to the temperature reduction which
decreases them. Therefore, sufficiently careful observations of the
non-linear shape of the power spectrum may help disentangle many of the
properties of the IGM.

%______________________figu______________________%
\begin{figure} [!htbp]
\centering
\includegraphics[width=1.0\textwidth]{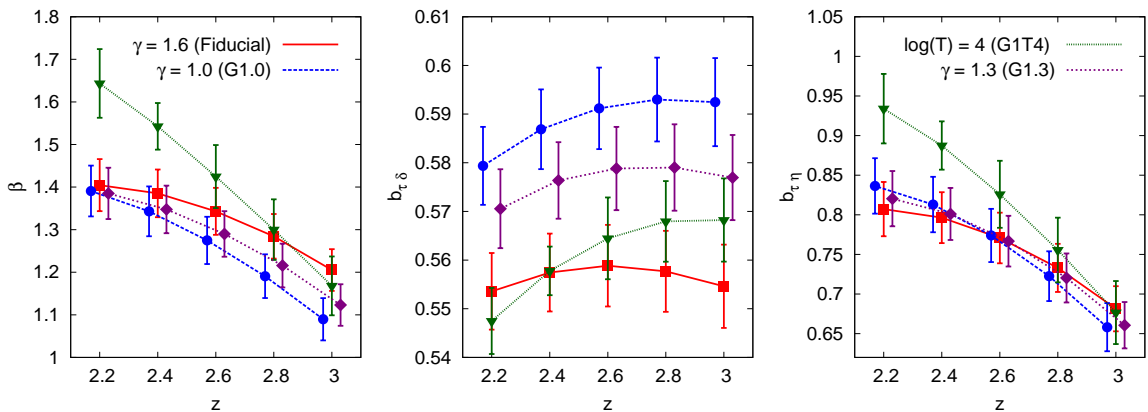}
\caption{\label{fig:tempb} Bias and redshift distortion factors for
models with varying temperature-density relation,
$T=T_0(\rho/\bar\rho)^{\gamma-1}$. Three models have a temperature at
the mean density $T_0=10^{4.3}$ K, and slopes $\gamma=1.6$ (fiducial,
red squares), $\gamma=1.3$ (G1.3, purple rhombi), and $\gamma=1$
(G1.0, blue circles). The model G1T4 (green triangles) also has
$\gamma=1$ and a lower temperature, $T_0=10^4$ K.}
\end{figure} 

  The redshift distortion and bias factors of these models are shown in
figure \ref{fig:tempb}. A fourth model of intermediate slope between the
fiducial and G1.0 models is added. The figure confirms the general
property we have mentioned above: the density bias factor increases with
the temperature $T_0$ and decreases with the slope $\gamma$, but the
variation is small. The general conclusion that $\beta$ drops with
redshift is valid for all models, but the steeper redshift evolution of
$\beta$ for the low-temperature model G1T4 seen in the left panel is not
so reliable because of the difference between the fits and the actual
simulation points at low $k$ described in figure \ref{fig:T0Mcd}. This
detailed question can only be resolved by examining more simulations of
different temperature-density relations on large boxes.

%______________________figu______________________%
\begin{figure} [!htbp]
\centering
\includegraphics[width=1.0\textwidth]{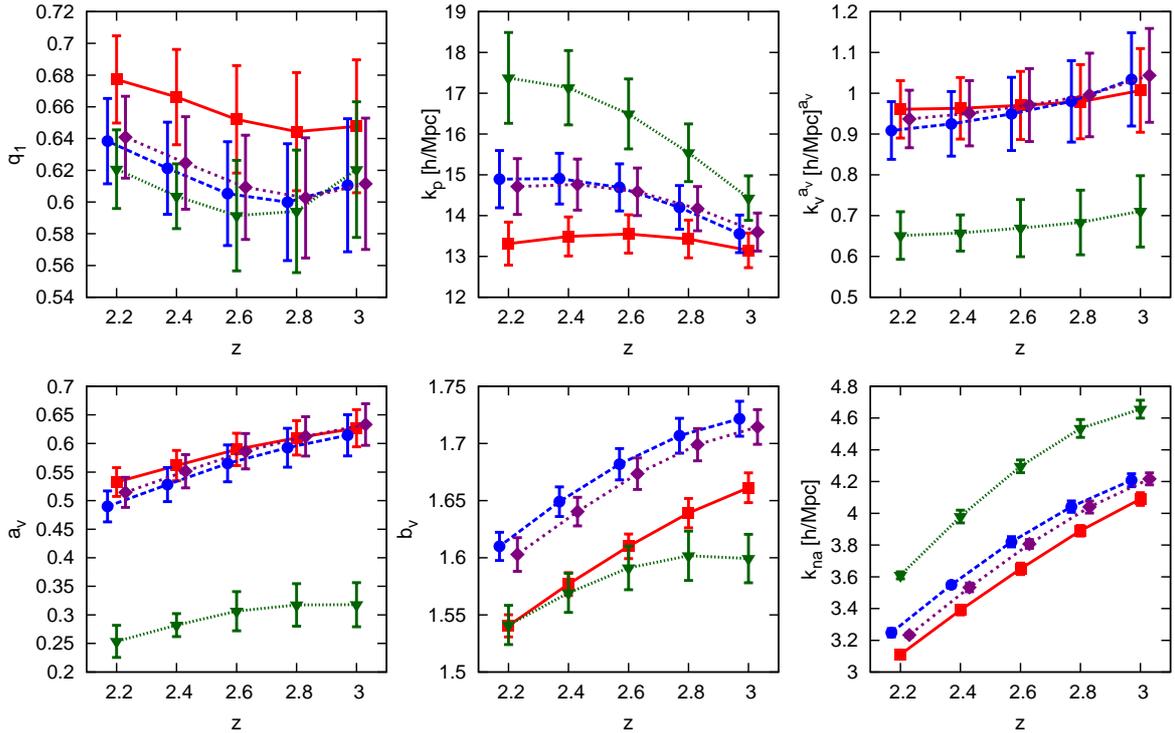}
\caption{\label{fig:flupar} Parameters of the non-linear function
$D_1(k,\mu)$ for the fits to the same four models with different
temperature-density relation as in figure \ref{fig:tempb}. }
\end{figure} 

  The non-linear parameters obtained from the same fits are shown in
figure \ref{fig:flupar}. The parameter $k_p$ increases as expected when
$T_0$ is reduced for the G1T4 model (green triangles), corresponding to
a decreasing Jeans scale, and the flattening of the slope also increases
$k_p$ even for the models of fixed $T_0$, as mentioned above. However,
this variation of $k_p$ with $T_0$ has a large dependence on redshift. A
much clearer effect of reducing $T_0$ is the reduction of $k_v^{a_v}$ and
$a_v$, which is not affected by the slope $\gamma$. The model dependence
of $k_{na}$, which has much smaller errors due to the much smaller
degeneracy with other parameters, shows clearly the effects that we have
referred to earlier: the wavenumber $k_{na}$ decreases with the
temperature $T_0$, and decreases also to a smaller extent when
the slope $\gamma$ is increased. There are also interesting changes of
the power-law index $b_v$ for the $\mu$-dependence with the slope
$\gamma$, suggesting again that a precise measurement of the multiple
characteristics of the non-linear power spectrum in future observational
studies can provide a precise diagnostic of the evolution of the IGM.

%% file: discussion.tex
\section{Discussion}
\label{sec:disc}

 The study of the \lya forest has reached an age of maturity. The
observations are providing precision measurements of the transmission
correlation function or power spectrum: the one-dimensional power
spectrum is measured to very high accuracy \citep{Palanque13}, and the
three-dimensional correlation function has been measured on large-scales
with a new method that treats the distortion introduced by quasar
continuum fitting \citep{Blomqvist15}. In this context, reliable
theoretical predictions are needed to confront the large amount of
observational results that are becoming available from large surveys
\citep{Lee13}.

  In this work, we have introduced a new formula for fitting the
numerical results on the \lya power spectrum from simulations of
structure formation, designated as $D_1$ in equation (\ref{eq:Dapm7}).
This formula was obtained from a guess for how the non-linear
correction should depend on the fluctuation amplitude in Fourier space,
$\Delta^2(k)$, but it is not derived from any perturbation theory. It
is based on trial-and-error experimentation to see which formula best
fits the numerical results, and on the previous work by M03,
where a $\mu$-dependence of the form $D \propto \exp[-A(k)\mu^{b_v} ]$
was proposed. We have found in this work that in fact, this
$\mu$-dependence of the non-linear correction is a surprisingly good fit
to the simulation results (figure \ref{fig:pmu}), but we have no
analytic explanation for this. In the future, it would be desirable to
attempt developing a second-order perturbation theory for modeling the
non-linear coupling of a given Fourier mode to all other modes, which
suggests that $D(k,\mu)$ should be given by an integral over all
possible wavevectors ${\bf k}'$ of the amplitude product
$\Delta({\bf k'})\Delta({\bf k} - {\bf k'})$ times a coupling strength
\citep{PMcD06}, instead of our simple term proportional to $\Delta^2(k)$.

  The form of the non-linear power spectrum is clearly predicted by our
simulations, and basically agrees with that found in M03. We
have characterized it in terms of the wavenumber, $k_{na}$, at which
non-linear anisotropy due to velocity dispersion and thermal broadening
starts dominating over the linear anisotropy of large-scale velocity
flows, and the curves of the ratio $P_F(k,\mu)/P_L(k)$ at different
$\mu$ cross each other. At higher wavenumbers the power falls fast for
high $\mu$ due to thermal broadening, and more slowly at low $\mu$ due
to the Jeans scale of gas at the densities of typical absorption
systems, with a characteristic smoothing scale given by our $k_p$
parameter. As the precision and reliability of the hydrodynamic
simulations improves, more accurate predictions on the precise shape of
the non-linear power spectrum should be possible. This requires only
analyzing enough simulations with large boxes and sufficient resolution
to reduce the systematic effects we discuss in \S 5. The details of the
shape of $P_F(k,\mu)$ should probably allow to measure independently the
amplitude and slope of the power spectrum $P_L(k)$ and the
temperature-density relation, as described in \S 6.

  The numerical results also make a prediction of the two bias factors
of the \lya forest. The form of the linear power spectrum is that in
equation (\ref{eq:lpow}), but the values of the two bias factors depend
on all the model parameters that affect the non-linear scales. We have
introduced the bias factors based on the effective optical depth
in equation (\ref{eq:btau}), which are physically interpretable in the
same way as the bias of galaxy populations. Our
predictions for these bias factors are not highly accurate because of
the sampling variance of our simulations with limited box size and, to a
smaller extent, the limited resolution. Moreover, their fitted values
depend on the formula that is used to fit the whole shape of the
non-linear power spectrum, as illustrated by the different results we
obtain when the $q_2$ parameter is left free or is set to zero. An
improvement over the work presented here should be to consider also the
cross-power spectrum of $\delta_F$ with the initial conditions of the
simulation, which can give an improved estimate of the bias factors.
Nevertheless, as explained in \S 5 we believe that our estimates of the
bias factors are reliable within $\sim 10\%$. The basic result is
that the density bias factor is $\sim 0.6$ for the standard CDM$\Lambda$
model, decreases with the power spectrum amplitude as $\sigma_8^{-1}$,
and is close to constant with redshift as a result of the opposite
effects of the decreasing power spectrum amplitude and decreasing
$\bar F$ with redshift. The peculiar velocity gradient
bias, which is unity for an isotropic tracer, is predicted to be close
to but below unity, and decreasing with redshift. The resulting redshift
distortion factor, $\beta$, is $\sim 1.4$ at $z=2.2$, decreases with
redshift and increases with the amplitude $\sigma_8$.

  The difference of $b_{\tau\eta}$ from unity is an important prediction
of the hydrodynamic simulations we are studying. If the absorption line
features in the \lya forest were arising from a series of discrete
clouds that are small compared to the scales where cosmological
structures are entering the non-linear regime, then a change in the
large-scale peculiar velocity gradient $\eta$ (see equation
\ref{eq:etadef}) should simply compress
or stretch the fixed absorption profiles into a smaller or larger
redshift range, depending on the sign of the change in $\eta$. Even
though the lines increase their blending when $\eta$ increases, the
effective optical depth still varies as $(1-\eta)^{-1}$, and so
$b_{\tau\eta}$ must be exactly unity.

  It is only when the absorbers have internal velocity flows that are
systematically aligned with the large-scale peculiar velocity gradients
that they can respond differently to a varying $\eta$ at fixed $\delta$.
This is the situation when the absorbers in the \lya forest correspond
to collapsing structures in the cosmic web of the IGM:
if $\eta$ increases, the absorption lines in the forest are compressed
into a narrower redshift range, but they also become internally narrower
and therefore more saturated owing to their own internal velocity
gradients, which are correlated with those on large-scales. As a
consequence, the effective optical depth they produce increases slower
than $\eta$, and $b_{\tau\eta}$ is less than one. This difference from
unity should be more pronounced when the absorbers are low-density
structures following the velocity gradients on large-scales, and less
pronounced when the dominant absorbing structures are due to
high-density gas in more virialized regions, where the velocity
gradients and dispersions are randomized and less correlated with the
surrounding large-scale structure. As $\bar F$ increases closer to
unity (i.e., there is less mean absorption), the dominant absorption
lines correspond to more overdense gas, and $b_{\tau\eta}$ should be
closer to one. This agrees with the behavior we find in our simulations
(see green points in right panel of figure \ref{fig:zevb}). We also
find the value of $b_{\tau\eta}$ is substantially larger than
$b_{\tau\Gamma}$ in equation (\ref{eq:bfgam}), as shown in the right
panel of figure \ref{fig:zplot_L120_1,2}, which would be the value
predicted if the internal peculiar velocities of the absorbers could
be modeled linearly \citep{seljak12}.

  We now compare our results on the non-linear power spectrum with the
ones obtained by M03. The general shape of $P_F/P_L$ is in
very good agreement, as can be seen by comparing Figure 9 in
M03 with several of our figures, for example, figure
\ref{fig:sigMcd}. For a more quantitative comparison, it is useful to
compute first the power spectrum normalization at $z=2.25$, the only
redshift at which results were shown in M03. Our Planck model, which
has a very similar spectral index and gas temperature as the M03 model,
has $\sigma_8(z=0)=0.834$ and $\Omega_{m0}=0.3175$, implying
$\sigma_8(z=2.25)=0.321$, and $\sigma_8(z=2.6)=0.291$.
The model used in M03 had $\sigma_8(z=0)=0.79$ and
$\Omega_{m0}=0.4$, implying $\sigma_8(z=2.25)=0.290$, very close to the
amplitude of our fiducial model at $z=2.6$.
Figure 9 of M03 also shows the feature that the curves at
different $\mu$ cross each other nearly at the same point, at
$k_{na}\simeq 4 h/{\rm Mpc}$. This is in fact nearly the same value
found for our Planck model at $z=2.6$, as seen in figure
\ref{fig:PlaMcd}, in agreement with our finding that this scale
depends mostly on the power spectrum amplitude (there is also a
dependence on gas temperature, which is similar in our Planck model
and the one in M03). The way the power drops at $k>k_{na}$ for the
two models is also fairly similar.

  At low $k$, the value found for the redshift distortion parameter in
M03 of $\beta=1.58$ is higher than ours. Taking into account the lower
power spectrum normalization and lower value of $\bar F(z=2.25)=0.8$
used in M03 compared to our fiducial model, our prediction for the model
used in M03 would be $\beta\simeq 1.32$ when using our $q_2=0$ fits, as
derived by interpolating results in our figures \ref{fig:sigb} and
\ref{fig:zevb}. Using our free $q_2$ fits generally reduces the
predicted value of $\beta$ by a further few percent. There is therefore
a substantial discrepancy in the redshift distortion factor. The density
bias factor given in Table 1 of M03 of $b_{F\delta}^2=0.0173$ is, on
the other hand, in very good agreement with our $q_2=0$ fits: the
implied $b_{\tau\delta}=0.59$ for the mean transmission $\bar F=0.8$
used in M03 is nearly equal to our predicted value for the Planck model
(see figure \ref{fig:sigb}), which has the same amplitude at $z=2.6$ and
a similar value of $\bar F$ at this redshift. This suggests that the
discrepancy in $\beta$ is related to a higher prediction for
$b_{\tau\eta}$ from M03 compared to our models. However, if we use our
free $q_2$ fits, then our value of $b_{\tau\delta}$ increases by
$\sim 12\%$, and then the discrepancy in $\beta$ is mostly due to the
low prediction for $b_{\tau\delta}$ from M03.

  The reason for this discrepancy and the correct prediction for $\beta$
and $b_{\tau\delta}$ will only be resolved with the analysis of more
simulations on large boxes. Understanding the systematic errors and the
accuracy of the prediction of $b_{\tau\eta}$ from hydrodynamic
simulations is especially important because if this value can be
predicted to an accuracy better than 1\%, then a precise measurement
of $\beta$ and $b_{\tau\delta}$ yields an observational determination of
the growth factor logarithmic derivative, $f(\Omega)$, from equation
(\ref{eq:rdp}), and therefore a fundamental test of the presence and
evolution of dark energy in the Universe at high redshift.

  We also find that the variations of the bias factors with the model
parameters specified in Table 1 of M03 are in broad agreement with our
results described in \S \ref{sec:get}.

\subsection{Comparison to observations}

  There are several observations of the \lya forest power spectrum that
our predictions can be compared to. First, on the linear regime, the
measurements of the \lya autocorrelation on large scales yielded a first
detection of redshift distortions in \cite{Slosar11} using BOSS. More
recently, an improved method that effectively corrects for distortions
introduced by continuum fitting has been presented in
\cite{Blomqvist15}, allowing for accurate measurements of the bias
factors. Apart from this, the one-dimensional power spectrum is a
projection of the full redshift space transmission power spectrum which
includes non-linear effects, and was measured first in \cite{McD06},
and then in \cite{Palanque13} with the higher accuracy allowed by BOSS.
The full redshift-space power spectrum down to small scales, where
non-linear effects are important, can also be measured from the BOSS
data and from close pairs of quasars where correlations can be measured
down to the smallest scales. Results have been presented constraining
the Jeans scale \cite{Kulkarni15}, but a complete analysis of the
available data in terms of the models for the full shape of the
three-dimensional power spectrum is yet to be done.

  Here, we make only a brief comparison with the measured bias factors
in \cite{Blomqvist15}, leaving the comparison with the one-dimensional
results and other measurements of the non-linear regime for future
papers. The result of \cite{Blomqvist15}, obtained at a mean redshift
$z=2.3$ from a fit restricted to large scales
(from $40 \hmpc$ to $160 \hmpc$) is $\beta=1.39 \pm 0.11$, and
$b_{F\delta}(1+\beta) = -0.374\pm 0.007$; the second quantity is chosen
because it has the smallest marginalized observational error.

  The measured value of $\beta$ is in excellent agreement with our
predictions. Our Planck model is the one that uses parameters consistent
with present constraints from the Cosmic Background Radiation, and
predicts $\beta=1.4$ at $z=2.3$ for the $q_2=0$ fit, which drops to
$1.3$ for free $q_2$. We note from figures \ref{fig:seedb} and
\ref{fig:boxes1,2}, however, that this prediction is subject to a
substantial systematic error due to box size and resolution effects.
This value also has some sensitivity to other
poorly determined parameters of the IGM, in particular to $\bar F$, but
we see in figure \ref{fig:zevb} (green triangles) that we expect a weak
variation of $\beta$ in the range $0.85 < \bar F < 0.78$.

  Comparing the predicted and measured values of $b_{\tau\delta}$ is
more model-dependent. The results of \cite{Blomqvist15} are obtained
using their fiducial model, with $\Omega_{m0}=0.27$ and
$\sigma_8(z=0)=0.79$, implying $\sigma_8(z=2.3)=0.307$. While the
measurement of $\beta$ is not affected by this fiducial model used to
fit the data, the density bias factor is affected because the \lya
transmission power that is actually observed is proportional to
$[\sigma_8(z)b_{F\delta}]^2$. Our Planck model has a slightly higher
amplitude, $\sigma_8(z=2.3)=0.313$. Correcting for this, if our Planck
model had been used for the data analysis, the \cite{Blomqvist15} result
would then be $b_{F\delta}(1+\beta) = -0.367\pm 0.007$. The theoretical
prediction of our Planck model at $z=2.3$ using our $q_2=0$ fit, and
our standard value of $\bar F(z=2.3)=0.836$, is $b_{\tau\delta}=0.586$
and $\beta=1.40$, implying $b_{F\delta}=-0.105$, and
$b_{F\delta}(1+\beta)=-0.253$. This is substantially smaller than the
measured value.

  There are two systematic errors that dominate the uncertainty in this
comparison. The first is the error of our theoretical predictions,
dominated by the uncertainty in fitting the bias parameters from the
low-$k$ modes. If we use the fit that includes the $q_2$ free parameter
in equation (\ref{eq:Dapm7}), the predicted density bias factor
increases to $b_{\tau\delta}=0.68$ (see Table 4 in Appendix
\ref{subs:tables}), and with $\beta=1.3$,
$b_{F\delta}(1+\beta)=-0.28$. This is still substantially below the
observed value. The second systematic error is related to the value of
$\bar F$, which is poorly determined by observations. Assuming that
$\bar F=0.8$ at $z=2.3$ (instead of the value $\bar F=0.836$ derived
from equation \ref{eq:meanF}), we find from the results in figure
\ref{fig:zevb} that the predicted bias for $q_2=0$ would change to
$b_{\tau\delta}\simeq 0.56$, implying $b_{F\delta}=-0.125$ and
$b_{F\delta}(1+\beta)=-0.300$. If we combine both systematic errors
(assuming the higher $b_{\tau\delta}$ from our free $q_2$ fits and a
low mean transmission $\bar F=0.8$ at $z=2.3$) yields a value
$b_{F\delta}(1+\beta)=-0.334$, closer but still not consistent with the
observational determination of \cite{Blomqvist15}.

  It remains to be seen if the mean transmission from the \lya forest is
in fact substantially lower than was found by \cite{Kim07}, or if there
is a large discrepancy between the predicted value of $b_{\tau\delta}$
from the simulations analyzed in this paper and the observations.
The recent determinations of $\bar F$ by \cite{Becker13} give a value
similar to that of \cite{Kim07}. It is also possible that the value of
the bias factors is altered by the presence of intensity fluctuations in
the ionizing background. The need to understand the observed values of
the bias factors highlights the importance of obtaining more precise
predictions from cosmological simulations.

%% file: conclusions.tex
\section{Conclusions}

  The analysis of a variety of simulations of the \lya forest we have
presented highlights the fact that the detailed form of the
\lya transmission power spectrum in redshift space is in principle
predictable from a well-defined theory of the IGM. The theory assumes
that photoionization by the cosmic ultraviolet background and the
gravitational evolution of structure that arises from primordial
fluctuations in the Cold Dark Matter scenario determine the properties
of the IGM and the \lya power spectrum. Detailed observations in the
future should provide tests of the validity of this simple theory, and 
investigate to what extent the effects of galactic winds, quasar jets,
fluctuations of the ionizing background intensity, and inhomogeneities
of the density-temperature relation arising from reionization, have an
impact on the observable power spectrum and other characteristics of
the \lya forest.

  Much work remains to be done both on the theoretical and
data analysis front to fully confront the model predictions with
observations. The existing data from BOSS and close quasar pairs that
have been individually observed should provide powerful measurements
of the full \lya power spectrum, beyond the constraints obtained from
the one-dimensional analysis \cite{Palanque13,nathalie15}. This will
also be complemented by new surveys of absorption spectra that will
improve on the BOSS results. At the same
time, the theory needs to be further developed with the analysis of
more simulations on large boxes, as they are made possible by modern
computational technology. An improved understanding of the effects of
variations on the physical model of the IGM is necessary before one
can use the non-linear \lya power spectrum to constrain fundamental
parameters in cosmology related, for example, to neutrino masses or
other modifications of the dark matter sector.

  To make the results of our simulations accessible for comparing to
future simulations or observations, we are providing in Appendix
\ref{subs:tables} the fit parameters to the most useful models that
are presented in this paper. At the same time, for more direct
comparisons to the results of our simulations, we are also providing
the results for $P_F(k,\mu)$ of all our models, with the bins that
we have used to obtain all our fits. These data are publicly available
at  {\small{GibHub}}{\footnote{\texttt{https://github.com/andreuandreu/3D\_Power\_spectrum\_modes\_tables}}} repositories
(\url{https://github.com/andreuandreu/3D_Power_spectrum_modes_tables}).

%% file: acknowledgments.tex
\begin{acknowledgments}

  We would like to thank Michael Blomqvist, David Kirkby,
Jos\'e O\~norbe and Nathalie Palanque-Delabrouille for discussions.
AA and JM are supported in part by Spanish grant AYA2012-33938.
MV is supported by the ERC Starting Grant "cosmoIGM", PRIN INAF and PRIN MIUR.
I would like to thank the universe for existing and letting me study it. 

\end{acknowledgments}

%% file: appendix.tex
\begin{appendices}
\label{appen}

\appendix
\section{Dependence of the Bias Factors on the Minimum Error parameter}
\label{subs:epsilon}

  The fits to the non-linear \lya power spectrum we have obtained depend
on several technical parameters for binning the power spectrum and
evaluating the $\chi^2$ function, which are described in \S
\ref{subs:bins}. We have checked that our results for the best fits are
not strongly dependent on the way the Fourier modes are binned. As long
as our parameter $k_t$ is not too small, the rebinning of Fourier modes
should not modify the fit. However, the $\epsilon$ parameter that is
inserted in the expression for the error in equation \ref{eq:pferr}
inevitably affects our fits. The introduction of this parameter is
necessary to reduce the weight of the high-$k$ modes for our fits,
because the number of modes below $k$ increases as $k^3$. We note that
if the total number of bins $n_b$ at $k>k_t$ to evaluate the power $P_F$
is modified (in our fits this was fixed to $n_b=256$), then our fits are
modified depending on the paremeter $\epsilon n_b^{-3/2}$, because the
fits depend on the effective weight assigned to modes in each fixed
region of Fourier space.

  The choice of the parameter $\epsilon$ cannot be made in a very
objective way. As $\epsilon$ is increased, an increasing discrepancy in
the high-$k$ modes is allowed, taking into account that our fitting
formula is not a perfect theory that should precisely match the results
obtained from numerical simulations, which have very small statistical
errors at high $k$ that are at some point smaller than the systematic
errors. Reducing the weight of the high-$k$ modes allows for a better
fit of the bias factors from the low-$k$ modes.

%______________________figu______________________%
\begin{figure}[!htbp]
\centering
\includegraphics[width=\textwidth]{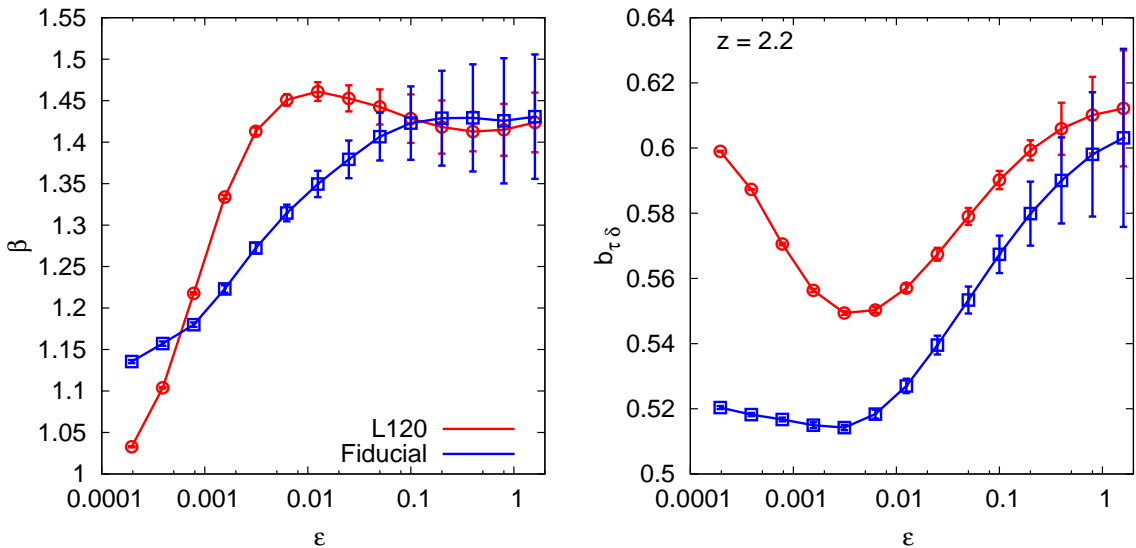}
\caption{\label{fig:eps} Dependence of the redshift distortion factor
$\beta$ and the density bias factor $b_{\tau\delta}$ on the parameter
$\epsilon$ determining a minimum error in our fit with the $D_1$
formula with $q_2=0$. Blue squares (red circles) show the result for our
fiducial (L120) model, at $z=2.2$.}
\end{figure} 

  Therefore, we look into the variation of the redshift distortion
factor and the density bias factor in the two panels of figure
\ref{fig:eps}. As $\epsilon$ is increased, the value of $\beta$
converges just near the value we have chosen in this paper
$\epsilon=0.05$. The value of $b_{\tau\delta}$ continues to increase
above this value of $\epsilon$, by $\sim 10\%$ as $\epsilon$ is
increased to 1. Interestingly, the bias factor reaches a value
$b_{\tau\delta}\simeq 0.61$ at very high $\epsilon$, very close to the
result we obtain in our fits with free $q_2$ for these models (see
figure \ref{fig:boxes1,2}). This suggests that our models with free
$q_2$ yield more accurate values of the bias factors, that are less
affected by the need to correctly fit the numerous data points for
$P_F$ at high $k$ from our simulations. Nevertheless, this will need
to be tested with results from more simulations on large boxes than
are analyzed in this paper.

\newpage

%______________________subsection______________________%
%\renewcommand{\thesection}{B.\arabic{section}}
\section{Tables}
\label{subs:tables}
\input{./tables/tests_table}

\newpage
\input{./tables/table_q2-free}

\newpage
\input{./tables/planck_table}

\input{./tables/physical_table}

\newpage
\input{./tables/flx_table}
\newpage

\end{appendices}

%% file: tables/tests_table.tex
\begin{table}[!htbp]
 \centering
 \caption{Bias parameters for simulations with different resolution and
box size, for the $q_2 = 0$ fit. 
  \vspace{2 mm}}{ 
  \begin{tabular}{c|*{3}c}
 z  & $\beta$    & $b_{\tau\delta}$  &  $b_{\tau\eta}$ \\ 
  \hline 
  \hline 
  \multicolumn{4}{l}{Fiducial} \\ 
 \hline 
3.0  & $ 1.205  \pm   0.049$  & $0.5546  \pm  0.0086$  & $ 0.681  \pm   0.028$ \\
2.8  & $ 1.284  \pm   0.052$  & $0.5577  \pm  0.0083$  & $ 0.733  \pm   0.030$ \\
2.6  & $ 1.343  \pm   0.055$  & $0.5588  \pm  0.0084$  & $ 0.771  \pm   0.032$ \\
2.4  & $ 1.385  \pm   0.056$  & $0.5574  \pm  0.0080$  & $ 0.796  \pm   0.032$ \\
2.2  & $ 1.405  \pm   0.061$  & $0.5536  \pm  0.0079$  & $ 0.807  \pm   0.034$ \\
\hline 
\multicolumn{4}{l}{P1024} \\ 
 \hline 
3.0  & $ 1.205  \pm   0.050$  & $0.5566  \pm  0.0979$  & $ 0.684  \pm   0.119$ \\
2.8  & $ 1.283  \pm   0.052$  & $0.5597  \pm  0.0084$  & $ 0.735  \pm   0.030$ \\
2.6  & $ 1.340  \pm   0.056$  & $0.5611  \pm  0.0084$  & $ 0.772  \pm   0.033$ \\
2.4  & $ 1.381  \pm   0.059$  & $0.5599  \pm  0.0082$  & $ 0.797  \pm   0.034$ \\
2.2  & $ 1.403  \pm   0.061$  & $0.5556  \pm  0.0080$  & $ 0.809  \pm   0.035$ \\
\hline 
\multicolumn{4}{l}{L120} \\ 
 \hline 
3.0  & $ 1.195  \pm   0.039$  & $0.5710  \pm  0.0057$  & $ 0.695  \pm   0.023$ \\
2.8  & $ 1.283  \pm   0.040$  & $0.5768  \pm  0.0058$  & $ 0.757  \pm   0.024$ \\
2.6  & $ 1.355  \pm   0.042$  & $0.5808  \pm  0.0057$  & $ 0.808  \pm   0.025$ \\
2.4  & $ 1.411  \pm   0.042$  & $0.5820  \pm  0.0059$  & $ 0.847  \pm   0.025$ \\
2.2  & $ 1.443  \pm   0.044$  & $0.5789  \pm  0.0051$  & $ 0.867  \pm   0.025$ \\
\hline 
\multicolumn{4}{l}{R384C} \\ 
 \hline 
3.0  & $ 1.210  \pm   0.060$  & $0.5661  \pm  0.0094$  & $ 0.698  \pm   0.035$ \\
2.8  & $ 1.290  \pm   0.060$  & $0.5714  \pm  0.0097$  & $ 0.754  \pm   0.036$ \\
2.6  & $ 1.360  \pm   0.060$  & $0.5732  \pm  0.0086$  & $ 0.800  \pm   0.035$ \\
2.4  & $ 1.420  \pm   0.050$  & $0.5715  \pm  0.0065$  & $ 0.837  \pm   0.029$ \\
2.2  & $ 1.455  \pm   0.064$  & $0.5588  \pm  0.0085$  & $ 0.844  \pm   0.037$ \\
\hline 
\multicolumn{4}{l}{R384} \\ 
 \hline 
3.0  & $ 1.180  \pm   0.060$  & $0.5708  \pm  0.0100$  & $ 0.687  \pm   0.036$ \\
2.8  & $ 1.270  \pm   0.060$  & $0.5743  \pm  0.0096$  & $ 0.746  \pm   0.036$ \\
2.6  & $ 1.330  \pm   0.060$  & $0.5760  \pm  0.0086$  & $ 0.787  \pm   0.035$ \\
2.4  & $ 1.380  \pm   0.070$  & $0.5758  \pm  0.0087$  & $ 0.820  \pm   0.041$ \\
2.2  & $ 1.420  \pm   0.070$  & $0.5703  \pm  0.0099$  & $ 0.841  \pm   0.041$ \\
\hline 
\multicolumn{4}{l}{R640} \\ 
 \hline 
3.0  & $ 1.251  \pm   0.052$  & $0.5409  \pm  0.0953$  & $ 0.690  \pm   0.120$ \\
2.8  & $ 1.316  \pm   0.054$  & $0.5456  \pm  0.0085$  & $ 0.735  \pm   0.031$ \\
2.6  & $ 1.364  \pm   0.054$  & $0.5482  \pm  0.0082$  & $ 0.768  \pm   0.031$ \\
2.4  & $ 1.397  \pm   0.056$  & $0.5487  \pm  0.0079$  & $ 0.791  \pm   0.032$ \\
2.2  & $ 1.421  \pm   0.061$  & $0.5461  \pm  0.0085$  & $ 0.805  \pm   0.034$ \\
\hline 
  \end{tabular}
 }
 \end{table}

\begin{table}[!htbp]
\centering
 \caption{Non-linear fit parameters for the same simulations as in Table 2.
  \vspace{2 mm}}
 \adjustbox{width=1.0 \textwidth}{ 
    \begin{tabular}{c|*{6}c}
 z  &  $q_{1}$  &  $k_p\,[h/{\rm Mpc}]$ &  $k_{v}^{a_v}\,[h/{\rm Mpc}]^{a_v}$  &  $a_v$  &  $b_v$ &  $k_{na}$ \\ 
  \hline 
   \multicolumn{7}{l}{Fiducial} \\ 
  \hline 
3.0  & $ 0.648  \pm   0.042$  & $  13.1  \pm     0.4$  & $ 1.007  \pm   0.103$  & $ 0.627  \pm   0.033$  & $  1.66  \pm    0.01$  & $  4.09  \pm    0.04$ \\
2.8  & $ 0.644  \pm   0.037$  & $  13.4  \pm     0.5$  & $ 0.979  \pm   0.091$  & $ 0.610  \pm   0.030$  & $  1.64  \pm    0.01$  & $  3.89  \pm    0.03$ \\
2.6  & $ 0.652  \pm   0.034$  & $  13.6  \pm     0.5$  & $ 0.970  \pm   0.084$  & $ 0.590  \pm   0.028$  & $  1.61  \pm    0.01$  & $  3.65  \pm    0.04$ \\
2.4  & $ 0.666  \pm   0.030$  & $  13.5  \pm     0.5$  & $ 0.963  \pm   0.076$  & $ 0.561  \pm   0.027$  & $  1.58  \pm    0.01$  & $  3.39  \pm    0.03$ \\
2.2  & $ 0.677  \pm   0.027$  & $  13.3  \pm     0.5$  & $ 0.961  \pm   0.070$  & $ 0.533  \pm   0.025$  & $  1.54  \pm    0.01$  & $  3.11  \pm    0.03$ \\
\hline 
\multicolumn{7}{l}{P1024} \\ 
 \hline 
3.0  & $ 0.644  \pm   0.042$  & $  13.2  \pm     0.4$  & $ 0.997  \pm   0.100$  & $ 0.634  \pm   0.032$  & $  1.66  \pm    0.02$  & $  4.09  \pm    0.04$ \\
2.8  & $ 0.642  \pm   0.037$  & $  13.5  \pm     0.6$  & $ 0.973  \pm   0.091$  & $ 0.617  \pm   0.030$  & $  1.63  \pm    0.02$  & $  3.89  \pm    0.03$ \\
2.6  & $ 0.650  \pm   0.035$  & $  13.7  \pm     0.5$  & $ 0.964  \pm   0.084$  & $ 0.597  \pm   0.029$  & $  1.60  \pm    0.02$  & $  3.65  \pm    0.04$ \\
2.4  & $ 0.665  \pm   0.031$  & $  13.6  \pm     0.5$  & $ 0.959  \pm   0.075$  & $ 0.567  \pm   0.026$  & $  1.57  \pm    0.02$  & $  3.39  \pm    0.03$ \\
2.2  & $ 0.679  \pm   0.028$  & $  13.4  \pm     0.5$  & $ 0.956  \pm   0.069$  & $ 0.537  \pm   0.025$  & $  1.53  \pm    0.01$  & $  3.11  \pm    0.03$ \\
\hline 
\multicolumn{7}{l}{L120} \\ 
 \hline 
3.0  & $ 0.661  \pm   0.032$  & $  11.3  \pm     0.2$  & $ 0.803  \pm   0.064$  & $ 0.460  \pm   0.026$  & $  1.55  \pm    0.01$  & $  4.01  \pm    0.03$ \\
2.8  & $ 0.635  \pm   0.028$  & $  12.0  \pm     0.3$  & $ 0.782  \pm   0.057$  & $ 0.471  \pm   0.024$  & $  1.56  \pm    0.01$  & $  3.80  \pm    0.03$ \\
2.6  & $ 0.626  \pm   0.026$  & $  12.7  \pm     0.4$  & $ 0.772  \pm   0.053$  & $ 0.469  \pm   0.023$  & $  1.56  \pm    0.01$  & $  3.56  \pm    0.02$ \\
2.4  & $ 0.633  \pm   0.023$  & $  13.2  \pm     0.4$  & $ 0.771  \pm   0.048$  & $ 0.456  \pm   0.021$  & $  1.54  \pm    0.01$  & $  3.29  \pm    0.02$ \\
2.2  & $ 0.643  \pm   0.020$  & $  13.5  \pm     0.6$  & $ 0.778  \pm   0.046$  & $ 0.439  \pm   0.020$  & $  1.52  \pm    0.01$  & $  3.00  \pm    0.02$ \\
\hline 
\multicolumn{7}{l}{R384C} \\ 
 \hline 
3.0  & $ 0.592  \pm   0.051$  & $  14.0  \pm     0.6$  & $ 0.786  \pm   0.104$  & $ 0.504  \pm   0.042$  & $  1.57  \pm    0.01$  & $  4.15  \pm    0.05$ \\
2.8  & $ 0.578  \pm   0.045$  & $  14.9  \pm     0.8$  & $ 0.774  \pm   0.095$  & $ 0.509  \pm   0.038$  & $  1.58  \pm    0.01$  & $  3.92  \pm    0.05$ \\
2.6  & $ 0.584  \pm   0.039$  & $  15.5  \pm     0.9$  & $ 0.773  \pm   0.083$  & $ 0.500  \pm   0.036$  & $  1.58  \pm    0.02$  & $  3.67  \pm    0.04$ \\
2.4  & $ 0.602  \pm   0.032$  & $  15.9  \pm     1.1$  & $ 0.778  \pm   0.064$  & $ 0.478  \pm   0.028$  & $  1.55  \pm    0.01$  & $  3.40  \pm    0.03$ \\
2.2  & $ 0.644  \pm   0.027$  & $  15.4  \pm     0.8$  & $ 0.822  \pm   0.060$  & $ 0.458  \pm   0.024$  & $  1.52  \pm    0.02$  & $  3.12  \pm    0.03$ \\
\hline 
\multicolumn{7}{l}{R384} \\ 
 \hline 
3.0  & $ 0.586  \pm   0.051$  & $  12.9  \pm     0.5$  & $ 0.921  \pm   0.120$  & $ 0.617  \pm   0.041$  & $  1.66  \pm    0.02$  & $  4.13  \pm    0.05$ \\
2.8  & $ 0.584  \pm   0.043$  & $  13.3  \pm     0.5$  & $ 0.894  \pm   0.104$  & $ 0.600  \pm   0.038$  & $  1.64  \pm    0.02$  & $  3.93  \pm    0.05$ \\
2.6  & $ 0.594  \pm   0.041$  & $  13.6  \pm     0.6$  & $ 0.885  \pm   0.097$  & $ 0.578  \pm   0.037$  & $  1.61  \pm    0.01$  & $  3.68  \pm    0.04$ \\
2.4  & $ 0.612  \pm   0.038$  & $  13.7  \pm     0.7$  & $ 0.877  \pm   0.089$  & $ 0.547  \pm   0.034$  & $  1.58  \pm    0.01$  & $  3.42  \pm    0.03$ \\
2.2  & $ 0.634  \pm   0.030$  & $  13.5  \pm     0.6$  & $ 0.880  \pm   0.079$  & $ 0.514  \pm   0.032$  & $  1.54  \pm    0.01$  & $  3.13  \pm    0.03$ \\
\hline 
\multicolumn{7}{l}{R640} \\ 
 \hline 
3.0  & $ 0.642  \pm   0.042$  & $  14.9  \pm     0.8$  & $ 1.023  \pm   0.111$  & $ 0.651  \pm   0.034$  & $  1.71  \pm    0.03$  & $  4.14  \pm    0.04$ \\
2.8  & $ 0.633  \pm   0.038$  & $  15.1  \pm     0.7$  & $ 0.998  \pm   0.094$  & $ 0.636  \pm   0.030$  & $  1.68  \pm    0.02$  & $  3.95  \pm    0.04$ \\
2.6  & $ 0.636  \pm   0.034$  & $  15.1  \pm     0.7$  & $ 0.984  \pm   0.085$  & $ 0.614  \pm   0.028$  & $  1.65  \pm    0.02$  & $  3.71  \pm    0.04$ \\
2.4  & $ 0.645  \pm   0.030$  & $  14.8  \pm     0.6$  & $ 0.969  \pm   0.077$  & $ 0.582  \pm   0.027$  & $  1.61  \pm    0.01$  & $  3.45  \pm    0.02$ \\
2.2  & $ 0.656  \pm   0.027$  & $  14.4  \pm     0.7$  & $ 0.950  \pm   0.071$  & $ 0.543  \pm   0.026$  & $  1.57  \pm    0.02$  & $  3.16  \pm    0.03$ \\
\hline 
  \end{tabular}
 }
 \end{table}

%% file: tables/table_q2-free.tex
\begin{table}[!htbp]
 \centering
  \caption{Bias parameters for $D_1$ with $q_2$ set free.
  \vspace{2 mm}}{
    \begin{tabular}{c|*{3}c}
 z  & $\beta$    & $b_{\tau\delta}$  &  $b_{\tau\eta}$ \\ 
    \hline 
  \hline 
  \multicolumn{4}{l}{Fiducial} \\ 
 \hline 
3.0  & $ 1.179  \pm   0.061$  & $0.6010  \pm  0.0126$  & $ 0.722  \pm   0.039$ \\
2.8  & $ 1.256  \pm   0.049$  & $0.6096  \pm  0.0132$  & $ 0.784  \pm   0.033$ \\
2.6  & $ 1.292  \pm   0.069$  & $0.6206  \pm  0.0146$  & $ 0.823  \pm   0.046$ \\
2.4  & $ 1.321  \pm   0.067$  & $0.6286  \pm  0.0130$  & $ 0.856  \pm   0.044$ \\
2.2  & $ 1.347  \pm   0.047$  & $0.6306  \pm  0.0099$  & $ 0.882  \pm   0.031$ \\
\hline 
\multicolumn{4}{l}{R384C} \\ 
 \hline 
3.0  & $ 1.200  \pm   0.067$  & $0.5981  \pm  0.0194$  & $ 0.732  \pm   0.045$ \\
2.8  & $ 1.288  \pm   0.068$  & $0.6043  \pm  0.0137$  & $ 0.797  \pm   0.044$ \\
2.6  & $ 1.355  \pm   0.070$  & $0.6096  \pm  0.0116$  & $ 0.848  \pm   0.044$ \\
2.4  & $ 1.399  \pm   0.078$  & $0.6147  \pm  0.0144$  & $ 0.887  \pm   0.050$ \\
2.2  & $ 1.426  \pm   0.074$  & $0.6144  \pm  0.0126$  & $ 0.909  \pm   0.047$ \\
\hline 
\multicolumn{4}{l}{L120} \\ 
 \hline 
3.0  & $ 1.204  \pm   0.032$  & $0.5808  \pm  0.0098$  & $ 0.713  \pm   0.021$ \\
2.8  & $ 1.296  \pm   0.028$  & $0.5962  \pm  0.0104$  & $ 0.791  \pm   0.021$ \\
2.6  & $ 1.365  \pm   0.053$  & $0.6092  \pm  0.0131$  & $ 0.854  \pm   0.036$ \\
2.4  & $ 1.409  \pm   0.052$  & $0.6202  \pm  0.0100$  & $ 0.901  \pm   0.034$ \\
2.2  & $ 1.431  \pm   0.036$  & $0.6249  \pm  0.0105$  & $ 0.928  \pm   0.026$ \\
\hline 
\multicolumn{4}{l}{L80} \\ 
 \hline 
3.0  & $ 1.120  \pm   0.040$  & $0.6169  \pm  0.0185$  & $ 0.704  \pm   0.032$ \\
2.8  & $ 1.210  \pm   0.030$  & $0.6251  \pm  0.0115$  & $ 0.774  \pm   0.023$ \\
2.6  & $ 1.280  \pm   0.030$  & $0.6344  \pm  0.0117$  & $ 0.834  \pm   0.024$ \\
2.4  & $ 1.340  \pm   0.030$  & $0.6375  \pm  0.0078$  & $ 0.881  \pm   0.021$ \\
2.2  & $ 1.350  \pm   0.030$  & $0.6474  \pm  0.0120$  & $ 0.907  \pm   0.024$ \\
\hline 
\multicolumn{4}{l}{Planck} \\ 
 \hline 
3.0  & $ 1.072  \pm   0.059$  & $0.6462  \pm  0.0185$  & $ 0.657  \pm   0.045$ \\
2.8  & $ 1.155  \pm   0.065$  & $0.6573  \pm  0.0167$  & $ 0.723  \pm   0.049$ \\
2.6  & $ 1.254  \pm   0.044$  & $0.6601  \pm  0.0109$  & $ 0.791  \pm   0.034$ \\
2.4  & $ 1.290  \pm   0.074$  & $0.6784  \pm  0.0154$  & $ 0.840  \pm   0.056$ \\
2.2  & $ 1.316  \pm   0.057$  & $0.6751  \pm  0.0141$  & $ 0.858  \pm   0.044$ \\
\hline 

  \end{tabular}
  }
\end{table}

\begin{table}[!htbp]
\centering
 \caption{Non-linear fit parameters for $D_1$ with $q_2$ set free.
  \vspace{2 mm}}
 \adjustbox{width=1.0 \textwidth}{ 
  \begin{tabular}{c|*{6}c}
z  &  $q_1$  &   $q_2$  &  $k_p\, [h/Mpc]$  &  $k_{v}^{a_v}\, [h/Mpc]^{a_v}$  &  $a_v$  &  $b_v$ \\ 
 \hline 
  \hline 
  \multicolumn{7}{l}{Fiducial} \\ 
 \hline 
3.0  & $0.104  \pm  0.121$  & $ 0.444  \pm   0.093$  & $  10.1  \pm     0.5$  & $ 0.516  \pm   0.114$  & $ 0.248  \pm   0.072$  & $  1.66  \pm    0.02$ \\
2.8  & $0.086  \pm  0.110$  & $ 0.417  \pm   0.078$  & $   9.9  \pm     0.5$  & $ 0.493  \pm   0.087$  & $ 0.217  \pm   0.069$  & $  1.63  \pm    0.02$ \\
2.6  & $0.068  \pm  0.101$  & $ 0.390  \pm   0.063$  & $   9.6  \pm     0.5$  & $ 0.483  \pm   0.052$  & $ 0.190  \pm   0.056$  & $  1.61  \pm    0.02$ \\
2.4  & $0.057  \pm  0.085$  & $ 0.368  \pm   0.049$  & $   9.2  \pm     0.3$  & $ 0.480  \pm   0.028$  & $ 0.156  \pm   0.045$  & $  1.57  \pm    0.02$ \\
2.2  & $0.090  \pm  0.052$  & $ 0.316  \pm   0.027$  & $   8.9  \pm     0.3$  & $ 0.493  \pm   0.016$  & $ 0.145  \pm   0.031$  & $  1.54  \pm    0.01$ \\
\hline 
\multicolumn{7}{l}{R384C} \\ 
 \hline 
3.0  & $0.207  \pm  0.208$  & $ 0.320  \pm   0.159$  & $  11.2  \pm     1.0$  & $ 0.468  \pm   0.165$  & $ 0.208  \pm   0.120$  & $  1.57  \pm    0.01$ \\
2.8  & $0.202  \pm  0.132$  & $ 0.289  \pm   0.094$  & $  11.4  \pm     0.8$  & $ 0.462  \pm   0.111$  & $ 0.207  \pm   0.093$  & $  1.57  \pm    0.01$ \\
2.6  & $0.203  \pm  0.073$  & $ 0.267  \pm   0.041$  & $  11.2  \pm     0.4$  & $ 0.467  \pm   0.035$  & $ 0.198  \pm   0.042$  & $  1.57  \pm    0.01$ \\
2.4  & $0.208  \pm  0.083$  & $ 0.246  \pm   0.035$  & $  11.0  \pm     0.3$  & $ 0.475  \pm   0.019$  & $ 0.182  \pm   0.027$  & $  1.54  \pm    0.01$ \\
2.2  & $0.233  \pm  0.062$  & $ 0.213  \pm   0.023$  & $  10.8  \pm     0.2$  & $ 0.499  \pm   0.015$  & $ 0.181  \pm   0.019$  & $  1.51  \pm    0.01$ \\
\hline 
\multicolumn{7}{l}{L120} \\ 
 \hline 
3.0  & $0.529  \pm  0.115$  & $ 0.117  \pm   0.102$  & $  10.6  \pm     0.6$  & $ 0.684  \pm   0.196$  & $ 0.362  \pm   0.085$  & $  1.54  \pm    0.01$ \\
2.8  & $0.398  \pm  0.112$  & $ 0.190  \pm   0.087$  & $  10.5  \pm     0.6$  & $ 0.583  \pm   0.167$  & $ 0.290  \pm   0.076$  & $  1.55  \pm    0.01$ \\
2.6  & $0.316  \pm  0.105$  & $ 0.226  \pm   0.070$  & $  10.3  \pm     0.7$  & $ 0.529  \pm   0.118$  & $ 0.233  \pm   0.074$  & $  1.54  \pm    0.01$ \\
2.4  & $0.264  \pm  0.091$  & $ 0.241  \pm   0.060$  & $   9.9  \pm     0.6$  & $ 0.501  \pm   0.069$  & $ 0.183  \pm   0.065$  & $  1.53  \pm    0.01$ \\
2.2  & $0.245  \pm  0.086$  & $ 0.232  \pm   0.051$  & $   9.6  \pm     0.6$  & $ 0.497  \pm   0.046$  & $ 0.153  \pm   0.059$  & $  1.51  \pm    0.01$ \\
\hline 
\multicolumn{7}{l}{L80} \\ 
 \hline 
3.0  & $0.144  \pm  0.232$  & $ 0.430  \pm   0.202$  & $   8.9  \pm     0.8$  & $ 0.514  \pm   0.286$  & $ 0.212  \pm   0.158$  & $  1.61  \pm    0.01$ \\
2.8  & $0.130  \pm  0.133$  & $ 0.394  \pm   0.108$  & $   9.0  \pm     0.6$  & $ 0.490  \pm   0.108$  & $ 0.190  \pm   0.093$  & $  1.59  \pm    0.01$ \\
2.6  & $0.117  \pm  0.110$  & $ 0.367  \pm   0.080$  & $   8.9  \pm     0.5$  & $ 0.480  \pm   0.059$  & $ 0.167  \pm   0.071$  & $  1.56  \pm    0.01$ \\
2.4  & $0.147  \pm  0.082$  & $ 0.320  \pm   0.054$  & $   8.8  \pm     0.4$  & $ 0.493  \pm   0.041$  & $ 0.161  \pm   0.054$  & $  1.53  \pm    0.01$ \\
2.2  & $0.120  \pm  0.083$  & $ 0.308  \pm   0.047$  & $   8.5  \pm     0.4$  & $ 0.484  \pm   0.021$  & $ 0.123  \pm   0.050$  & $  1.50  \pm    0.01$ \\
\hline 
\multicolumn{7}{l}{Planck} \\ 
 \hline 
3.0  & $0.0004  \pm  0.2241$  & $ 0.787  \pm   0.220$  & $  11.8  \pm     1.2$  & $ 0.552  \pm   0.201$  & $ 0.178  \pm   0.108$  & $  1.64  \pm    0.03$ \\
2.8  & $-0.0168  \pm  0.1686$  & $ 0.711  \pm   0.143$  & $  12.0  \pm     1.0$  & $ 0.548  \pm   0.126$  & $ 0.203  \pm   0.073$  & $  1.65  \pm    0.03$ \\
2.6  & $0.0004  \pm  0.1241$  & $ 0.651  \pm   0.113$  & $  11.8  \pm     1.0$  & $ 0.545  \pm   0.115$  & $ 0.195  \pm   0.073$  & $  1.64  \pm    0.02$ \\
2.4  & $-0.0020  \pm  0.1108$  & $ 0.623  \pm   0.069$  & $  10.9  \pm     0.5$  & $ 0.517  \pm   0.027$  & $ 0.152  \pm   0.035$  & $  1.62  \pm    0.02$ \\
2.2  & $0.0405  \pm  0.0890$  & $ 0.534  \pm   0.049$  & $  10.6  \pm     0.3$  & $ 0.534  \pm   0.022$  & $ 0.139  \pm   0.029$  & $  1.60  \pm    0.01$ \\
\hline 
  \end{tabular}
 }
 \end{table}

%% file: tables/planck_table.tex
\begin{table}[!htbp]
 \centering
 \caption{Bias parameters for Eulerian, Lagrangian and Planck simulations, for $q_2=0$.
  \vspace{2 mm}}{ 
  \begin{tabular}{c|*{3}c}
 z  & $\beta$    & $b_{\tau\delta}$  &  $b_{\tau\eta}$ \\ 
  \hline 
  \hline 
  \multicolumn{4}{l}{Fiducial} \\   
 \hline 
3.0  & $1.21  \pm  0.049$  & $0.555  \pm  0.0086$  & $0.681  \pm  0.028$ \\
2.8  & $1.28  \pm  0.052$  & $0.558  \pm  0.0083$  & $0.733  \pm  0.030$ \\
2.6  & $1.34  \pm  0.055$  & $0.559  \pm  0.0084$  & $0.771  \pm  0.032$ \\
2.4  & $1.39  \pm  0.056$  & $0.557  \pm  0.0080$  & $0.796  \pm  0.032$ \\
2.2  & $1.40  \pm  0.061$  & $0.554  \pm  0.0079$  & $0.807  \pm  0.034$ \\
\hline 
\multicolumn{4}{l}{Eulerian} \\ 
 \hline 
2.9  & $1.29  \pm  0.073$  & $0.557  \pm  0.011$  & $0.737  \pm  0.042$ \\
2.6  & $1.51  \pm  0.084$  & $0.561  \pm  0.010$  & $0.870  \pm  0.048$ \\
2.3  & $1.63  \pm  0.083$  & $0.557  \pm  0.011$  & $0.941  \pm  0.047$ \\
\hline 
\multicolumn{4}{l}{Lagrangian} \\ 
 \hline 
3.0  & $1.03  \pm  0.055$  & $0.555  \pm  0.011$  & $0.583  \pm  0.032$ \\
2.8  & $1.14  \pm  0.059$  & $0.561  \pm  0.011$  & $0.653  \pm  0.034$ \\
2.6  & $1.24  \pm  0.064$  & $0.563  \pm  0.011$  & $0.720  \pm  0.037$ \\
2.4  & $1.33  \pm  0.069$  & $0.558  \pm  0.010$  & $0.769  \pm  0.039$ \\
2.2  & $1.47  \pm  0.070$  & $0.556  \pm  0.0097$  & $0.854  \pm  0.04$ \\
  \hline 
  \multicolumn{4}{l}{Planck} \\ 
 \hline 
3.0  & $1.12  \pm  0.056$  & $0.588  \pm  0.0096$  & $0.627  \pm  0.036$ \\
2.8  & $1.22  \pm  0.055$  & $0.592  \pm  0.0094$  & $0.689  \pm  0.036$ \\
2.6  & $1.30  \pm  0.057$  & $0.591  \pm  0.0093$  & $0.736  \pm  0.037$ \\
2.4  & $1.39  \pm  0.058$  & $0.591  \pm  0.0095$  & $0.788  \pm  0.038$ \\
2.2  & $1.42  \pm  0.058$  & $0.581  \pm  0.0092$  & $0.796  \pm  0.038$ \\
  \hline 
  \end{tabular}
 }
 \end{table}

\begin{table}[!htbp]
\centering
 \caption{Non-linear fit parameters for the Eulerian, Lagrangian and Planck
 simulation, for $q_2=0$.
  \vspace{2 mm}}
 \adjustbox{width=1.0 \textwidth}{ 
  \begin{tabular}{c|*{6}c}
 z  &  $q_{1}$  &  $k_p\,[h/Mpc]$ &  $k_{v}^{a_v}\,[h/Mpc]^{a_v}$  &  $a_v$  &  $b_v$ &  $k_{na}$ \\ 
  \hline 
  \hline 
  \multicolumn{7}{l}{Fiducial} \\ 
 \hline 
3.0  & $0.648  \pm  0.042$  & $13.1  \pm  0.42$  & $1.01  \pm  0.10$  & $0.627  \pm  0.033$  & $1.66  \pm  0.013$  & $4.09  \pm  0.040$ \\
2.8  & $0.644  \pm  0.037$  & $13.4  \pm  0.46$  & $0.979  \pm  0.091$  & $0.610  \pm  0.030$  & $1.64  \pm  0.013$  & $3.89  \pm  0.033$ \\
2.6  & $0.652  \pm  0.034$  & $13.6  \pm  0.47$  & $0.970  \pm  0.084$  & $0.590  \pm  0.028$  & $1.61  \pm  0.011$  & $3.65  \pm  0.036$ \\
2.4  & $0.666  \pm  0.030$  & $13.5  \pm  0.48$  & $0.963  \pm  0.076$  & $0.561  \pm  0.027$  & $1.58  \pm  0.010$  & $3.39  \pm  0.030$ \\
2.2  & $0.677  \pm  0.027$  & $13.3  \pm  0.53$  & $0.961  \pm  0.07$  & $0.533  \pm  0.025$  & $1.54  \pm  0.0097$  & $3.11  \pm  0.028$ \\
\hline 
\multicolumn{7}{l}{Eulerian} \\ 
 \hline 
2.9  & $0.719  \pm  0.064$  & $16.1  \pm  0.73$  & $0.734  \pm  0.10$  & $0.413  \pm  0.043$  & $1.65  \pm  0.038$  & $5.09  \pm  0.071$ \\
2.6  & $0.71  \pm  0.054$  & $18.0  \pm  1.2$  & $0.702  \pm  0.085$  & $0.394  \pm  0.037$  & $1.60  \pm  0.031$  & $4.76  \pm  0.064$ \\
2.3  & $0.73  \pm  0.046$  & $19.2  \pm  1.8$  & $0.711  \pm  0.082$  & $0.370  \pm  0.034$  & $1.55  \pm  0.028$  & $4.22  \pm  0.065$ \\
 \hline 
\multicolumn{7}{l}{Lagrange} \\ 
 \hline 
3.0  & $0.847  \pm  0.070$  & $16.4  \pm  0.95$  & $1.39  \pm  0.20$  & $0.664  \pm  0.049$  & $1.63  \pm  0.035$  & $4.87  \pm  0.07$ \\
2.8  & $0.822  \pm  0.063$  & $17.4  \pm  1.3$  & $1.31  \pm  0.18$  & $0.649  \pm  0.046$  & $1.61  \pm  0.032$  & $4.70  \pm  0.059$ \\
2.6  & $0.824  \pm  0.056$  & $18.1  \pm  1.4$  & $1.24  \pm  0.15$  & $0.619  \pm  0.043$  & $1.59  \pm  0.028$  & $4.46  \pm  0.054$ \\
2.4  & $0.836  \pm  0.050$  & $18.1  \pm  1.5$  & $1.16  \pm  0.13$  & $0.574  \pm  0.038$  & $1.56  \pm  0.026$  & $4.16  \pm  0.051$ \\
2.2  & $0.907  \pm  0.043$  & $15.8  \pm  0.8$  & $0.981  \pm  0.085$  & $0.463  \pm  0.029$  & $1.49  \pm  0.019$  & $3.57  \pm  0.034$ \\
\hline 
\multicolumn{7}{l}{Planck} \\ 
 \hline 
3.0  & $0.792  \pm  0.055$  & $17.1  \pm  1.1$  & $1.16  \pm  0.15$  & $0.578  \pm  0.045$  & $1.63  \pm  0.035$  & $4.79  \pm  0.054$ \\
2.8  & $0.773  \pm  0.050$  & $19.2  \pm  1.4$  & $1.16  \pm  0.14$  & $0.608  \pm  0.040$  & $1.65  \pm  0.031$  & $4.50  \pm  0.053$ \\
2.6  & $0.781  \pm  0.044$  & $21.1  \pm  1.8$  & $1.15  \pm  0.12$  & $0.611  \pm  0.035$  & $1.64  \pm  0.027$  & $4.19  \pm  0.051$ \\
2.4  & $0.851  \pm  0.040$  & $19.5  \pm  1.9$  & $1.06  \pm  0.093$  & $0.548  \pm  0.029$  & $1.61  \pm  0.023$  & $3.63  \pm  0.031$ \\
2.2  & $0.867  \pm  0.035$  & $19.4  \pm  1.5$  & $1.06  \pm  0.081$  & $0.514  \pm  0.027$  & $1.60  \pm  0.021$  & $3.35  \pm  0.029$ \\
\hline
  \end{tabular}
 }
 \end{table}

%% file: tables/physical_table.tex
\begin{table}[!htbp]
 \centering
 \caption{Values of the bias parameters for different physical properties and for $q_2 = 0$. 
  \vspace{2 mm}}{ 
  \begin{tabular}{c|*{3}c}
 z  & $\beta$    & $b_{\tau\delta}$  &  $b_{\tau\eta}$ \\ 
  \hline 
 \hline 
    \multicolumn{4}{l}{Fiducial $(\sigma = 0.88, \log{T} = 4.3$) } \\ 
 \hline 
3.0  & $ 1.205  \pm  0.049$  & $0.5546  \pm  0.0086$  & $ 0.681  \pm  0.028$ \\
2.8  & $ 1.284  \pm  0.052$  & $0.5577  \pm  0.0083$  & $ 0.733  \pm  0.030$ \\
2.6  & $ 1.343  \pm  0.055$  & $0.5588  \pm  0.0084$  & $ 0.771  \pm  0.032$ \\
2.4  & $ 1.385  \pm  0.056$  & $0.5574  \pm  0.0080$  & $ 0.796  \pm  0.032$ \\
2.2  & $ 1.405  \pm  0.061$  & $0.5536  \pm  0.0079$  & $ 0.807  \pm  0.034$ \\
\hline 
\multicolumn{4}{l}{S0.76 ($\sigma = 0.76$)} \\ 
 \hline 
3.0  & $ 1.086  \pm  0.042$  & $0.6319  \pm  0.0096$  & $ 0.700  \pm  0.028$ \\
2.8  & $ 1.161  \pm  0.045$  & $0.6373  \pm  0.0096$  & $ 0.757  \pm  0.030$ \\
2.6  & $ 1.219  \pm  0.047$  & $0.6409  \pm  0.0098$  & $ 0.803  \pm  0.032$ \\
2.4  & $ 1.257  \pm  0.050$  & $0.6428  \pm  0.0095$  & $ 0.833  \pm  0.033$ \\
2.2  & $ 1.284  \pm  0.050$  & $0.6401  \pm  0.0094$  & $ 0.853  \pm  0.033$ \\
\hline 
\multicolumn{4}{l}{S0.64 ($\sigma = 0.64$)} \\ 
 \hline 
3.0  & $ 0.965  \pm  0.033$  & $0.7287  \pm  0.011$  & $ 0.717  \pm  0.025$ \\
2.8  & $ 1.035  \pm  0.035$  & $0.7366  \pm  0.011$  & $ 0.780  \pm  0.028$ \\
2.6  & $ 1.094  \pm  0.038$  & $0.7429  \pm  0.011$  & $ 0.835  \pm  0.030$ \\
2.4  & $ 1.137  \pm  0.039$  & $0.7478  \pm  0.011$  & $ 0.877  \pm  0.031$ \\
2.2  & $ 1.163  \pm  0.041$  & $0.7490  \pm  0.011$  & $ 0.904  \pm  0.032$ \\
\hline 
\multicolumn{4}{l}{G1.3 ($\gamma = 1.3$)} \\ 
 \hline 
3.0  & $ 1.123  \pm  0.049$  & $0.5770  \pm  0.0088$  & $ 0.661  \pm  0.029$ \\
2.8  & $ 1.216  \pm  0.051$  & $0.5790  \pm  0.0089$  & $ 0.720  \pm  0.031$ \\
2.6  & $ 1.290  \pm  0.054$  & $0.5788  \pm  0.0086$  & $ 0.767  \pm  0.032$ \\
2.4  & $ 1.348  \pm  0.056$  & $0.5764  \pm  0.0079$  & $ 0.801  \pm  0.033$ \\
2.2  & $ 1.385  \pm  0.060$  & $0.5705  \pm  0.0081$  & $ 0.820  \pm  0.035$ \\
\hline 
\multicolumn{4}{l}{G1.0 ($\gamma = 1.0$)} \\ 
 \hline 
3.0  & $ 1.090  \pm  0.050$  & $0.5925  \pm  0.0091$  & $ 0.658  \pm  0.030$ \\
2.8  & $ 1.191  \pm  0.051$  & $0.5930  \pm  0.0086$  & $ 0.723  \pm  0.031$ \\
2.6  & $ 1.275  \pm  0.055$  & $0.5912  \pm  0.0084$  & $ 0.774  \pm  0.034$ \\
2.4  & $ 1.343  \pm  0.059$  & $0.5869  \pm  0.0082$  & $ 0.813  \pm  0.035$ \\
2.2  & $ 1.391  \pm  0.060$  & $0.5794  \pm  0.0080$  & $ 0.836  \pm  0.035$ \\
\hline 
\multicolumn{4}{l}{G1T4 ($\gamma = 1.0, \log{T} = 4.0$)} \\ 
 \hline 
3.0  & $ 1.168  \pm  0.069$  & $0.5682  \pm  0.0086$  & $ 0.677  \pm  0.040$ \\
2.8  & $ 1.300  \pm  0.071$  & $0.5679  \pm  0.0083$  & $ 0.756  \pm  0.041$ \\
2.6  & $ 1.425  \pm  0.074$  & $0.5645  \pm  0.0084$  & $ 0.826  \pm  0.042$ \\
2.4  & $ 1.543  \pm  0.055$  & $0.5578  \pm  0.0050$  & $ 0.887  \pm  0.031$ \\
2.2  & $ 1.643  \pm  0.081$  & $0.5475  \pm  0.0068$  & $ 0.934  \pm  0.044$ \\
\hline 

  \end{tabular}
 }
 \end{table}

\begin{table}[!htbp]
\centering
 \caption{Values of the fitting parameters for different physical properties and for $q_2 = 0$.
  \vspace{2 mm}}
 \adjustbox{width=1.05 \textwidth}{ 
  \begin{tabular}{c|*{6}c}
z  &  $q_1$  &  $k_p\, [h/Mpc]$  &  $k_{v}^{a_v}\, [h/Mpc]^{a_v}$  &  $a_v$  &  $b_v$ & $k_{na}$ \\ 
  \hline 
  \hline 
  \multicolumn{7}{l}{Fiducial ($\sigma = 0.88, \log{T} = 4.3$) } \\ 
 \hline 
3.0  & $ 0.648  \pm   0.042$  & $  13.1  \pm     0.4$  & $ 1.007  \pm   0.103$  & $ 0.627  \pm   0.033$  & $  1.66  \pm    0.01$  & $  4.09  \pm    0.04$ \\
2.8  & $ 0.644  \pm   0.037$  & $  13.4  \pm     0.5$  & $ 0.979  \pm   0.091$  & $ 0.610  \pm   0.030$  & $  1.64  \pm    0.01$  & $  3.89  \pm    0.03$ \\
2.6  & $ 0.652  \pm   0.034$  & $  13.6  \pm     0.5$  & $ 0.970  \pm   0.084$  & $ 0.590  \pm   0.028$  & $  1.61  \pm    0.01$  & $  3.65  \pm    0.04$ \\
2.4  & $ 0.666  \pm   0.030$  & $  13.5  \pm     0.5$  & $ 0.963  \pm   0.076$  & $ 0.561  \pm   0.027$  & $  1.58  \pm    0.01$  & $  3.39  \pm    0.03$ \\
2.2  & $ 0.677  \pm   0.027$  & $  13.3  \pm     0.5$  & $ 0.961  \pm   0.070$  & $ 0.533  \pm   0.025$  & $  1.54  \pm    0.01$  & $  3.11  \pm    0.03$ \\
\hline 
\multicolumn{7}{l}{G1.3 ($\gamma = 1.3$)} \\ 
 \hline 
3.0  & $ 0.612  \pm   0.041$  & $  13.6  \pm     0.5$  & $ 1.044  \pm   0.115$  & $ 0.633  \pm   0.036$  & $  1.71  \pm    0.02$  & $  4.22  \pm    0.04$ \\
2.8  & $ 0.603  \pm   0.038$  & $  14.2  \pm     0.5$  & $ 0.996  \pm   0.103$  & $ 0.612  \pm   0.034$  & $  1.70  \pm    0.01$  & $  4.04  \pm    0.04$ \\
2.6  & $ 0.609  \pm   0.033$  & $  14.6  \pm     0.6$  & $ 0.971  \pm   0.089$  & $ 0.586  \pm   0.031$  & $  1.67  \pm    0.01$  & $  3.81  \pm    0.03$ \\
2.4  & $ 0.625  \pm   0.029$  & $  14.8  \pm     0.6$  & $ 0.951  \pm   0.080$  & $ 0.551  \pm   0.029$  & $  1.64  \pm    0.01$  & $  3.53  \pm    0.03$ \\
2.2  & $ 0.641  \pm   0.026$  & $  14.7  \pm     0.7$  & $ 0.937  \pm   0.071$  & $ 0.514  \pm   0.026$  & $  1.60  \pm    0.01$  & $  3.23  \pm    0.01$ \\
\hline 
\multicolumn{7}{l}{G1.0 ($\gamma = 1.0$) }  \\ 
 \hline 
3.0  & $ 0.611  \pm   0.042$  & $  13.6  \pm     0.5$  & $ 1.034  \pm   0.114$  & $ 0.614  \pm   0.036$  & $  1.72  \pm    0.02$  & $  4.21  \pm    0.04$ \\
2.8  & $ 0.600  \pm   0.037$  & $  14.2  \pm     0.5$  & $ 0.980  \pm   0.100$  & $ 0.593  \pm   0.034$  & $  1.71  \pm    0.02$  & $  4.04  \pm    0.04$ \\
2.6  & $ 0.605  \pm   0.033$  & $  14.7  \pm     0.6$  & $ 0.949  \pm   0.090$  & $ 0.565  \pm   0.032$  & $  1.68  \pm    0.01$  & $  3.82  \pm    0.03$ \\
2.4  & $ 0.621  \pm   0.029$  & $  14.9  \pm     0.6$  & $ 0.925  \pm   0.079$  & $ 0.528  \pm   0.030$  & $  1.65  \pm    0.01$  & $  3.55  \pm    0.02$ \\
2.2  & $ 0.638  \pm   0.027$  & $  14.9  \pm     0.7$  & $ 0.909  \pm   0.071$  & $ 0.490  \pm   0.027$  & $  1.61  \pm    0.01$  & $  3.25  \pm    0.03$ \\
\hline 
\multicolumn{7}{l}{G1T4 ($\gamma = 1.0, \log{T} = 4.0$) }\\ 
 \hline 
3.0  & $ 0.620  \pm   0.043$  & $  14.4  \pm     0.5$  & $ 0.711  \pm   0.087$  & $ 0.318  \pm   0.039$  & $  1.60  \pm    0.02$  & $  4.66  \pm    0.06$ \\
2.8  & $ 0.594  \pm   0.039$  & $  15.5  \pm     0.7$  & $ 0.683  \pm   0.079$  & $ 0.317  \pm   0.037$  & $  1.60  \pm    0.02$  & $  4.53  \pm    0.06$ \\
2.6  & $ 0.591  \pm   0.035$  & $  16.5  \pm     0.9$  & $ 0.669  \pm   0.070$  & $ 0.306  \pm   0.034$  & $  1.59  \pm    0.02$  & $  4.30  \pm    0.04$ \\
2.4  & $ 0.604  \pm   0.020$  & $  17.1  \pm     0.9$  & $ 0.658  \pm   0.044$  & $ 0.282  \pm   0.020$  & $  1.57  \pm    0.02$  & $  3.98  \pm    0.04$ \\
2.2  & $ 0.621  \pm   0.025$  & $  17.4  \pm     1.1$  & $ 0.651  \pm   0.058$  & $ 0.254  \pm   0.028$  & $  1.54  \pm    0.02$  & $  3.61  \pm    0.02$ \\
\hline 
\multicolumn{7}{l}{S0.76 ($\sigma = 0.76$)}  \\ 
 \hline 
3.0  & $ 0.762  \pm   0.055$  & $  13.7  \pm     0.6$  & $ 1.383  \pm   0.173$  & $ 0.788  \pm   0.041$  & $  1.66  \pm    0.03$  & $  4.58  \pm    0.05$ \\
2.8  & $ 0.758  \pm   0.050$  & $  14.1  \pm     0.7$  & $ 1.314  \pm   0.139$  & $ 0.763  \pm   0.035$  & $  1.64  \pm    0.03$  & $  4.37  \pm    0.04$ \\
2.6  & $ 0.767  \pm   0.045$  & $  14.4  \pm     0.6$  & $ 1.270  \pm   0.122$  & $ 0.734  \pm   0.032$  & $  1.61  \pm    0.02$  & $  4.12  \pm    0.04$ \\
2.4  & $ 0.783  \pm   0.040$  & $  14.6  \pm     0.7$  & $ 1.242  \pm   0.109$  & $ 0.700  \pm   0.031$  & $  1.58  \pm    0.02$  & $  3.83  \pm    0.04$ \\
2.2  & $ 0.804  \pm   0.035$  & $  14.5  \pm     0.6$  & $ 1.211  \pm   0.096$  & $ 0.659  \pm   0.028$  & $  1.55  \pm    0.02$  & $  3.52  \pm    0.03$ \\
\hline 
\multicolumn{7}{l}{S0.64 ($\sigma = 0.64$)}  \\ 
 \hline 
3.0  & $ 0.958  \pm   0.074$  & $  13.6  \pm     0.6$  & $ 2.191  \pm   0.277$  & $ 0.994  \pm   0.044$  & $  1.63  \pm    0.03$  & $  5.17  \pm    0.05$ \\
2.8  & $ 0.949  \pm   0.069$  & $  14.2  \pm     0.5$  & $ 2.015  \pm   0.238$  & $ 0.958  \pm   0.041$  & $  1.61  \pm    0.02$  & $  4.93  \pm    0.04$ \\
2.6  & $ 0.958  \pm   0.061$  & $  14.7  \pm     0.6$  & $ 1.879  \pm   0.203$  & $ 0.915  \pm   0.038$  & $  1.59  \pm    0.03$  & $  4.67  \pm    0.04$ \\
2.4  & $ 0.979  \pm   0.054$  & $  15.1  \pm     0.8$  & $ 1.772  \pm   0.172$  & $ 0.868  \pm   0.035$  & $  1.56  \pm    0.02$  & $  4.37  \pm    0.04$ \\
2.2  & $ 1.004  \pm   0.051$  & $  15.3  \pm     0.8$  & $ 1.677  \pm   0.151$  & $ 0.818  \pm   0.032$  & $  1.53  \pm    0.02$  & $  4.03  \pm    0.04$ \\
\hline 
  \end{tabular}
 }
 \end{table}

%% file: tables/flx_table.tex
\begin{table}[!htbp]
 \centering
  \setlength\extrarowheight{2.5pt}
 \caption{Bias parameters when varying $\bar{F}$ and $z$ independently. 
  \vspace{2 mm}}{ 
  \begin{tabular}{c|*{3}c}
 z, $\bar{F}$  & $\beta$    & $b_{\tau\delta}$  &  $b_{\tau\eta}$ \\ 
  \hline 
  \hline 
  \multicolumn{4}{l}{Fiducial} \\ 
 \hline 
3.0  & $ 1.205  \pm   0.049$  & $0.5546  \pm  0.0086$  & $ 0.681  \pm   0.028$ \\
2.8  & $ 1.284  \pm   0.052$  & $0.5577  \pm  0.0083$  & $ 0.733  \pm   0.030$ \\
2.6  & $ 1.343  \pm   0.055$  & $0.5588  \pm  0.0084$  & $ 0.771  \pm   0.032$ \\
2.4  & $ 1.385  \pm   0.056$  & $0.5574  \pm  0.0080$  & $ 0.796  \pm   0.032$ \\
2.2  & $ 1.405  \pm   0.061$  & $0.5536  \pm  0.0079$  & $ 0.807  \pm   0.034$ \\
\hline 
\multicolumn{4}{l}{\small{$\bar{F}$ fixed to 0.781} } \\ 
 \hline 
3.0  & $ 1.274  \pm   0.048$  & $0.6038  \pm  0.0061$  & $ 0.790  \pm   0.021$ \\
2.8  & $ 1.310  \pm   0.052$  & $0.5813  \pm  0.0068$  & $ 0.782  \pm   0.026$ \\
2.6  & $ 1.344  \pm   0.056$  & $0.5588  \pm  0.0077$  & $ 0.721  \pm   0.031$ \\
2.4  & $ 1.383  \pm   0.060$  & $0.5356  \pm  0.0093$  & $ 0.768  \pm   0.039$ \\
2.2  & $ 1.429  \pm   0.065$  & $0.5121  \pm  0.0110$  & $ 0.751  \pm   0.052$ \\
\hline 
\multicolumn{4}{l}{$z$ fixed to 2.6} \\ 
 \hline 
0.696  & $ 1.275  \pm   0.057$  & $0.517  \pm  0.008$  & $0.676  \pm  0.030$ \\
0.740  & $ 1.318  \pm   0.058$  & $0.536  \pm  0.008$  & $0.640  \pm  0.036$ \\
0.781  & $ 1.343  \pm   0.057$  & $0.558  \pm  0.008$  & $0.769  \pm  0.032$ \\
0.818  & $ 1.347  \pm   0.054$  & $0.581  \pm  0.008$  & $0.804  \pm  0.032$ \\
0.852  & $ 1.332  \pm   0.069$  & $0.609  \pm  0.008$  & $0.834  \pm  0.032$ \\
\hline 
  \hline 
  \end{tabular}
 }
 \end{table}
 
 \begin{table}[!htbp]
 \centering
  \setlength\extrarowheight{2.5pt}
 \caption{Non-linear fit parameters when varying $\bar{F}$ and $z$ independently. 
  \vspace{2 mm}}
   \adjustbox{width=1.0 \textwidth}{
  \begin{tabular}{c|*{6}c}
 z  &  $q_{1}$  &  $k_p\,[h/Mpc]$ &  $k_{v}^{a_v}\,[h/Mpc]^{a_v}$  &  $a_v$  &  $b_v$ &  $k_{na}$ \\ 
  \hline 
  \hline 
  \multicolumn{7}{l}{Fiducial} \\ 
 \hline 
3.0  & $ 0.648  \pm   0.042$  & $  13.1  \pm     0.4$  & $ 1.007  \pm   0.103$  & $ 0.627  \pm   0.033$  & $  1.66  \pm    0.01$  & $  4.09  \pm    0.04$ \\
2.8  & $ 0.644  \pm   0.037$  & $  13.4  \pm     0.5$  & $ 0.979  \pm   0.091$  & $ 0.610  \pm   0.030$  & $  1.64  \pm    0.01$  & $  3.89  \pm    0.03$ \\
2.6  & $ 0.652  \pm   0.034$  & $  13.6  \pm     0.5$  & $ 0.970  \pm   0.084$  & $ 0.590  \pm   0.028$  & $  1.61  \pm    0.01$  & $  3.65  \pm    0.04$ \\
2.4  & $ 0.666  \pm   0.030$  & $  13.5  \pm     0.5$  & $ 0.963  \pm   0.076$  & $ 0.561  \pm   0.027$  & $  1.58  \pm    0.01$  & $  3.39  \pm    0.03$ \\
2.2  & $ 0.677  \pm   0.027$  & $  13.3  \pm     0.5$  & $ 0.961  \pm   0.070$  & $ 0.533  \pm   0.025$  & $  1.54  \pm    0.01$  & $  3.11  \pm    0.03$ \\
\hline 
\multicolumn{7}{l}{$\bar{F}$ fixed to 0.781} \\ 
 \hline 
3.0  & $ 0.710  \pm   0.041$  & $  16.0  \pm     0.8$  & $ 1.195  \pm   0.114$  & $ 0.703  \pm   0.033$  & $  1.66  \pm    0.02$  & $  4.12  \pm    0.04$ \\
2.8  & $ 0.683  \pm   0.036$  & $  14.7  \pm     0.7$  & $ 1.070  \pm   0.093$  & $ 0.644  \pm   0.029$  & $  1.64  \pm    0.02$  & $  3.88  \pm    0.04$ \\
2.6  & $ 0.652  \pm   0.033$  & $  13.6  \pm     0.5$  & $ 0.970  \pm   0.083$  & $ 0.590  \pm   0.029$  & $  1.61  \pm    0.02$  & $  3.65  \pm    0.03$ \\
2.4  & $ 0.620  \pm   0.030$  & $  12.5  \pm     0.5$  & $ 0.876  \pm   0.071$  & $ 0.530  \pm   0.027$  & $  1.58  \pm    0.02$  & $  3.43  \pm    0.03$ \\
2.2  & $ 0.585  \pm   0.028$  & $  11.6  \pm     0.4$  & $ 0.793  \pm   0.064$  & $ 0.473  \pm   0.026$  & $  1.56  \pm    0.02$  & $  3.22  \pm    0.04$ \\
\hline 
\multicolumn{7}{l}{$z$ fixed to 2.6} \\ 
 \hline 
0.696  & $ 0.585  \pm   0.034$  & $  11.7  \pm     0.3$  & $ 0.811  \pm   0.074$  & $ 0.517  \pm   0.030$  & $  1.62  \pm    0.02$  & $  3.61  \pm    0.04$ \\
0.740  & $ 0.614  \pm   0.034$  & $  12.5  \pm     0.4$  & $ 0.885  \pm   0.080$  & $ 0.555  \pm   0.029$  & $  1.61  \pm    0.02$  & $  3.65  \pm    0.04$ \\
0.781  & $ 0.652  \pm   0.034$  & $  13.6  \pm     0.5$  & $ 0.969  \pm   0.085$  & $ 0.590  \pm   0.029$  & $  1.61  \pm    0.02$  & $  3.65  \pm    0.03$ \\
0.818  & $ 0.698  \pm   0.034$  & $  14.8  \pm     0.7$  & $ 1.063  \pm   0.090$  & $ 0.620  \pm   0.029$  & $  1.61  \pm    0.02$  & $  3.60  \pm    0.03$ \\
0.852  & $ 0.751  \pm   0.044$  & $  16.4  \pm     1.1$  & $ 1.160  \pm   0.118$  & $ 0.646  \pm   0.035$  & $  1.60  \pm    0.02$  & $  3.52  \pm    0.03$ \\ 
\hline 
  \end{tabular}
 }
 \end{table}